\documentclass[11pt,a4paper,preprintnumbers]{article}
\pdfoutput=1
\usepackage{jheppub, hyperref, cleveref}
\usepackage{amsfonts, amsmath, amssymb, latexsym}
\usepackage{graphicx, caption, subcaption}
\usepackage[normalem]{ulem}
\usepackage[dvipsnames]{xcolor}
\usepackage{multirow}
\usepackage{arydshln}

\newcommand{\green}

\title{Gravitational waves from cosmic strings in LISA: reconstruction pipeline and physics interpretation}

\author[a,b,c]{Jose J.\,Blanco-Pillado,}
\affiliation[a\,]{IKERBASQUE, Basque Foundation for Science, 48011, Bilbao, Spain}
\affiliation[b\,]{EHU Quantum Center, University of the Basque Country, UPV/EHU, Bilbao, Spain} 
\affiliation[c\,]{Department of Theoretical Physics, UPV/EHU, 48080, Bilbao, Spain}

\author[d]{Yanou Cui,}
\affiliation[d\,]{Department of Physics and Astronomy, University of California, Riverside, CA 92521, USA}

\author[e,f]{Sachiko Kuroyanagi,}
\affiliation[e\,]{Instituto de F\'isica Te\'orica UAM-CSIC, Universidad Aut\'onoma de Madrid, Cantoblanco 28049 Madrid, Spain}
\affiliation[f\,]{Department of Physics, Nagoya University, Chikusa, Nagoya 464-8602, Japan}

\author[g,1]{Marek Lewicki,
 \note{Project coordinator: \href{mailto:marek.lewicki@fuw.edu.pl}{marek.lewicki@fuw.edu.pl}}
}
\affiliation[g\,]{Faculty of Physics, University of Warsaw, ul. Pasteura 5, 02-093 Warsaw, Poland}

\author[h,2]{Germano Nardini,
 \note{Project coordinator: \href{mailto:germano.nardini@uis.no}{germano.nardini@uis.no}}
}
\affiliation[h\,]{Department of Mathematics and Physics, University of Stavanger, NO-4036 Stavanger, Norway}

\author[i]{Mauro Pieroni,\note{Corresponding author: \href{mailto:mauro.pieroni@cern.ch}{mauro.pieroni@cern.ch}}
}
\affiliation[i\,]{Department of Theoretical Physics, CERN, 1211 Geneva 23, Switzerland}

\author[j,k,l]{Ivan Yu. Rybak,}
\affiliation[j]{CAPA \& Departamento de F\'{\i}sica Te\'{o}rica, Universidad de Zaragoza, Pedro Cerbuna, 12, 50009 Zaragoza, Spain}{}

\author[k,l]{Lara Sousa,\note{Corresponding author: \href{mailto:lara.sousa@astro.up.pt}{lara.sousa@astro.up.pt}}
}
\affiliation[k\,]{Instituto de Astrof\'{\i}sica e Ci\^encias do Espa{\c c}o, Universidade do Porto, CAUP, Rua das Estrelas, PT4150-762 Porto, Portugal}
\affiliation[l\,]{Centro de Astrof\'{\i}sica da Universidade do Porto, Rua das Estrelas, PT4150-762 Porto, Portugal}
\affiliation{Departamento de F\'{\i}sica e Astronomia, Faculdade de Ci\^encias, Universidade do Porto, Rua do Campo Alegre 687, PT4169-007 Porto, Portugal}

\author[\,m]{Jeremy M.~Wachter\note{Corresponding author: \href{mailto:wachterj@wit.edu}{wachterj@wit.edu}}}
\affiliation[m\,]{School of Sciences \& Humanities, Wentworth Institute of Technology, Boston, MA 02115}

\author[]{\\ \centering \texttt{(For the LISA Cosmology Working Group)}}

\abstract{
We initiate the LISA template databank for stochastic gravitational wave backgrounds sourced by cosmic strings. We include two templates, an analytical template, which enables more flexible searches,
and a numerical template derived directly from large Nambu-Goto simulations of string networks. Using searches based on these templates, we forecast the parameter space within the reach of the experiment and the precision with which their parameters will be reconstructed, provided a signal is observed. The reconstruction permits probing the Hubble expansion and new relativistic DoF in the early universe.
We quantify the impact that astrophysical foregrounds can have on these searches. Finally, we discuss the impact that these observations would have on our understanding of the fundamental models behind the string networks.
Overall, we prove that LISA has great potential for probing cosmic string models and may reach tensions as low as $G\mu =10^{-16} - 10^{-17} $, which translates into energy scales of the order $10^{11}~\text{GeV}$.
}

\begin{document}

\begin{figure}
\begin{flushright}
\href{https://lisa.pages.in2p3.fr/consortium-userguide/wg_cosmo.html}{\includegraphics[width = 0.2 \textwidth]{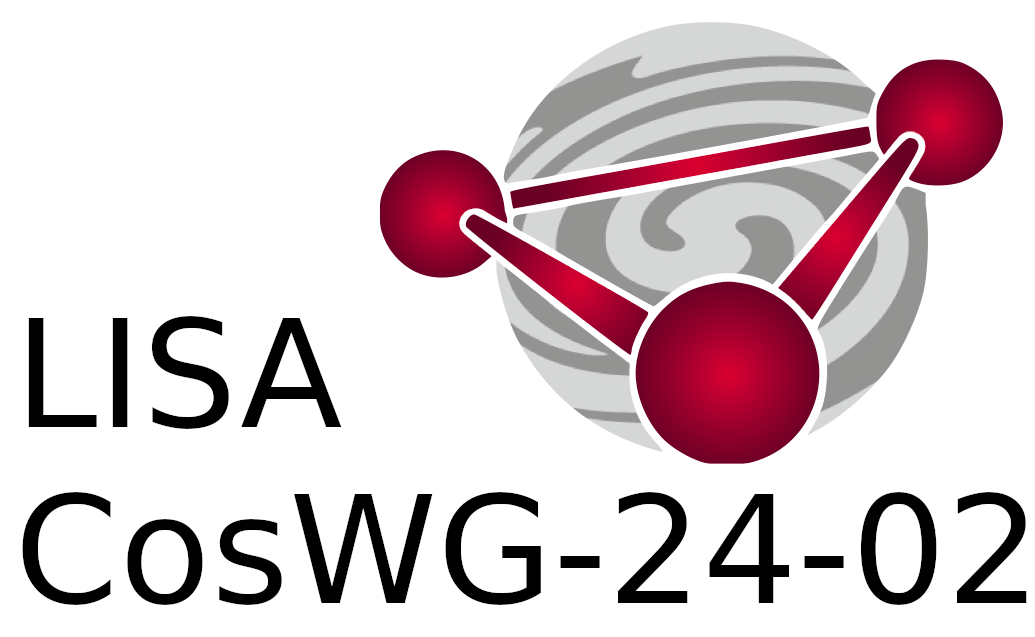}}\\[5mm]
\end{flushright}
\end{figure}

\preprint{CERN-TH-2024-085}

\maketitle

\section{Introduction}
\label{sec:intro}

We are currently witnessing a new era of rapid development in astrophysics and cosmology thanks to Gravitational Wave (GW) observations by the LIGO/Virgo/Kagra collaboration~\cite{Abbott:2016blz,Abbott:2016nmj,Abbott:2017vtc,Abbott:2017gyy,Abbott:2017oio,LIGOScientific:2018mvr,LIGOScientific:2020ibl,LIGOScientific:2021usb,LIGOScientific:2021djp}. Moreover, recent observations of multiple Pulsar Timing Array (PTA) observatories have shown compelling evidence for the presence of a Stochastic Gravitational Wave Background (SGWB) in the nanohertz frequency band~\cite{NANOGrav:2023gor,EPTA:2023sfo,Reardon:2023gzh,Xu:2023wog}. While all observations to date point towards compact-object binaries as a likely origin for these signals, they also clearly prove we have a new messenger allowing for observations of processes from the early universe~\cite{LISACosmologyWorkingGroup:2022jok, NANOGrav:2023hvm, EPTA:2023xxk}.

Symmetry-breaking phase transitions in the early universe may lead to the formation of stable vortex-like configurations
known as cosmic strings~\cite{Kibble:1976sj} (for a review on cosmic strings see, for example refs.~\cite{Vilenkin:2000jqa,Hindmarsh:1994re}). These strings are expected to survive throughout cosmic history and their interactions should lead to the copious production of closed loops that decay by emitting gravitational radiation \cite{Vilenkin:1981bx}. The superposition of these uncorrelated contributions of GWs from loops is expected to give rise to an SGWB. In this paper, we provide templates for the SGWB generated by the loops created throughout the evolution of a cosmic string network. The energy per unit length of the cosmic strings $\mu$~\footnote{In the simplest string models, $\mu$ coincides with the string tension.}, which parameterizes their gravitational interactions, is determined by the characteristic energy scale of the string-forming phase transition $\eta$:
\begin{equation}
    G\mu \sim 10^{-6}\left(\frac{\eta}{10^{16} {\rm GeV}}\right)^2\,,
\end{equation}
where $G=1/M_p^2$ is the gravitational constant and $M_p=1.22\times 10^{19} {\rm GeV}$ is the Planck mass. Here, we study the ability of LISA to reconstruct the dimensionless parameter $G\mu$ with these templates.

In this paper, we will focus on models in which loops decay by emitting GWs only, corresponding to the maximal SGWB. This corresponds to assuming that cosmic strings may be well described by the Nambu-Goto action or, in order words, that cosmic strings have no internal degrees of freedom and their thickness $\delta\sim 1/\eta$ is much smaller than their typical curvature length. Field theory cosmic strings, however, may also decay into excitations of their constituent fields, which could result in a weakening of their GW signals. For Abelian-Higgs strings --- local strings that are generally expected to be well described by the Nambu-Goto action --- the potential role of these other decay mechanisms is still a matter of debate~\cite{Vachaspati:1984yi, Vincent:1997cx, Moore:1998gp, Olum:1998ag, Olum:1999sg, Moore:2001px, Hindmarsh:2008dw, Hindmarsh:2017qff, Matsunami:2019fss, Hindmarsh:2021mnl, Blanco-Pillado:2022rad, Hindmarsh:2022awe, Blanco-Pillado:2023sap}. For the majority of this work, we will consider only energy loss into gravitational waves, consistent with prior work within the LISA Consortium~\cite{Auclair:2019wcv}. We only relax this assumption in Sec.~\ref{sec:fng_model} where we consider situations in which only a part of the energy of loops goes into GWs, with the remainder being lost into other particle radiation.

Our analysis, however, does not apply to other more complex cosmic string models, such as
cosmic superstrings~\cite{Sarangi:2002yt, Dvali:2003zj, Copeland:2003bj}, metastable cosmic strings~\cite{Preskill:1992ck,Leblond:2009fq, Buchmuller:2020lbh, Buchmuller:2023aus,Chitose:2023dam}, 
superconducting strings~\cite{Witten:1984eb, Nielsen:1987fy}, or models whose energy is not localized, such as global strings~\cite{Vilenkin:1982ks}. Superconducting string loops, for instance, carry current and are then expected to emit significant electromagnetic radiation~\cite{Vilenkin:1986zz, Blanco-Pillado:2000nbp} and this current was shown to lead to a suppression of the gravitational radiation power emitted too~\cite{Babichev:2003au, Rybak:2022sbo}. For global or axion strings, although an SGWB is inevitable and potentially detectable~\cite{Chang:2019mza, Chang:2021afa, Gorghetto:2021fsn, Figueroa:2020lvo}, the emission of GWs by loops is, in general, subdominant to the emission of Goldstone bosons/axions which would result in significant suppression of the amplitude of this background~\cite{Vilenkin:1986ku, Drew:2019mzc, Drew:2022iqz, Baeza-Ballesteros:2023say, Drew:2023ptp}. 

The forthcoming LISA mission~\cite{LISA:2017pwj, Colpi:2024xhw} will open a new window for observations of GWs. Its adoption on 25 January 2024 signals that construction of the experiment will commence. With the performance established, the scientific community can prepare the data analysis pipelines for signal searches and their scientific interpretation. 
LISA will be a signal-dominated experiment which means that at any time, its data stream will contain instrumental noise, unresolved astrophysical sources (collectively treated as stochastic foregrounds) as well as numerous transient events that overlap in time and are recorded in the data. This makes searches for any primordial signal in the form of an SGWB much more complicated.
Definitive conclusions on the LISA capabilities for the detection and reconstruction of the SGWB will require data analysis pipelines for all potential sources. In particular, quantifying such capabilities requires a global-fit pipeline \cite{Vallisneri:2008ye, MockLISADataChallengeTaskForce:2009wir} exploiting all possible means to isolate the primordial SGWB from the other sources, e.g., discriminators based on the signals' statistical properties, anisotropic characteristics, etcetera. So far, feasibility studies have proven that the global fit is achievable, but such analyses only marginally tackle the subtleties that may arise with a primordial SGWB~\cite{Littenberg:2023xpl}. The plan is to achieve the whole global-fit pipeline at the prototypical stage in around four years~\cite{Colpi:2024xhw}. The output of the present study constitutes a primitive SGWB module that, once linked to the other modules dedicated to other LISA sources, allows the global fit to deal with a cosmic-string SGWB properly. There is, however, no doubt that the SGWB module (and its template libraries) presented in this paper are just some of the first steps toward the final objective.

The first goal of this project is to initiate a LISA template databank of SGWBs associated with cosmic strings.
We begin with two such templates for an analytical model~\cite{Sousa:2013aaa,Sousa:2020sxs} and a model-inferred from Nambu-Goto simulations~\cite{Blanco-Pillado:2013qja}. For both templates, we have built a dedicated LISA SGWB data analysis pipeline reconstructing the SGWB signal as well as the expected astrophysical foregrounds and LISA instrumental noise.
Our second goal is to perform the reconstruction of the parameters of our templates using the new pipeline in order to forecast the precision LISA will be capable of in this task. We also verify what part of the parameter space of our models will be within reach of the experiment. This is particularly important for the theory community, as it provides maps that the community can use to focus their efforts on improving the theoretical predictions on areas that will come under experimental scrutiny.

Transient gravitational waves are also of significant interest theoretically and experimentally. LISA will be sensitive to many individual sources, including possible ``bursts'' from cosmic strings~\cite{LISAConsortiumWaveformWorkingGroup:2023arg}. We will not consider such bursts in this paper, as our focus is on SGWB.

As mentioned earlier, PTA observatories have shown evidence of an SGWB in the nanohertz range. While the most likely culprit is the spectrum produced by supermassive black holes~\cite{Phinney:2001di,NANOGrav:2023hfp,Ellis:2023owy}, a possible connection of PTA observations to cosmic strings has been pointed out earlier in the literature ~\cite{Ellis:2020ena,Blasi:2020mfx,Blanco-Pillado:2021ygr} and has recently been considered in the analysis of the new  data~\cite{NANOGrav:2023hvm, EPTA:2023hof, Ellis:2023tsl, Figueroa:2023zhu, Servant:2023mwt, Ellis:2023oxs, Kume:2024adn}. Even if these observations are ultimately confirmed to be due to causes other than cosmic strings, they will somewhat limit the cosmic-string detection prospects in the LISA band. On the other hand, the confirmation of a cosmic-string-induced SGWB in the nanohertz band would help set important priors for the LISA searches. Given the current interpretation uncertainties, we will not use the low-frequency information in our pipeline.

This paper is laid out as follows. In section~\ref{sec:models}, we review SGWB production by cosmic strings and discuss the two models used in this study. The reader familiar with cosmic string SGWB and the models used (Models I and II of ref.~\cite{Auclair:2019wcv}) can likely skip much of this section, but may wish to stop at section~\ref{subsec:modelI}, where we introduce a novel way to include all relevant emission modes in the computation of the template for Model I, and in section~\ref{subsec:modelI-II-nonstandard}, where we discuss the effects of non-standard cosmological histories on the SGWB. In section~\ref{sec:SGWBinner}, we go into further details about how we implement the cosmic-string SGWB search and reconstruction in the 
\texttt{SGWBinner}.
This involves reconstructing signals using templates and incorporating elements such as the LISA noise curve and GWs from other sources. In section~\ref{sec:results}, we use 
\texttt{SGWBinner} to forecast LISA's ability
to reconstruct the cosmic string SGWB for both models under various circumstances, most notably in the presence of astrophysical foregrounds from binary systems. In section \ref{sec:science-interpretation}, we comment on the physics interpretation of the potential results obtained by the LISA observatory. We conclude in section~\ref{sec:conclusions}.
We finally devote \cref{sec:app_modelI} and \cref{sec:app-prebbn} to some original, technical results on the templates investigated in our analysis.

\section{SGWB templates from cosmic strings}
\label{sec:models}

In this section, we discuss the templates we use in our analysis. We start with a review of the computation of the SGWB generated by cosmic strings. We then discuss the network modelling and the resulting loop population. We include a semi-analytical model as well as a model derived directly from numerical Nambu-Goto simulations of cosmic string networks. Finally, we discuss the modification of these templates in the case that the early universe undergoes a non-standard expansion.

\subsection{Review of SGWB production by cosmic strings}

Following the string-forming phase transition, the universe is filled with a cosmic string network. The shape and amplitude of the SGWB it generates is determined by the properties of the network and so computing this spectrum requires a characterization of these properties. There are two kinds of string in the network: closed loops of cosmic string, and long strings which span the observable universe. We will neglect the GWs produced by long strings, as their contribution to the SGWB is expected to be subdominant~\cite{Vilenkin:2000jqa,Figueroa:2012kw,Matsui:2019obe,daCunha:2021wyy,CamargoNevesdaCunha:2022mvg}, and focus on the spectrum generated by the population of cosmic string loops. However, long strings remain important to model the SGWB, as it is their self-intersections that lead to the continuous production of new loops.

Because we focus on closed, oscillating cosmic string loops, gravitational radiation will be emitted at discrete harmonic modes determined by their length $l$. The spectral energy density in GWs generated by cosmic strings, in units of the critical energy density of the universe, is given by (see, e.g., refs.~\cite{Blanco-Pillado:2013qja,Sousa:2013aaa,Auclair:2019wcv})
\begin{equation}\label{eqn:Om-gw-general}
    \Omega_{\rm gw}(f) = \frac{8\pi }{3}\left(\frac{G\mu}{H_0}\right)^2 f\sum_{j=1}^\infty P_j \mathrm{\Omega}_{\rm gw}^j(f)\;,
\end{equation}
where $H_0$ is the value of the Hubble parameter at the present time $t_0$ and the sum is taken over harmonic modes of emission $j$. Here, $P_j$ is the spectrum of emission of a typical cosmic string loop, which describes how their gravitational radiation is distributed in the different harmonic modes.
The power spectrum may either be measured in numerical simulations~\cite{Blanco-Pillado:2017oxo} or modelled analytically~\cite{Vachaspati:1984gt,Burden:1985md,Allen:1991bk} as a  power law of the form
\begin{equation}
    P_j = \frac{\Gamma}{\zeta(q)} j^{-q}\,,
    \label{eq:powerspectrum}
\end{equation}
where $\Gamma=\sum_j P_j$ is the total power emitted by the loop in units of $G\mu^2$, $\zeta$ is the Riemann-zeta function, and the spectral index $q$ depends on what gravitational-wave emission process dominates. Typical values taken are $q=4/3$, when the emission is dominated by cusps (points on the loop that move instantaneously at the speed of light); $q=5/3$, if it is dominated by discontinuities on the loop tangent known as kinks; or $q=2$, if it is dominated by kink--kink collisions~\cite{Vachaspati:1984gt,Binetruy:2009vt}. Here, we will take $\Gamma\sim 50$, as suggested by numerous analytical and numerical studies~\cite{Vachaspati:1984gt,Burden:1985md,Scherrer:1990pj,Allen:1991bk,Blanco-Pillado:2017oxo}. Note however that a more precise characterization of $P_j$ would require the knowledge of the effect of gravitational backreaction on the shape of the strings and on their radiation spectrum. Although some progress was made recently on this subject~\cite{Wachter:2016hgi,Wachter:2016rwc,Blanco-Pillado:2018ael,Chernoff:2018evo,Blanco-Pillado:2019nto} (see also ref.~\cite{Quashnock:1990wv}), the effects of backreaction remain one of the highest sources of uncertainty in the computation of the SGWB generated by Nambu-Goto cosmic strings.

The quantity $\mathrm{\Omega}_{\rm gw}^j$ is a coefficient characterizing the contribution of the $j$-th harmonic mode of emission to the SGWB, which we define as

\begin{equation}\label{eq:omegaj}
    \mathrm{\Omega}_{\rm gw}^j(f)=\frac{2j}{f^2}\int_{t_i}^{t_0} \mathbf{n}\left(\frac{2j}{f}\frac{a(t)}{a_0},t\right)\left(\frac{a(t)}{a_0}\right)^5 dt\,,
\end{equation}
where $t$ is the physical time, $a$ is the cosmological scale factor, $t_i\sim t_{p}/(G\mu)^2$ represents the instant of 
time at which significant GW emission by cosmic string loops starts, and $t_{p}$ is the Planck time. 
As it is manifest in this expression, besides the power spectrum of emission of cosmic string loops, there are two other crucial ingredients in the computation of the SGWB generated by cosmic string loop. The first is the loop number density, $\mathbf{n}(l,t)$, describing the number density of loops with lengths between $l$ and $l+dl$ that exist at a time $t$.\footnote{Often written as $\mathbf{n}(x)$ with $x=l/t$.} We will consider three contributors to the loop number density: loops created in the radiation era and emitting in the radiation era, $\mathbf{n}_{r}$; loops created in the radiation era and emitting in the matter era, $\mathbf{n}_{rm}$; and loops created in the matter era and emitting in the matter era, $\mathbf{n}_m$. Thus, the total SGWB is the superposition of contributions from these three populations:
\begin{equation}
\mathrm{\Omega}_{\rm gw}(f)=\mathrm{\Omega}_{\rm gw}^r(f)+\mathrm{\Omega}_{\rm gw}^{rm}(f)+\mathrm{\Omega}_{\rm gw}^m(f)\,.
\label{approx}
\end{equation}
The second necessary ingredient is the expansion history of the universe. Since cosmic strings are produced at early times and persist until the present time, the model of cosmic expansion chosen influences the predicted SGWB. Following ref.~\cite{Auclair:2019wcv} (see section~2.3 of that work), we assume a standard flat $\Lambda$CDM model and use the Planck-2015 fiducial parameters~\cite{Planck:2015fie} 
to set the dimensionless Hubble constant and density parameters. 
The post-inflationary evolution of the Hubble parameter in this scenario is encapsulated in the first Friedmann equation 
describing the expansion rate of the scale factor $a(t)$:
\begin{equation}\label{eq:friedmannI}
H^2 \equiv \left( \frac{\dot{a}}{a} \right)^2 = 
H_0^2\left[ 
\Delta_{r}(a)\,\Omega_r\left(\frac{a}{a_0}\right)^{-4}
+ {\Omega_m}\left(\frac{a}{a_0}\right)^{-3} 
+\Omega_\Lambda \right] \ ,
\end{equation} 
where $H_0 =100 h\,\rm{km/s/Mpc} \simeq 1.44\times 10^{-42} \ {\rm GeV} $ 
is the expansion rate measured today, $h=0.678$,
$\Omega_r\simeq 9.2\times 10^{-5}$ for radiation, 
$\Omega_m \simeq 0.31$ for matter, 
and  $\Omega_\Lambda\simeq 0.69$ for dark energy.
The correction factor
\begin{equation}
    \label{eq:deltarhoR}
\Delta_{r}(a)=\frac{g_*(a)}{g_*(a_0)}\left(\frac{g_{*S}(a_0)}{g_{*S}(a)}\right)^{4/3}
\end{equation} 
addresses the deviation from the naive $T\propto a^{-1}$ relation dictated by
entropy conservation, and represents the dependence on the effective number of
relativistic degrees of freedom (DoF): $g_*$ for energy and $g_{*S}$ for entropy.
Here we will adopt the parametrization of the time evolution of $g_*$ and $g_{*S}$ in the Standard Model (SM) of particle physics using the Gondolo--Gelmini equation of state~\cite{GONDOLO1991145} as implemented in \texttt{micrOMEGAs5.0}~\cite{Belanger:2018ccd}. While it is possible to make other choices for the various inputs to the expansion history while remaining within ``standard'' cosmology, these choices have only a minor impact on the SGWB generated by cosmic string; see, however, sections~\ref{subsec:modelI-II-nonstandard}, \ref{ssec:mod-pre-bbn} for the effects of more significant changes to this standard history.

The following two subsections outline two commonly-used approaches to constructing a string GWB template: one based on analytical modeling, and one based on simulation-inferred modeling. These models agree closely in their predictions for the shape and amplitude of resulting string GWBs, and thus in their predictions on LISA's ability to detect or constrain strings. There is also a simulation-inferred model of the loop number density which has been included in LISA Consortium work (see Refs.~\cite{Auclair:2019wcv,LISACosmologyWorkingGroup:2022jok}), but which we do not consider here; we direct the reader to Refs.~\cite{Ringeval:2005kr, Lorenz:2010sm, Auclair:2019zoz, Auclair:2020oww,Blanco-Pillado:2019vcs,Blanco-Pillado:2019tbi,Auclair:2021jud} for details on this model. 
That model predicts a very large population of small loops that can lead to significantly different predictions for the GWB than in the models we consider here~\cite{Ringeval:2017eww, LIGOScientific:2017ikf}. Due to this additional loop population, LVK analysis~\cite{LIGOScientific:2021nrg} resulted in stringent constraints on cosmic string tension in this scenario, limiting it to $G\mu \lesssim 4\times10^{-15}$, and consequently LISA is only expected to allow us to probe this model in a narrow range of tensions of $G\mu \sim \mathcal{O}(10^{-15})-\mathcal{O}(10^{-16})$~\cite{Boileau:2021gbr}. In this range of tensions, as was shown in~\cite{Auclair:2019wcv}, the effects of this additional loop population are not significant in the LISA frequency range and therefore the predictions of this model closely resemble that of the other models discussed here in the LISA window.

\subsection{Template based on analytical modeling (Model I) }\label{subsec:modelI}

Model I~\cite{Sousa:2013aaa, Sousa:2014gka} uses a semi-analytical approach to characterize the SGWB spectrum generated by cosmic string loops. This model relies on three main assumptions:
\begin{itemize}
    \item The production of closed loops of string is, aside from Hubble damping, the main energy loss mechanism in the evolution of cosmic string networks;
    \item Loops decay mainly by emitting gravitational radiation;
    \item Loops are created with a length $l$ that is a fixed fraction $\alpha_L\le 1$ of the characteristic length $L$ of the cosmic string network: $l=\alpha L$.
\end{itemize}
Under these assumptions, a description of the evolution of the energy density of the network is sufficient to fully characterize the loop distribution function throughout cosmic history and to compute the SGWB they generate. \footnote{Note that the results obtained under these assumptions may be used to construct the SGWB for any distribution of loop length at the moment of creation, as described in refs.~\cite{Sanidas:2012ee,Sousa:2020sxs}.} Here, as in refs.~\cite{Sousa:2013aaa, Sousa:2014gka}, we use the Velocity-dependent One-scale Model~\cite{Martins:1996jp} to describe the evolution of the network.

In this model, the loop-size parameter $\alpha=\alpha_L L/t$, which characterizes the loop length at formation, is treated as a free parameter and a power spectrum of the form in eq.~(\ref{eq:powerspectrum}) is assumed. This model then has, besides cosmic string tension $G\mu$, three additional parameters: the loop-size parameter $\alpha$, the spectral index $q$ (assuming a power spectrum of the form of eq.~(\ref{eq:powerspectrum})), and a fuzziness parameter $\mathcal{F}$. The fuzziness parameter can capture a variety of effects, such as situations in which loops are created with a distribution of lengths (and kinetic energies) instead of being created with exactly the same size~\cite{Blanco-Pillado:2013qja,Sousa:2020sxs}.

The contribution of the $j$-th harmonic mode of emission to the SGWB is such that $j \mathrm{\Omega}_{\rm gw}^j(jf)=\mathrm{\Omega}_{\rm gw}^1(f)$ (cf. eq.~(\ref{eq:omegaj})) and therefore determining the contribution of the fundamental mode of emission is sufficient to fully characterize the SGWB generated by cosmic string loops. In ref.~\cite{Sousa:2020sxs}, the authors derived accurate analytical approximations to the three contributions $\Omega^r_{\rm gw}$, $\Omega^{rm}_{\rm gw}$, and $\Omega^m_{\rm gw}$ in the fundamental mode of emission. Here, these will serve as the basis to construct a novel fully analytical template for Model I, $\Omega_{\rm gw}(f,q,G \mu, \alpha)$, including all relevant modes of emission in summation in eq.~(\ref{eqn:Om-gw-general}). Since this derivation is quite lengthy and technical, here we just provide a brief outline of the method used. We refer the reader to ref.~\cite{Sousa:2020sxs} and appendix~\ref{sec:app_modelI} for a detailed derivation.

Note that for frequencies smaller than that emitted, in the fundamental mode, by the last loop created, $f_{min,m}$, $\mathrm{\Omega}_{\rm gw}^m=0$. Similarly, $\mathrm{\Omega}_{\rm gw}^r$ and $\mathrm{\Omega}_{\rm gw}^{rm}$ do not contribute for frequencies smaller than the minimum frequency of the last loops created in the radiation era $f_{min,r}$. The $j$-th mode of emission of loops created in matter (or radiation) era will then only contribute to the SGWB at frequencies $f>j f_{min,m}$ (or $f>j f_{min,r}$). This means that, for any given frequency $f$, there is finite number of harmonic modes that contribute to the SGWB and, in practice, the infinite summation in eq.~(\ref{eqn:Om-gw-general}) can be substituted, for each frequency $f$, by a finite sum up to the $N$-th term, with
\begin{equation}
N(f)=\frac{f}{f_{min,i}}\,,
\end{equation}
where $i=r$ for $\mathrm{\Omega}_{\rm gw}^r$ and $\mathrm{\Omega}_{\rm gw}^{rm}$ and $i=m$ for $\mathrm{\Omega}_{\rm gw}^m$. Nevertheless, since computing this summation explicitly is computationally costly, we resort to the Euler–Maclaurin summation formula~\cite{knopp1990theory} to perform it, as detailed in Appendix~\ref{sec:app_modelI}.

Here, we have also included the impact of the decrease of the number of relativistic DoF that occurs during the radiation era in $\mathrm{\Omega}_{\rm gw}^r$. This decrease, encoded in the function $\Delta_r(a)$ in \eqref{eq:deltarhoR}, affects the expansion rate of the universe and thus may leave clear signatures on the spectrum. Although the change in $\Delta_r(a)$ is continuous, here we model it as a piecewise constant function based on the default implementation of \texttt{micrOMEGAS5.0}~\cite{Belanger:2018ccd}, and follow the approach in ref.~\cite{Sousa:2020sxs} to derive an analytical correction to $\mathrm{\Omega}_{\rm gw}^r$ to account for this effect. Note however that, when there is a sudden change in the rate of expansion caused by a change in the number of relativistic DoF, the evolution of the string network temporarily deviates from the linear scaling regime and the network takes some time to relax back into this regime. This effect is non-linear and complex to model analytically and, for this reason, we develop a fit using the full numerical computation of the SGWB (obtained using the approach in ref.~\cite{Sousa:2013aaa}). We also describe this fit in more detail in appendix~\ref{sec:app_modelI}.

\begin{figure}
    \centering
    \includegraphics[scale=0.5]{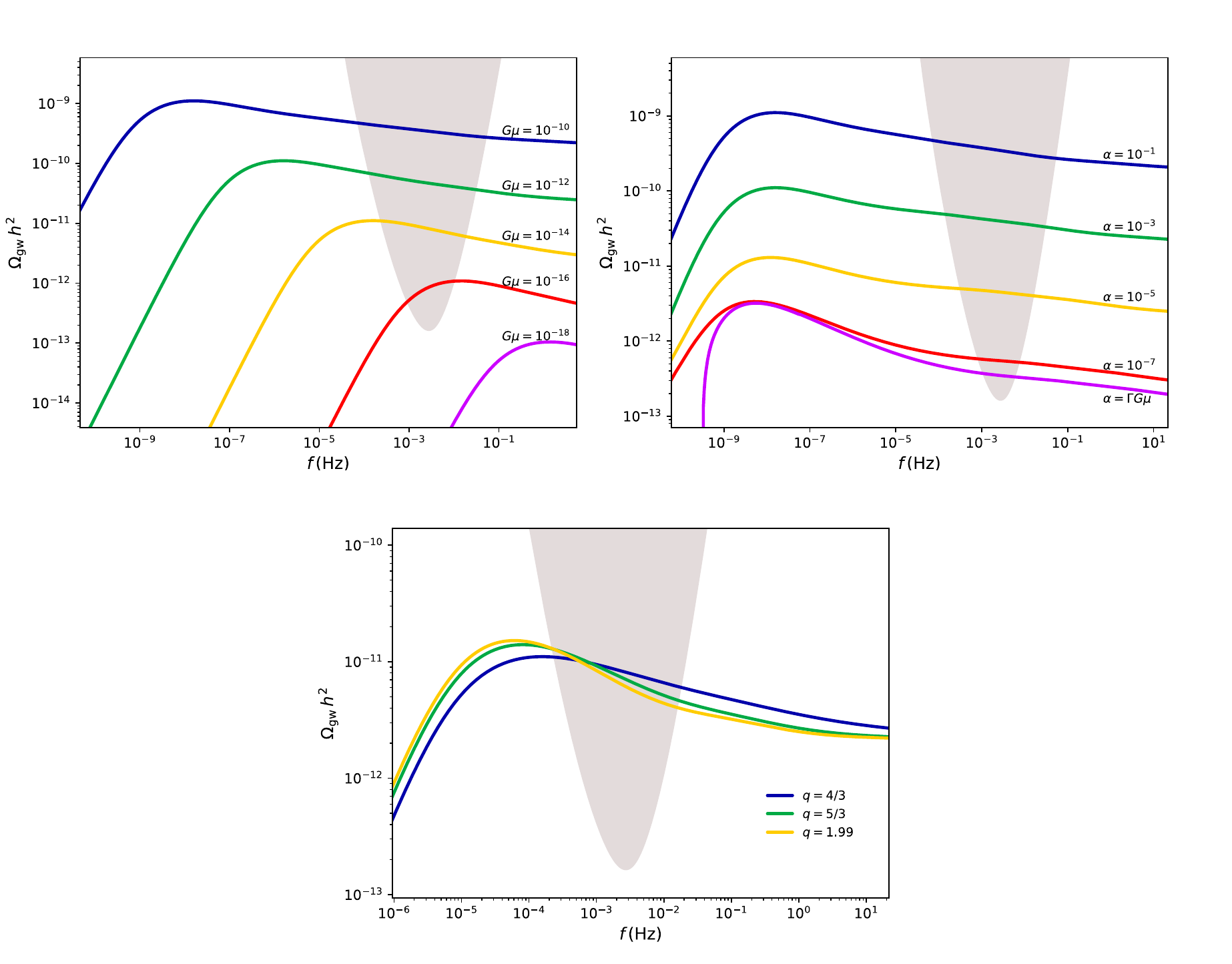}
    \caption{Analytical template for Model I for different values of the free parameters of the model. The gray shaded area represents the LISA power law sensitivity window. The top left panel displays the SGWB for different values of cosmic string tension $G\mu$, for $\alpha=10^{-1}$ and $q=4/3$. The top right panel shows the effects of varying the loop-size parameter $\alpha$, for $G\mu=10^{-10}$ and $q=4/3$. The spectra for different values of the spectral index $q$ for $G\mu=10^{-14}$ and $\alpha=10^{-1}$ is displayed in the bottom panel.}
    \label{fig:parvariation}
\end{figure}

The resulting template is displayed in figure~\ref{fig:parvariation} for different values of the free parameters of the model $G\mu$, $\alpha$ and $q$ spanning the relevant range for reconstructions in the LISA band.

\subsection{Template based on simulation-inferred modeling (Model II)}\label{subsec:modelII}

Model II~\cite{Blanco-Pillado:2011egf,Blanco-Pillado:2013qja,Blanco-Pillado:2015ana,Blanco-Pillado:2019tbi} uses results from numerical simulations of cosmic string networks to construct the SGWB from loops. The loop number densities are obtained directly from a scaling population of non-self-intersecting loops in the simulations of refs.~\cite{Blanco-Pillado:2011egf,Blanco-Pillado:2013qja,Blanco-Pillado:2019vcs,Blanco-Pillado:2019tbi}. The results for the loop number densities in the different cosmological eras are given by the following expressions:
\begin{subequations}\label{eqn:model-II-ns}\begin{align}
    \mathbf{n}_r (l,t) &= \frac{0.18}{t^{3/2} (l + \Gamma G \mu t)^{5/2}}\,,\\
    \mathbf{n}_{rm} (l,t) &= \frac{0.18 (2\sqrt{\Omega_r})^{3/2}} {(l + \Gamma G \mu t)^{5/2}} \left( \frac{a_0}{a} \right)^3\,,\\
    \mathbf{n}_{m} (l,t) &= \frac{0.27 - 0.45 (l/t)^{0.31}} {t^2 (l + \Gamma G \mu t)^{2}}\,.
\end{align}\end{subequations}
We characterize $P_j$ by what is known as the BOS spectrum \cite{Blanco-Pillado:2015ana,Blanco-Pillado:2017oxo}, instead of using the power-law approximation in~(\ref{eq:powerspectrum})~\footnote{The literature often uses ``Model II'' to mean a cosmic string SGWB model which adopts the loop number densities of eq.~\eqref{eqn:model-II-ns} without specifying the $P_j$, i.e., without distinguishing between the power-law and BOS approaches. Here, by ``Model II'', we will always mean the one which uses the BOS $P_j$. We will continue to specify this approach in the text to avoid any confusion with the literature.}  This spectrum is computed from a set of around $1000$ non-self-intersecting loops extracted from the scaling population of loops obtained from the cosmological network simulations \cite{Blanco-Pillado:2017oxo}. The calculation of this spectrum takes into account backreaction by using a toy model based on a smoothing procedure.
The result of this approach leads to a cusp-like spectrum at high $j$ and approximately matches with Model I when that model sets $\alpha=0.1$, $q=4/3$, and $\mathcal{F}=0.1$ (see Appendix~\ref{sec:app_modelI}). The spectra are thus similar to those in the top left panel of figure~\ref{fig:parvariation}.\footnote{For a more comprehensive comparison of the spectra predicted in model I and model II we refer the reader to~\cite{Auclair:2019wcv}.}

Creating an SGWB using Model II involves the use of data tables, most notably when taking the BOS power spectrum, but also in the DoF $\Delta_r(a)$, which ref.~\cite{Blanco-Pillado:2017oxo} indicates is an implementation courtesy of Masaki Yamada. As such, there is no closed-form expression for this $\Omega_{\rm gw}$ (even in the instance of analytic, single-index spectra such as the cusp or kink models). Instead, for the template subroutine, we make use of a comprehensive data table, plus interpolation, to determine the SGWB injected and recovered.

The table is a two-dimensional grid of $\log_{10}(\Omega_{\rm gw}h^2)$ values, with the axes being $\log_{10}(f)$ (from $-5$ to $0$ in steps of $1/20$) and $\log_{10}(G\mu)$ (from $-18$ to $-9.5$ in steps of $1/10$). The frequency range was chosen to cover the full LISA band; the tension range was chosen to run from above the current bounds on $G\mu$ from PTA non-detection \cite{NANOGrav:2023hvm}, to below the bounds on $G\mu$ predicted for Model II SGWB in ref.~\cite{Auclair:2019wcv}. The additional range beyond these bounds was given to allow for the template subroutine to be well-defined in a broader parameter space and, in turn, for a more complete exploration of the likelihood. If inputs outside of the ranges of the data tables are requested, the subroutine issues a warning.

\subsection{Effects of non-standard cosmologies on the SGWB}\label{subsec:modelI-II-nonstandard}

We will now discuss the impact of modifying cosmic history on the cosmic-string-induced SGWB, focusing on the essential physics and simple analytical results. The discussion here lays the foundation for the development of templates for searches for these signals using Models I and II. These templates will serve as the basis for the numerical analysis in section~\ref{ssec:mod-pre-bbn}, where we investigate \texttt{SGWBinner}'s ability to recover information about these modifications.

The templates presented so far assume that the universe evolves according to the SM of cosmology~\cite{Planck:2018vyg}, with the particle content as given by the SM of particle physics. 
In this paradigm, the universe undergoes a period of inflation, followed by a prolonged epoch of radiation domination, then transits to a period of matter domination, and very recently enters a phase of accelerated expansion driven by dark energy.

However, with our current observational data, very little is known about the era before the Big Bang Nucleosynthesis (BBN) --- the so-called \textit{the primordial dark age} \cite{Boyle:2005se, Boyle:2007zx} --- and the actual cosmic evolution may be very different from the commonly assumed standard scenario as outlined above. Meanwhile, non-standard cosmologies are well motivated and have attracted increasing interest in recent years. Any modification to the standard $H(a)$ in eq.~\eqref{eq:friedmannI} would induce changes in the GW spectrum originated by the cosmic string network. Moreover, modifications to the large-scale dynamics of cosmic string networks --- caused, for instance, by a period of inflationary expansion --- may also leave specific imprints on the SGWB spectrum~\cite{Guedes:2018afo,Cui:2019kkd}. In the following, we consider representative cases of non-standard cosmology and string evolution and their impact on the GW signals from strings. Detecting such non-standard features in the GW spectrum from cosmic strings provides a unique way to unveil pre-BBN history, enabling us to perform \textit{Cosmic Archaeology} \cite{Caldwell:1996en,Cui:2017ufi,Cui:2018rwi, Caldwell:2018giq,Gouttenoire:2019kij}.
\\
\subsubsection{Modified equation of state}\label{subsubsec:eos}
The first type of cosmological modification we consider is an early period, preceding the recent radiation dominated era, during which the universe is dominated by a new energy component, leading to a non-standard equation of state for the cosmos~\cite{Allahverdi:2020bys}. For example, an early matter domination era with, $\rho \propto a^{-3}$, may result from an appreciable abundance of a long-lived massive particle, oscillations of scalar fields or primordial black holes~\cite{Moroi:1999zb, Gouttenoire:2019rtn, Bernal:2021yyb, Ghoshal:2023sfa}. Such an era ends with a reheating-like process as the long-lived species decay into SM particles. Another example is kination domination, where the cosmic energy is dominated by a scalar field evolving in a potential $V(\phi) \propto \phi^{N_{\phi}}$, which results in $\rho \propto a^{-n}$, with $n = 6N_{\phi}/(N_{\phi}+2)$. In the limit of large $N_{\phi}$, $N_{\phi}\to \infty$, we have $n\to 6$ and the oscillation energy is dominated by the kinetic energy of the scalar. Kination may arise in certain theories related to inflation, quintessence, dark energy, and axions~\cite{Salati:2002md,Chung:2007vz, Poulin:2018dzj, Gouttenoire:2021jhk}. While we will not consider these cases directly, more complicated histories including combinations of these mentioned above are also possible~\cite{Gouttenoire:2021jhk,Co:2021lkc,Servant:2023tua}.
In order to retain the successful predictions of BBN theory~\cite{Hannestad:2004px}, these new phases in cosmic evolution must transit to the standard radiation domination before $T \sim 5 ~\rm MeV$.

Since Nambu-Goto simulations are mostly restricted to the radiation and matter eras (and Minkowski backgrounds as well), here we will resort to Model I to developed templates to describe these scenarios --- that we will label as model Ia.  Let us then assume that, before the recent radiation era, at temperatures $T>T_{\rm rd}$, the energy density of the universe is dominated by a component whose energy density scales as $\rho\propto a^{-3(1+w)}$, where $w$ is the equation-of-state parameter for this period. There are then two additional free parameters in this model, $T_{\rm rd}$ and $w$, besides the $3+1$ free parameters of Model I. In this scenario, for large enough frequencies $f>f_{\rm rd}$, the contribution of the fundamental mode of emission to the SGWB is modified as~\cite{Cui:2017ufi,Cui:2018rwi,Sousa:2020sxs}:

\begin{equation} \label{eq:modomega1}
\begin{gathered}
    {\Omega_{\rm gw}^*}^1 (f,\alpha,G\mu, f_{\rm rd}, d) = 
\begin{cases}
\Omega_{\rm gw}^1 (f_{\rm rd},\alpha,G\mu) \left(\frac{f_{\rm rd}}{f}\right)^d
 & \ {,} \ \ f > f_{\rm rd}
\\
\Omega_{\rm gw}^1 (f,\alpha,G\mu)& \ {,} \ \ f \leq f_{\rm rd}
\end{cases}
\,,\\
\quad \mbox{with}\quad d=
\begin{cases}
-2\frac{3w-1}{3w+1} & \ {,} \ \ w > \frac{1}{9}
\\
1 & \ {,} \ \ w \le \frac{1}{9}
\end{cases}
\,,
\end{gathered}
\end{equation}
where $\Omega_{\rm gw}^1 (f,\alpha,G\mu)$ is given by eqs.~(\ref{Rad})-(\ref{Mat}), and the characteristic frequency above which the spectrum is modified is related to the temperature of the universe at the onset of the recent radiation era,  $T_{\rm rd}$, through the following analytical approximation
\begin{equation} \label{eqn:fdeltaforlargealpha}
f_{\rm rd}=
  (8.67\times 10^{-3} \, {\rm Hz})\,
\left(\frac{T_{\rm rd}}{\rm GeV} \right)
\left(\frac{ 10^{-12}}{\alpha G\mu}\right)^{1/2}
  \left[\frac{g_*^{\rm SM}(T_{\rm rd})}{g_*^{\rm SM}(T_0)}\right]^\frac{8}{6} \left[\frac{g_s^{\rm SM}(T_0)}{g_s^{\rm SM}(T_{\rm rd})}\right]^{-\frac{7}{6}}\, ,
\end{equation}
for Nambu-Goto strings~\cite{Cui:2018rwi}, which we focus on here~\footnote{For other types of cosmic strings the $f_{rd}$-$T_{rd}$ relation and the GW spectrum would generally take different forms (e.g., see refs.~\cite{Chang:2019mza, Gouttenoire:2019kij,Chang:2021afa} in case of global/axion strings).}.

\begin{figure}
    \centering
    \includegraphics[scale=0.45]{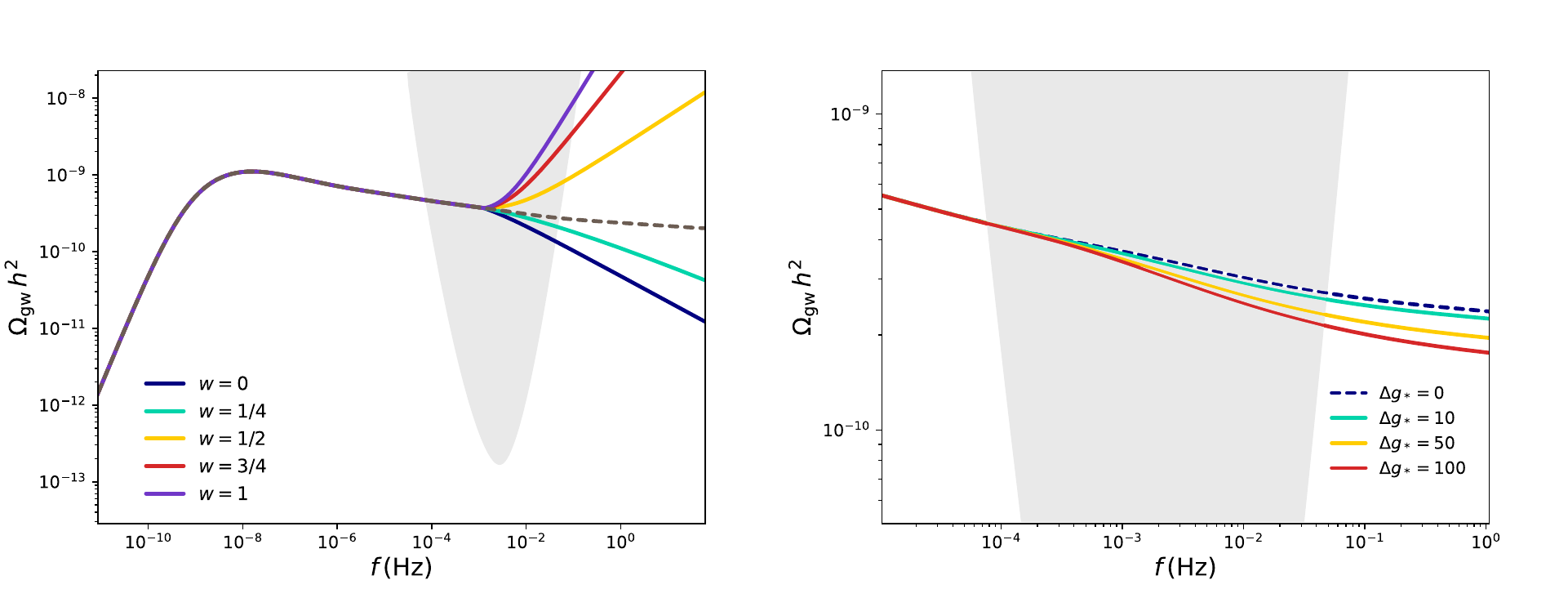}
    \caption{Examples of the full SGWB spectrum generated by cosmic strings in modified pre-BBN scenarios. The dashed lines represent the spectra generated by cosmic string evolving according to the SM of Cosmology and assuming the particle content of the SM of Particle Physics. Here we took $G\mu=10^{-10}$, $\alpha=10^{-1}$, $q=4/3$ and $\mathcal{F}=0.1$, and the shaded area corresponds to the LISA power-law sensitivity window. The left panel displays the results for various benchmark scenarios in which the universe underwent a period with a modified equation of state $w$ before the recent radiation dominated era (Model Ia), starting at $T_{\rm rd}=1\, \rm GeV$ (represented by the colored solid lines). The right panel shows the results with various benchmark numbers of new relativistic DoF, $\Delta g_*$, based on Model Ib, that decouple around $T_\Delta= 10\, \rm MeV$ (the colored solid lines). }
    \label{fig:prebbn}
\end{figure}

The full SGWB spectrum may again be obtained by summing over harmonic modes. We find that for $f\lesssim f_{\rm rd}$ the full SGWB spectrum is completely unaffected by this modification and may simply be computed as described in section~\ref{subsec:modelI} and in appendix~\ref{sec:app_modelI}. For $f>f_{\rm rd}$, however, we find

\begin{equation}
\label{eq:modeos-app}
  \Omega_{\rm gw}^*(f\gg f_{\rm rd},q,G\mu,\alpha, f_{\rm rd}, d)  \propto \left(\frac{f_{\rm rd}}{f}\right)^{d_*}, \quad\mbox{with}\quad d_*=
  \begin{cases}
    d  & \ {,} \ \ w> \frac{1}{3}\frac{3-q}{q+1}
    \\
    q-1 & \ {,} \ \ w\le\frac{1}{3}\frac{3-q}{q+1}
  \end{cases}
\,.
\end{equation}
Note that for all values of $w\le(3-q)/(q+1)/3$, the predicted spectrum slope is identical. This means that, in this limit, a full reconstruction of the equation-of-state parameter cannot be performed and we can only derive an upper constraint for $w$. This equation also shows that, for $f>f_{\rm rd}$, parameters $q$ and $w$ are degenerate. In appendix~\ref{sec:app-prebbn}, we provide the full template for the SGWB generated by cosmic string networks that have undergone a period with a modified expansion rate before the onset of the recent radiation-dominated era and more details regarding the derivation of the results in eq.~(\ref{eq:modeos-app}). The SGWB predicted by our template for different values of $w$ is displayed in the left panel of figure~\ref{fig:prebbn}.

\subsubsection{String network diluted by inflation}\label{subsubsec:inflation}
Modifications to the large-scale evolution of the cosmic string network may also induce modifications to the slope of the SGWB of the form discussed above. A particularly well-motivated example of such a scenario is that of cosmic strings created during the inflationary era~\cite{Shafi:1984tt,Vishniac:1986sk,Yokoyama:1988zza,Yokoyama:1989pa,Jeannerot:2003qv} (for a review of later works, see~\cite{Auclair:2019wcv,LISACosmologyWorkingGroup:2022jok}). In this case, cosmic strings are diluted by the accelerated expansion of the universe and thus their correlation length is much larger than the horizon at the end of inflation. The network is then frozen in a non-relativistic conformal stretching regime after inflation and loop production is suppressed as a result. Significant loop production and GW emission only start once the cosmic string network re-enters the horizon and attains relativistic velocities and, as a result, the SGWB is suppressed at large frequencies~\cite{Guedes:2018afo,Gouttenoire:2019kij,Cui:2019kkd}. In this case, the fundamental mode of the spectrum exhibits a characteristic $f^{-1}$ cut-off in this frequency range~\cite{Guedes:2018afo}, corresponding to a slope of $q-1$ in the full SGWB~\cite{Cui:2019kkd}. The frequency where this cut-off starts is determined by the value of the scale factor at the time when the strings re-enter the Hubble volume, $a_e$, and given by~\cite{Guedes:2018afo}
\begin{equation}
    f_{\rm rd}\sim \frac{H_0\Omega_r^{1/2}}{\Gamma G\mu(\epsilon_r+1/2)^{1/2}}\frac{a_0}{a_e}\,,
\end{equation}
where $\epsilon_r= \alpha/(\Gamma G\mu)$ and we have assumed that this happens before the end of the radiation era, or equivalently for $f_{\rm rd} > \left(9.4\times 10^{-18}/G\mu\right)\,{\rm Hz}$ (see ref.~\cite{Guedes:2018afo} for the complete expression). The modification to the spectrum caused by this deviation from the standard evolution of cosmic string networks is then also of the form of eq.~(\ref{eq:modeos-app}), with $d=1$. In this case, the predicted spectrum is then also identical to that predicted for the scenarios in which there is a modified equation of state with $w\le(3-q)/(q+1)/3$, which means that these two scenarios may be indistinguishable.\footnote{Except for the fact that, in this case, the spectrum may be modified up to lower frequencies since cosmic strings may re-enter the horizon at late cosmological times.} Template Ia may then also be used to probe this scenario by setting $w\le(3-q)/(q+1)/3$ and, in this case, the only additional free parameter would be $a_e$.

\subsubsection{New particle species}\label{subsubsec:newdofs}
Another type of well-motivated modification to the standard scenario is the inclusion of additional particle species that are relativistic in the early universe and contribute to $g_*$ and $g_{*S}$. Such new states are ubiquitously predicted in the extensions to the SM. 
In some cases, the increase in $g_*$ can be dramatic:
for instance, the minimal supersymmetric extension of the SM predicts roughly a hundred new DoF beyond the $g_* = 106.75$ present in the SM, and in mirror dark sector scenario the number of new DOFs equals that of the SM \cite{Kolb:1985bf, Berezhiani:1995am}.

As we can see from eq.~(\ref{eq:deltarhoR}), a deviation from the standard $g_*, g_{*s}$ evolution would induce a change in the Hubble expansion history during the radiation era and thus affect the shape of the SGWB generated by cosmic strings. Identifying these imprints in the GW signal would allow us then to probe new massive particles that may be well beyond the reach of other experiments such as particle accelerators and CMB probes. 

The precise evolution of $g_*, g_{*s}$ depends on the specific dynamics of the underlying Beyond the Standard Model (BSM) theory. To illustrate the generic effect of new relativistic DoFs on the string GW spectrum, we take a simple approach and model the change as a rapid decrease in $g_*$ as the temperature falls below the 
threshold $T_\Delta$:\footnote{For new DoFs coupled to the SM plasma, $T_\Delta$ sets the temperature scale at which the new particles are Boltzmann suppressed, i.e.~their mass scale.  In the presence of dark sectors, the associated dark plasma may have its own temperature independent of that of the SM. In this case, the connection between $T_\Delta$ and the mass scale of the new DoFs is loose.}
\begin{equation}
    \label{eq:dofmod}
g_*(T) \approx
 \begin{cases}
 g_*^{\rm SM}(T)  \quad \quad \quad \quad {\rm ;} \quad T<T_{\Delta}
 \\ 
g_*^{\rm SM}(T) + \Delta g_* \quad {\rm ;} \quad T>T_{\Delta}
\end{cases} \ ,
\end{equation} 
where $\Delta g_*$ is the number of additional relativistic DoF in relation to the SM predictions encoded in $ g_*^{\rm SM}(T)$. An identical modification is assumed for $g_{*S}$, and entropy conservation may be used to derive the dependence of temperature on the scale factor during the decoupling process. 
The change in the amplitude of the spectrum at high frequencies can be approximated analytically as~\cite{Cui:2018rwi}
\begin{equation}\label{eq:dof_plat}
\Omega_{\rm gw}(f\gg f_{\Delta})\simeq \Omega_{\rm gw}^{\rm SM}(f)   \left( \frac{g_*^{\rm SM}}{g_*^{\rm SM} + \Delta g_*} \right)^{\frac13},
\end{equation}
where the $f_\Delta$-$T_\Delta$ relation is the same as the $f_{\rm rd}$-$T_{\rm rd}$ as defined in eq.~(\ref{eqn:fdeltaforlargealpha}). 
Equation~(\ref{eq:dof_plat}) provides a good analytic estimate for the decrease in the magnitude of $\Omega_{\rm gw}$ in the high frequency plateau. However, depending on $T_\Delta$, the LISA band may not be able to capture such a plateau region. Instead, the transitional region may lie within LISA reach. The spectral shape for this transitional region is more subtle, as it depends on $T_\Delta$ and the details of how these new DoFs enter/decouple from the radiation thermal bath. Moreover, the sudden change in the expansion rate caused by this decrease has a non-linear impact on the large-scale dynamics of cosmic string networks, which is not straightforward to model but should also affect the shape of the spectrum in this transitional region. Here, for simplicity, we assume that all additional DoFs decouple instantaneously around the same temperature $T_\Delta$ and that cosmic strings remain in a linear scaling regime during the process. The templates for Models I and II are then recalculated, using the same methods described in the previous sections, for this modified thermal history to develop Models Ib and IIb, which now include two additional free parameters ($T_\Delta$ and $\Delta g_*$). Model I as always involves the analytical description of the evolution of the network throughout the transition. At the same time in Model II we assume an instantaneous relaxation of the network and the effect of the modification comes mostly through modified redshift of the signal. The impact of these additional DoFs on the cosmic SGWB from Model I is illustrated in the right panel of figure~\ref{fig:prebbn}.

\section{Template-based reconstruction with the SGWBinner pipeline}
\label{sec:SGWBinner}

Initially the \texttt{SGWBinner} code was developed as a tool for LISA reforming agnostic searches and parameter reconstruction of an SGWB~\cite{Caprini:2019pxz, Flauger:2020qyi, Seoane:2021kkk, Colpi:2024xhw}. The code included templates for instrument noise and astrophysical foregrounds while the primordial signal was treated as power-law within each bin whose width was optimally computed. This approach is very flexible due to its minimal assumptions, however, is outmatched by template-based searches when we have a theoretical understanding of the spectral shape of our source.

In this work, as well as in the papers~\cite{Caprini:2024hue, LISACosmologyWorkingGroupInflation}, we introduce the option for template-based analysis of a primordial SGWB signal within \texttt{SGWBinner}. This paper's focus is on cosmic string SGWBs.\footnote{The same development is presented in refs.~\cite{Caprini:2024hue, LISACosmologyWorkingGroupInflation} for the implementation of the first-order phase transition and inflationary templates.}
 To elucidate this new option, we provide a brief overview of the foundational aspects of the code and LISA measurements.

\subsection{LISA measurement channels}

LISA comprises three satellites, which we label as $i=1,2,3$. The satellites are positioned at relative distances $\ell_{ij}$ in a triangular constellation. Each of them emits a laser beam to the other two and vice versa. The starting and arrival phases of the beam at every satellite are recorded. The goal is to measure GWs via interferometry but,  
unfortunately, the simplest techniques are unfeasible due to the presence of large laser noise.
Such noise can, however, be largely suppressed by performing the so-called  Time-Delay Interferometry (TDI)~\cite{Prince:2002hp, Shaddock:2003bc, Shaddock:2003dj, Tinto:2003vj, Vallisneri:2005ji, Tinto:2020fcc, Muratore:2020mdf, Muratore:2021uqj,Hartwig:2021mzw}.

Among the different TDI approaches, the so-called ``first generation'' approach leads to an excellent suppression of the noise when, in good approximation, the LISA satellites have identical noise levels and are in an equilateral configuration (i.e., $\ell_{ij}=\ell$ for any $i,j$). In this configuration, the laser phase emitted from the satellite $j$ at time $t-L/c$ reaches the satellite $i$ at time $t$, with $c$ being the speed of light. We dub these phases $\eta_{ij}(t)$ and define the one-arm length delay operator acting on them as $D_{ij} \eta_{lk}(t) \equiv \eta_{lk}(t - \ell_{ij}/c)$. By combining the phases at different times, the following three interferometric measurements can be performed~\cite{Prince:2002hp}:
\begin{equation}
    {\rm A} = \frac{{\rm Z} - {\rm X}}{\sqrt{2}}\;, \qquad  
    {\rm E} = \frac{{\rm X} - 2 {\rm Y} + {\rm Z}}{\sqrt{6}} \;, \qquad  
    {\rm T}=\frac{{\rm X} + {\rm Y} + {\rm Z}}{\sqrt{3}}  \; ,
    \label{eq:AETbasis}
\end{equation}
where 
\begin{equation}\label{eq:tdi-definition}
	{\rm X}  = (1 - D_{13}D_{31})(\eta_{12} + D_{12} \eta_{21}) + (D_{12}D_{21} - 1)(\eta_{13} + D_{13} \eta_{31}) \; ,
\end{equation}
and Y and Z are defined as in~\cref{eq:tdi-definition} with cyclic permutations of the indices. 

Within the above approximations, casting the phase measurements in terms of the channels A, E, and T gives important advantages as they virtually constitute three independent GW interferometers.\footnote{See, e.g.,~\cite{Hartwig:2023pft} for an analysis of how the non-orthogonality of TDI variables, induced by non-equilateral constellations and unequal noise levels, affects signal parameter reconstruction.} The A and E channels exhibit identical properties while the sensitivity of channel T to the signal is significantly lower and it can be treated as a quasi-null channel~\cite{Prince:2002hp}. Due to these properties, the \texttt{SGWBinner} code generates mock data by simulating these three channel's measurements.

\subsection{Data in each channel}

The computational cost of the global fit~\cite{Cornish:2005qw, Vallisneri:2005ji, MockLISADataChallengeTaskForce:2009wir, Littenberg:2023xpl, Colpi:2024xhw} is prohibitive for the multiple runs we carry out in our study. For this reason, we simplify our study by analyzing the data that the global fit, or an analogous data analysis procedure, would have ascribed to the overall stochastic signal plus noise if it had been able to perfectly separate the transient noises and resolvable events. Thus, our time-domain data $\tilde d_i$ in the channel $i=\textrm{A},\textrm{E},\textrm{T}$ are
\begin{equation}
 \tilde{d}_i(t) = \sum_\nu \tilde{n}^\nu_i(t) + \sum_\sigma \tilde{s}^\sigma_i(t) \; ,
 \label{eq:data_time_domain}
\end{equation}
where $\nu$ indexes all the $i$-channel stochastic noise sources, $\tilde{n}^{\nu}_i$, and $\sigma$ labels each $i$-channel stochastic signal contribution, $\tilde{s}_i^\sigma$. 

We approximate the noise sources $\tilde{n}^{\nu}_i$ as uncorrelated, stationary, Gaussian variables with zero mean, where $\tilde{n}^{\nu}_A$ and $\tilde{n}^{\nu}_E$ only differ by statistical realizations. Moreover, in our ideal case, we assume no transient noise contaminating the data and all non-stationarities are absent. Finally, for identical satellites in equilateral configuration, every satellite exhibits statistically equivalent noise sources, and A and E have equivalent response functions, which are different from the response function of T. (For the same reason, $\tilde{n}^\nu_A(t)$  and $\tilde{n}^\nu_T(t)$ have radically different behaviours.)

We assume $\tilde{s}^\sigma_i(t)$ to be independent, stationary Gaussian variables with zero mean. While this simplification streamlines the simulation and analysis of the data, it is a suboptimal approach for isolating the Galactic foreground contribution. Specifically, it is expected not to introduce biases but to reduce the accuracy of the signal reconstruction. 
The reason is that the Galactic foreground originates from a region of the sky that, in the LISA frame, follows a yearly-periodic trajectory across regions with different LISA response functions. LISA thus records the Galactic signal with a yearly modulation (c.f.~\cref{sec:injected}). To reach the stationary approximation, we average the data, which includes the Galactic signal, over multiple years. This operation transforms the yearly-modulated Gaussian signal component into an effectively stationary Gaussian component with a higher variance, with a consequent reduction of the accuracy at which this signal is reconstructed. In this respect, our assumption can be seen as conservative.

With the above assumptions, we can Fourier-transform the data in \cref{eq:data_time_domain} in every time interval $\tau$ during which LISA records $\tilde d_i$ without any interruption:
\begin{equation}
	d_i(f) = \int_{-\tau/2}^{\tau/2} \tilde d_i(t) \; \textrm{e}^{- 2 \pi i f t }  \; \textrm{d} t\; .
\end{equation}
These data have the statistical properties
\begin{equation}
    \label{eq:data_statistics}
	\langle  d_i(f) \rangle  = 0  \; ,  \qquad 
    \langle d_i(f) d^{*}_j(f^{\prime})  \rangle  = \delta_{ij} \frac{\delta(f -f')}{2} \left[  P_{N,ii}(f)  +  P_{S,ii}(f) \right]   \; , 
\end{equation}
where the brackets denote the ensemble average, while $P_{N,ii}$  and $P_{S,ii}$ are the noise power spectra given by 
\begin{equation}
    P_{N,ii}(f)=\sum_\nu T^{\nu}_{ii}(f) S_N^{\nu}(f)      \;,
\end{equation}
\begin{equation}
     P_{S,ii}
    = 
    R_{ii}(f) \sum_\sigma S_S^{\sigma}(f) 
    =
    \frac{3H_0^2}{4\pi^2 f^3} R_{ii}(f) \sum_\sigma h^2 \Omega_{\rm gw}^\sigma (f)\,,
    \label{eq:P_Sii}
\end{equation}
where $T^{\nu}_{ii}$ denotes the $i$-channel response function relative to the strain $S^{\nu}$ of the noise source $\nu$, and  $R^{\sigma}_{ii}= R_{ii}$ denotes the $i$-channel response function to the SGWB signal generated by the source $\sigma$. 

Hereafter, we focus on an effective description of all the noise sources~\cite{LISA:2017pwj, Colpi:2024xhw}. In this case, the index $\nu$ only runs over two noise sources, called ``Optical Measurement System" (OMS) and ``Test Mass" (TM) noises (see~\cref{sec:noise}). The explicit expressions for  $T^{\rm OMS}_{ii}$, $T^{\rm TM}_{ii}$ and  $R_{ii}$ that we adopt, can be found in ref.~\cite{Flauger:2020qyi}.

\subsection{Signals and noises injected and reconstructed}
\label{sec:injected}

We now proceed to the description of signals and noises we implement in the \texttt{SGWBinner} to carry out our analysis. In such an implementation, we make some choices. Firstly, we implement the SGWB from cosmic strings while assuming any other cosmological SGWB source is negligible. Secondly, we only include the astrophysical foregrounds which are expected to be non-negligible in the LISA data based on existing observations.

\subsubsection{Instrument noise}
\label{sec:noise}

As in any interferometer, LISA measures the variation in the distances of its TMs. The sought-after GW signal causes a geodesic deviation between their positions; however, many other effects can influence it, disrupting the measurement. The most important of these effects is errors in optical systems measuring the distance and deviation from free-fall in TM trajectories. These are modelled by the well-known OMS and TM effective strains given by~\cite{LISA:2017pwj, Colpi:2024xhw}
\begin{align}
\label{eq:OMS_noise_def}
S^\text{OMS}_{ii}(f) & = P^2 \; \left(1 + \left(\frac{2\times10^{-3} \,\textrm{Hz}}{f}\right)^4 \right) \times \left(\frac{2 \pi f}{c}\right)^2 \; \times \left(\frac{ \mathrm{pm}^2 }{ \mathrm{Hz} }\right) \;,
\\
\label{eq:Acc_noise_def}
S^\text{TM}_{ii}(f) & = A^2 \;  \left(1 + \left(\frac{0.4 \,\textrm{mHz}}{f}\right)^2\right)\left(1 + \left(\frac{f}{8 \,  \textrm{mHz}}\right)^4\right) \left(\frac{1}{2 \pi f c}\right)^2 \;  \left( \frac{\mathrm{fm}^2 }{ \mathrm{s}^3 } \right) \;.   
\end{align}

We set the \texttt{SGWBinner} to simulate these noises with fiducial values $P=15$ and $A=3$. For the noise reconstruction, we adopt Gaussian priors with widths of $20\%$ and centered on the fiducial values. It is worth mentioning that this noise model is rather minimal and cannot accommodate unexpected contributions that may be present in the real data. Efforts to develop more flexible noise models are currently ongoing within the LISA collaboration~\cite{Baghi:2023qnq, Muratore:2023gxh, Pozzoli:2023lgz}. These models are typically based on weaker assumptions on the noise shape and can thus accommodate for deviations from the functional shapes provided in eq.~\eqref{eq:OMS_noise_def} and eq.~\eqref{eq:Acc_noise_def}. The impact the relaxation of this assumption would have on the parameter reconstruction is a subject for ongoing/future studies.

\subsubsection{Extragalactic foreground}
This foreground is sourced by the compact object binaries outside of the Milky Way that LISA will not be able to resolve individually. The main contributions come from neutron star and white dwarf binaries, as well as binaries of stellar-origin black holes emitting in their inspiral phase. Due to the limited angular resolution of LISA and relatively even distribution of the sources, this foreground can be modelled as an isotropic SGWB signal with the following spectral shape~\cite{Regimbau:2011rp, Perigois:2020ymr, Babak:2023lro, Lehoucq:2023zlt}:
\begin{equation}
\label{eq:Ext_template}
h^2 \Omega^{\textrm{Ext}}_{\rm gw}  =  (h^2 \Omega_{\rm Ext}) \left(\frac{f}{{\textrm{mHz}}}\right)^{2/3} \; . 
\end{equation}

For the injection of this SGWB signal, we adopt $h^2 \Omega_{\rm Ext} = 10^{-12.38} $ as fiducial value~\cite{Babak:2023lro}. For the reconstruction, we run the analysis on the parameter  $\log_{10} (h^2 \Omega_{\rm Ext})$ and set its prior as a Gaussian distribution with $\sigma=0.17$ and mean equal to -12.38, which approximates the more complex posterior obtained in ref.~\cite{Babak:2023lro}. This contribution is sourced by black holes; in principle, neutron star and white dwarf binaries can significantly contribute as well~\cite{Mapelli:2019bnp}.

\subsubsection{Galactic foreground}
This foreground is associated with binaries, mostly of white dwarfs, present in our galaxy~\cite{Nissanke:2012eh}. Some of those can be resolved and subtracted from the data. However, the remaining unresolved events will produce a strong foreground. The sources are distributed in the galactic disc and, due to the yearly orbit of LISA around the Sun, the entire foreground signal will have a modulation with a period of one year. This feature is in principle a clear signature that helps separate the Galactic foreground component from the primordial SGWB component, which is stationary.  In practice, however, this procedure would require reconstruction strategies that are more sensitive to uncertainties due to, e.g.,~periodic noise instability and data gaps, which we have assumed to be negligible in our analysis exactly because their effects should be smoothed out when integrated over several years~\footnote{Reaching a robust conclusion on the matter would, however, require progress on the currently available reconstruction codes. In particular, \texttt{SGWBinner} does not include yet the option of a periodic-signal reconstruction.}. We then expect to be conservative by not leveraging on the yearly modulation signature and instead analyze this foreground as it results after averaging its signal over the whole observation time $T_{\rm obs}$~\cite{Karnesis:2021tsh}:
\begin{equation}
\label{eq:SGWB_gal}
h^2\Omega^{\textrm{Gal}}_{\rm gw}(f)= \frac{1}{2}
h^2 \Omega_{\rm Gal} \left(\frac{f}{\textrm{Hz}}\right)^{\frac{2}{3}}   
\left[1+\tanh \left({\frac{f_{\textrm{knee}}-f}{f_2}} \right) \right] e^{-(f/f_1)^\upsilon}  \;,
\end{equation}
with
\begin{eqnarray}
\log_{10} \left( f_1 / {\rm Hz}\right) &=& a_1 \log_{10} \left( T_{\mathrm obs}/{\rm year} \right) + b_1\,,\nonumber\\
\log_{10} \left( f_{\textrm{knee}}/ {\rm Hz} \right) &=& a_k \log_{10} \left( T_{\mathrm obs}/{\rm year} \right) + b_k\,.
\end{eqnarray}

For the injection, we adopt the fiducial values 
$a_1 = -0.15,~b_1=-2.72,  ~a_k = -0.37,   ~b_k=-2.49$, $\upsilon = 1.56, ~ f_2 = 6.7 \times 10^{-4}$\,Hz and $h^2 \Omega_{\rm Gal}=10^{-7.84}$~\cite{Karnesis:2021tsh}. For the parameter reconstruction, we assume only the latter to be unknown. To reconstruct it, we adopt a Gaussian prior on the parameter $\log_{10} \left( h^2 \Omega_{\rm Gal} \right)$, with $\sigma = 0.21$ and mean at $-7.84$.

\subsubsection{Cosmic string SGWB}

For an SGWB expected for a Nambu-Goto string coupled only to gravity in the standard cosmology scenario, we implement the Model I and Model II  templates described in sections~\ref{subsec:modelI} and \ref{subsec:modelII}. We also include the template deformations suitable for the representative non-standard cosmological scenarios discussed in section~\ref{subsec:modelI-II-nonstandard}. These templates are used for both the injections and the parameter reconstructions.

In standard cosmology, the free parameters of Model I are $G\mu$, $q$, and $\alpha$, whereas Model II only depends on $G\mu$. The additional free parameters $\log_{10}(T_\Delta)$, $\Delta g$, $\log_{10}(T_{rd})$ and $w$ are introduced when we consider non-standard cosmology scenarios. In the signal injections, we vary the values of these parameters as we will detail in each case. For the parameter reconstruction, we adopt flat priors on the intervals given in Table~\ref{tbl:parameter-ranges}.

\begin{table}[]
    \centering
    \begin{tabular}{llll}
       Scenario & Parameter name & Symbol & Range \\\hline\hline
       Model I & String tension & $\log_{10}(G\mu)$ & $[-18.0,-7.0]$\\
       & Loop size & $\log_{10}(\alpha)$ & $[-3,0]$\\
       & Power spectral index & $q$ & $[1.10,1.99]$\\
       & Fuzziness & $\log_{10}(\mathcal{F})$ & $[-5,1]$\\\hdashline
       \;~+ Modified e.o.s. (Ia) & Equation-of-state & $w$ & $[0,1]$ \\
       & Rad. domination temp. & $\log_{10}(T_{rd}/\text{GeV})$ & $[-2.3,1.7]$ \\\hdashline
       \;~+ New particle species (Ib) & New deg. of freedom & $\Delta g$ & $[0,120]$ \\
       & Decoupling temp. & $\log_{10}(T_\Delta/\text{GeV})$ & $[-2.3,1.7]$ \\\hline
       Model II & String tension & $\log_{10}(G\mu)$ & $[-18.0,-9.5]$\\\hdashline
       \;~+ New particle species (IIb) & New deg. of freedom & $\Delta g$ & $[0,120]$ \\
       & Decoupling temp. & $\log_{10}(T_\Delta/\text{GeV})$ & $[-2.3,1.7]$ \\
    \end{tabular}
    \caption{The parameters used in reconstructing signals in sec.~\ref{sec:results}, along with their associated ranges. All parameters listed here are drawn from flat priors during reconstruction. Note that Model I's fuzziness parameter is set to $\mathcal{F}=0.1$ in all scenarios except for the reconstructions of sec.~\ref{sec:fng_model}.}
    \label{tbl:parameter-ranges}
\end{table}

\subsection{Data generation and analysis}

LISA will operate for a period between 4.5 and 10 years. This period includes time dedicated to maintenance activities during which LISA will not take GW data, corresponding to a duty cycle of around 82\%~\cite{Colpi:2024xhw}. For concreteness, we consider the option where such activities are scheduled with a cadence of about two days out of two weeks.  We consequently set \texttt{SGWBinner} to work with  $N_d=126$ data segments with each lasting $\tau=11.4$ days, equivalent to $T_{\rm obs}=4$ effective years of data. This is likely a conservative scenario since the mission may last more years, as just said.  

The \texttt{SGWBinner} code operates in the frequency domain, thus we define the Fourier transformed data $d^s_i(f_\textrm{k})$ with bin frequency $f_{\rm k}$ and spacing $\Delta f = 1/\tau\simeq 10^{-6}$ for each segment $s=1, ..., N_{\rm d}$.  To generate the data, 
for every frequency bin $f_\textrm{k}$ of the LISA band $[0.03, 500]\,$mHz, the code produces $N_d$ Gaussian realizations of each of the signal and noise sources following the templates of (and the fiducial values specified in)  section~\ref{sec:injected}. The data are then analysed in a stationary approximation which involves averaging all the $N_d$ realizations of the data at a given frequency $f_{\rm k}$. This yields the averaged data $D^\textrm{k}_{ii} \equiv \sum_{s = 1}^{N_d} d^s_i(f_\textrm{k}) d^{s \, *}_j(f_\textrm{k}) / N_d$. 
Working with the full set of such data would be very cumbersome due to its very high resolution $\Delta f$, which is not necessary to accurately describe the SGWBs we analyse. The next step is thus coarse-graining the frequency resolution~\cite{Caprini:2019pxz, Flauger:2020qyi} which results in a smaller set of frequency bins $\bar{f}^{k}$. While doing this operation, a weight $w^{\rm k}_{ii}$ is assigned to each data point; this weight takes into account the statistical properties of the initial and coarse-grained data and assures the new set is statistically equivalent to the original one.  

The next step in the code is the calculation of the likelihood that the data $D^{\mathrm Th}_{ii} (f_{\rm k}, \vec{\theta})$ predicted in our theoretical model, which involves the noise and signal templates presented in~\cref{sec:injected}, describes the generated data. Here $\vec{\theta}=\{
\vec{\theta}_N,
\vec{\theta}_{S}
\}$ 
denotes the free parameters of the model theoretical model depends on the parameters, namely
\begin{equation}\label{eq:injectionvec}
\vec{\theta}_N=\{A,P\} \;, \qquad \qquad  \vec{\theta}_S=\{
\log_{10} (h^2 \Omega_{\rm Gal}),
\log_{10} (h^2 \Omega_{\rm Ext}),
\log_{10} (G\mu), \dots
\}\;,
\end{equation}
where the dots represent the additional parameters arising in certain models: $q$ and $\log_{10}(\alpha)$ for Model I, $\Delta g$ and $\log_{10}(T_\Delta)$ for additional DoF; and $w$ and $\log_{10}(T_{rd})$ for modified equations of state (see section~\ref{sec:injected}).
The likelihood reads as
\begin{equation}
\label{eq:likelihood}
\ln \mathcal{L} (\vec{\theta} ) = \frac{1}{3} \ln \mathcal{L}_{\mathrm G} (\vec{\theta} | D^{\rm k}_{ii}) +  \frac{2}{3} \ln \mathcal{L}_{\mathrm{LN}} (\vec{\theta} | D^{\rm k}_{ii}) \; ,
\end{equation}
where
\begin{equation}
\ln \mathcal{L}_{\mathrm G} (\vec{\theta} | D^{\rm k}_{ii}) = -\frac{N_d}{2} \sum_{k} \sum_{i} w^{\rm k}_{ii} \left[ 1 - D^k_{ii} / D^{\mathrm Th}_{ii} (f_{\rm k}, \vec{\theta}) \right]^2  \; ,
\end{equation}
\begin{equation}
\ln \mathcal{L}_{\mathrm LN} (\vec{\theta} | D^{\rm k}_{ii}) = -\frac{N_d}{2} \sum_{k} \sum_{i} w^{\rm k}_{ii} \ln^2 \left[ D^{\mathrm Th}_{ii} (f_{\rm k}, \vec{\theta}) / D^{\rm k}_{ij}   \right] \; .
\end{equation}
In the above, $\ln \mathcal{L}_{\mathrm G}$ is the standard Gaussian likelihood while $\ln \mathcal{L}_{\mathrm LN}$ is a Lognormal likelihood added to correctly take into account the non-Gausainity resulting from the data compression procedure~\cite{Bond:1998qg, Sievers:2002tq, WMAP:2003pyh, Hamimeche:2008ai}. After introducing priors for all the components of eq.~\eqref{eq:injectionvec}, the code uses \texttt{Polychord}~\cite{Handley:2015vkr, Handley:2015fda}, \texttt{Cobaya}~\cite{Torrado:2020dgo} and \texttt{GetDist}~\cite{Lewis:2019xzd} in order to explore the likelihood, and compute and plot the posterior distributions.

Finally, let us define the Signal-to-Noise ratio (SNR) which is often used as a simple indicator of the reach of LISA and which we will also report in our results. The SNR reads~\cite{Romano:2016dpx}
\begin{equation}
\label{eq:SNR_def}
\textrm{SNR} = \sqrt{T_{\mathrm obs} \;\sum_{i} \int \left( \frac{S_{\mathrm GW}}{S_{i, \mathrm N}} \right)^2 \; \textrm{d}f }  \; ,
\end{equation}
where $S_{\mathrm GW}$ is related to our cosmic string spectrum $h^2\Omega_{\mathrm GW}$ as in~\cref{eq:P_Sii}, the noise contributions correspond to~\cref{eq:OMS_noise_def,eq:Acc_noise_def} and the integral is over the LISA frequency band.

\section{Results on the reconstruction of the signal}\label{sec:results}

In this section, we use  \texttt{SGWBinner} to forecast the LISA's capabilities in the reconstruction of the SGWB signal generated by cosmic strings. We use the code to repeatedly produce signal and noise realizations for different values of the cosmic-string template parameters and then obtain reconstruction posteriors.
For Model I, the parameters are (logarithmic) tension $\log_{10}(G\mu)$, spectral index $q$, and loop size $\alpha$. 

For Model II we only sample the likelihood over the logarithmic tension. We will discuss that value directly in most cases; e.g., ``a logarithmic tension of $-11.5$'' indicates $G\mu = 10^{-11.5}$. Prior work~\cite{Auclair:2019wcv} suggests that when only the cosmic string SGWB is included, \texttt{SGWBinner} should recover the injected parameters with good accuracy and precision down to tensions in the range of $-16.5$ to $-17.0$, around which the spectrum drops below the LISA noise curve. However, we expect more than just the cosmic string SGWB to be seen in LISA; it will be necessary to subtract additional sources to measure the string background. Here, we investigate the impact of astrophysical foregrounds on \texttt{SGWBinner}'s reconstruction of the string signal. At high tensions, the effect of foregrounds is minimal, but they become significant in the range of $-14.0$ to $-16.0$, and the quality of reconstructions degrades rapidly for tensions below about $-16.0$.

\subsection{Results for Model I}

For Model I, we use the semi-analytic template discussed in~\cref{subsec:modelI}.
Its free parameters are
$G\mu$, $\alpha$, and $q$, although other parameters may also be included.\footnote{For example, in section~\ref{sec:fit-I-II}, we also include the ``fuzziness'' parameter, $\mathcal{F}$, which stands in for several effects, such as the amount of kinetic energy carried out by the loops produced by the network as well as the spread in size.} Here, we will follow the prescription in~\cite{Auclair:2019wcv} to study Nambu-Goto strings and take $\mathcal{F}=0.1$ as the standard normalization, as this allows us to match the amplitude of the loop number density inferred from the Nambu--Goto simulations of ref.~\cite{Blanco-Pillado:2013qja}.

\begin{figure}
    \centering
    \includegraphics[width=\linewidth]{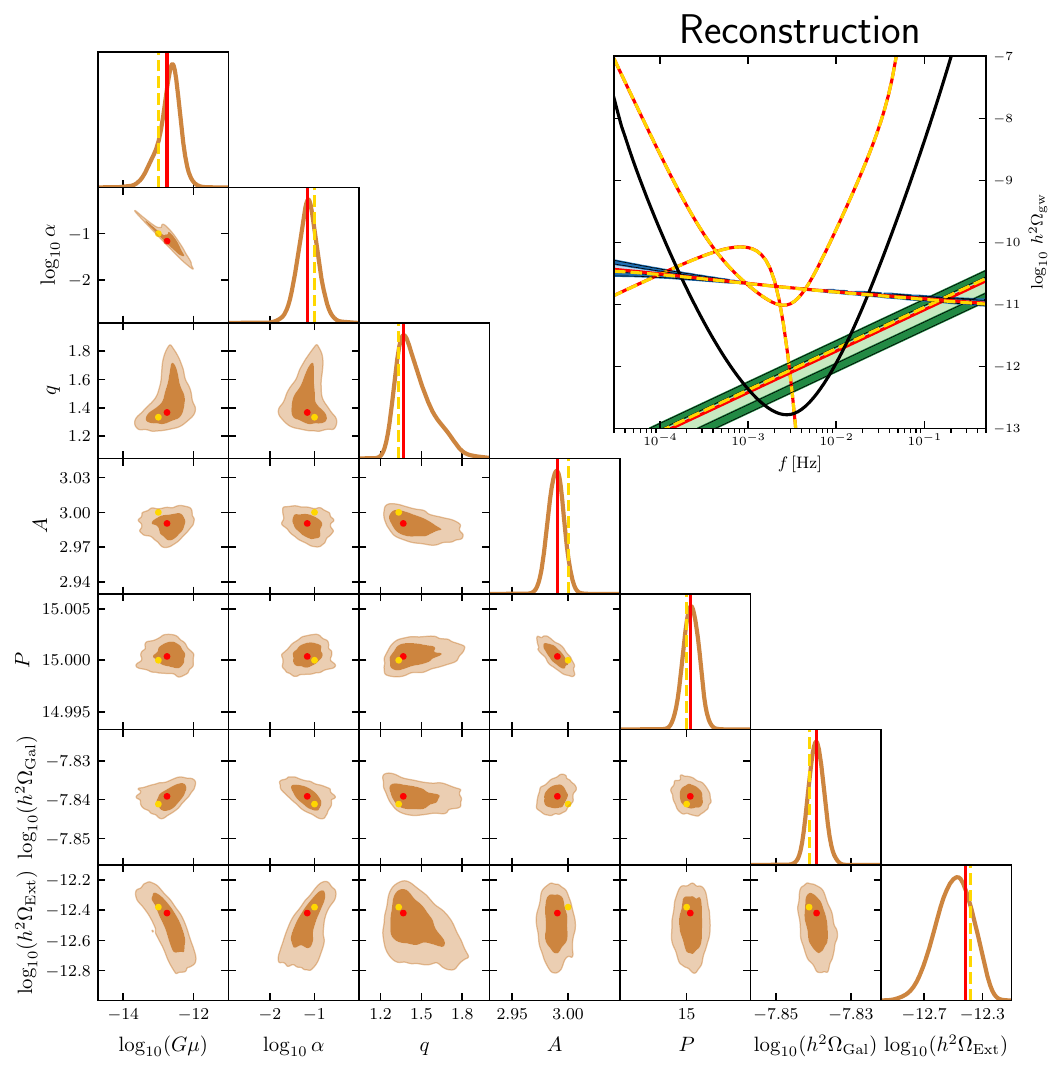}
    \caption{Corner plot of the parameter posterior distributions for Model I. The $1\sigma$ and $2\sigma$ allowed regions are shaded in dark and light orange colors, respectively. The fiducial parameters are $\log_{10}(G\mu)=-13.0$, $\log_{10}\alpha=-1.0$, $q=4/3$, and foregrounds are also included. The injected and recovered signals are displayed inset in the upper-right corner. The spectra for the injected signal, galactic foreground, and extragalactic foreground are shown with the yellow curves. The recovered spectrum is depicted with the red solid line, with $1$ and $2\sigma$ error uncertainties shaded in light green and dark green. We also plot the sensitivity of the AA channel (yellow) and the power-law sensitivity~\cite{Thrane:2013oya} (black). 
    }
    \label{fig:csI-sample-reconstruction}
\end{figure}

Figure~\ref{fig:csI-sample-reconstruction} illustrates an example of parameter reconstruction for the signal based on the Model I template, assuming $\log_{10} (G\mu)=-13.0$, $\log_{10} \alpha=-1.0$, and $q=4/3$ as fiducial values. 
A characteristic feature of Model I is that there are degeneracies among its three free parameters, that may be strong in some limits. In this figure, we can observe a strong correlation between $\log_{10} (G\mu)$ and $\log_{10} \alpha$. Therefore, we must always ensure that we have a sufficient number of sampling points and enough precision.

\begin{figure}[htbp]
\begin{center}
\includegraphics[width=0.99\textwidth]{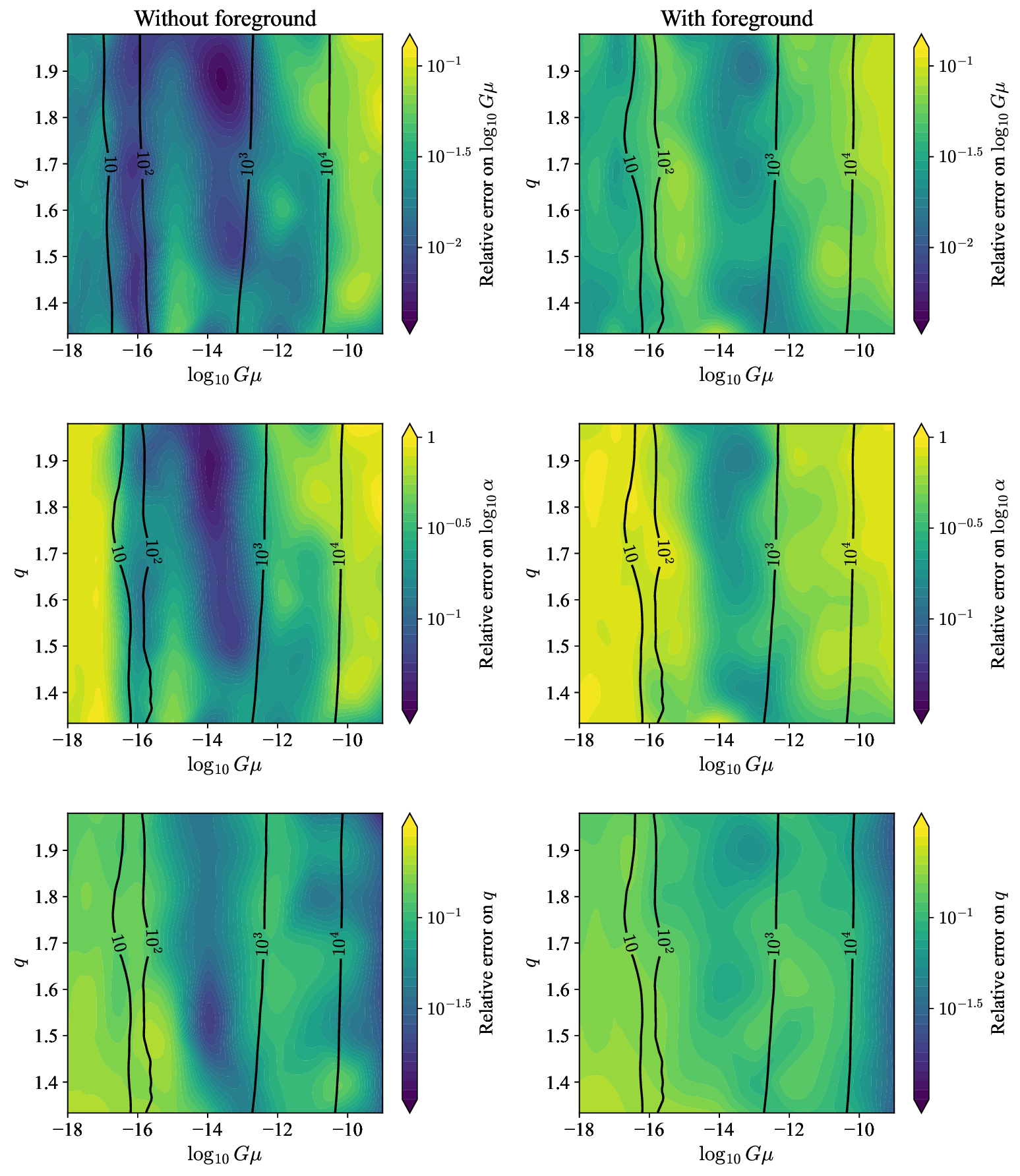}
\end{center}
\caption{ Relative error in the reconstruction of the parameters of Model I without (left) and with (right) injections of the Galactic and extra-Galactic foregrounds.
The panels show the marginalized $1\sigma$ relative errors on $\log_{10} G\mu$, $\log_{10} \alpha$, $q$ (top, middle, bottom), explored in the $G\mu - q$ parameter space. The fiducial value of $\alpha$ is fixed at $0.1$. The solid black lines represent the SNR contours at the labeled values.
\label{fig:cs-I-error-contours}}
\end{figure}

\begin{figure}[htbp]
\begin{center}
\includegraphics[width=0.99\textwidth]{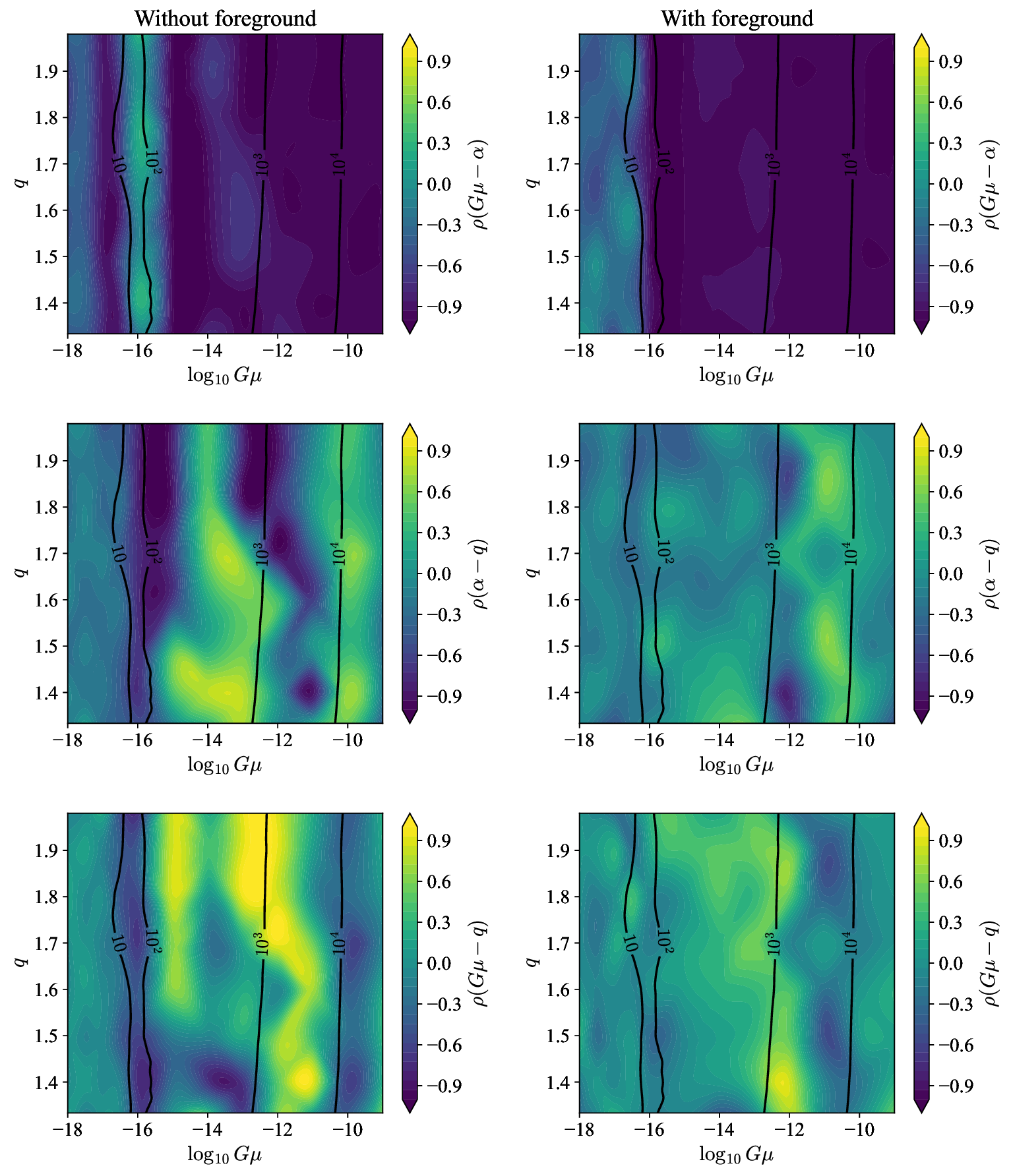}
\end{center}
\caption{\label{fig:cs-I-correlations}
Non-diagonal elements $\rho=C_{ij}/\sqrt{C_{ii}C_{jj}}$ in the reconstruction of the parameters of Model I. The value of $\rho$
varies from $0$ (independent) to $1$ (completely correlated). The left and right panels represent the cases without and with the foregrounds, respectively. The solid black lines represent the SNR contours at the labelled values.}
\end{figure}

Figure~\ref{fig:cs-I-error-contours} presents the exploration of parameter space for 1\,$\sigma$~relative errors in the $\log_{10} G\mu - q$ space, with $\log_{10} \alpha$ fixed at $-1.0$.\footnote{For lower values of $\alpha$, the amplitude of the SGWB is lower and, as a result, the lowest tension that LISA may probe is increased (see, e.g., ref.~\cite{Auclair:2019wcv}).} 
For a given parameter, the relative error is derived from diagonal entry in the (inversed) reconstructed covariance matrix; this corresponds to the error on the given parameter after marginalizing over all the others. In the figure
we display the cases with and without foregrounds, as the behavior related to the cosmic string model is more prominent in the former case, while the latter represents a more realistic situation.\footnote{Possibly, by leveraging the yearly modulation of the galactic foreground, one could obtain an intermediate result.}
We observe a general trend where the error is smallest at intermediary values of the tension (between $G \mu \sim 10^{-16}$ and $10^{-13}$). 
The increase in error at low tension ($G \mu < 10^{-17}$) is easily explained by the SGWB signal dropping much below the LISA power-law sensitivity curve. On the other hand, at large tensions, the increase in error is due to correlations in the parameters. 

To understand this behavior, in figure~\ref{fig:cs-I-correlations}, we plot the correlation coefficient $\rho=C_{ij}/\sqrt{C_{ii}C_{jj}}$, where $C_{ij}$ is the covariance matrix. The top panel shows that the correlation is indeed very strong when $G \mu$ is large, even if the SNR is quite large in this regime. This occurs because, for this range of tensions, the LISA band coincides with the radiation-era plateau of the spectrum.

As shown in figure~\ref{fig:parvariation}, varying both $G\mu$ and $\alpha$ leads to a change of the amplitude of this plateau. Although a decrease of $G\mu$ is also accompanied by a shift of the peak of the spectrum towards higher frequencies, there is no other spectral information to resolve this degeneracy when LISA only probes this plateau, which is the origin of the large errors for large tension seen in figure~\ref{fig:cs-I-error-contours}. This tendency is consistent both with and without the existence of foregrounds. Another thing to note is that the location of the features on the plateau caused by the decrease in the effective number of relativistic DoF depends on the ratio $\alpha/G\mu$ and their slope is determined by $q$ (see Appendix~\ref{sec:app_modelI} for details). This helps to partially resolve the degeneracy between $G\mu$ and $\alpha$ for tensions between $G \mu \sim 10^{-14}$ and $10^{-12}$, since in this case the feature caused by the latest decrease in the effective number of DoF is fully within the LISA sensitivity band.

In figure~\ref{fig:cs-I-error-contours}, we also see that the errors in the reconstruction of the parameters are less dependent on $q$ than they are on $G\mu$, but we can see a tendency for larger errors when $q$ is larger in the high $G\mu$ range. This can be understood by referring to figure~\ref{fig:parvariation}: the width of the ``bump'' in the spectrum varies inversely with $q$, and so for larger $q$ this bump is entirely outside of the LISA band for large tensions. Conversely, the smaller-$q$ SGWB have part of the peak inside the LISA band, and so $G\mu$ and $\alpha$ are not as strongly degenerate. For lower tensions ($G\mu<10^{-14}$), this tendency is inverted and the error in the determination of $G\mu$ and $\alpha$ are somewhat (but not significantly) larger for lower values of $q$. This is again caused by the increase in the correlation between the 3 parameters in this region of parameter space, since a smaller portion of the peak of the spectrum coincides with the LISA window when $q$ is smaller.

The inclusion of foregrounds generally increases the errors on the recovered parameters. Another significant change is at the lowest tensions: the smallest tension at which we generally recover low errors is increased due to the inclusion of foregrounds. Now, rather than the SGWB dropping below the LISA PLS, it is the SGWB dropping below the combined galactic and extragalactic foregrounds that leads to higher errors. This can be seen, e.g., in the top row of figure~\ref{fig:cs-I-error-contours}, where adding foregrounds shifts the lines representing a fixed SNR to the right and in some cases gives the mostly-vertical lines more horizontal variation.

In general, we can reconstruct the logarithmic tension of an injected Model I signal to within $\approx10\%$ relative error across a wide range of parameter-space. Excluding foregrounds, this error can be as good as $\approx 0.3\%$ at moderate to low tensions; including foregrounds raises the best relative error by about a factor of ten, to $\approx 3\%$. Note that because the largest relative errors on the logarithmic tension are at high tensions, adding foregrounds does not much change them.

The highest relative errors on the logarithmic loop size, $\log_{10}(\alpha)$, and the spectral index, $q$, tend to be larger than those on the logarithmic tension, at roughly $100\%$ and $30\%$, respectively. However, the best relative errors on both parameters, again at moderate tensions, are comparable to the best relative errors on the logarithmic tension. Note that the largest errors on $q$ happen at very low tensions, where we are unlikely to make robust reconstructions; as we can see from figure~\ref{fig:cs-I-correlations}, there is little degeneracy between $q$ and either $G\mu$ or $\alpha$ at largest tensions, explaining why we do not see the large relative errors in $q$ at largest tensions which we do for $\log_{10}(G\mu)$ and $\log_{10}(\alpha)$.

\subsection{Results for Model II}\label{ssec:model-ii-results}

\begin{figure}
    \centering
    \includegraphics[scale=0.75]{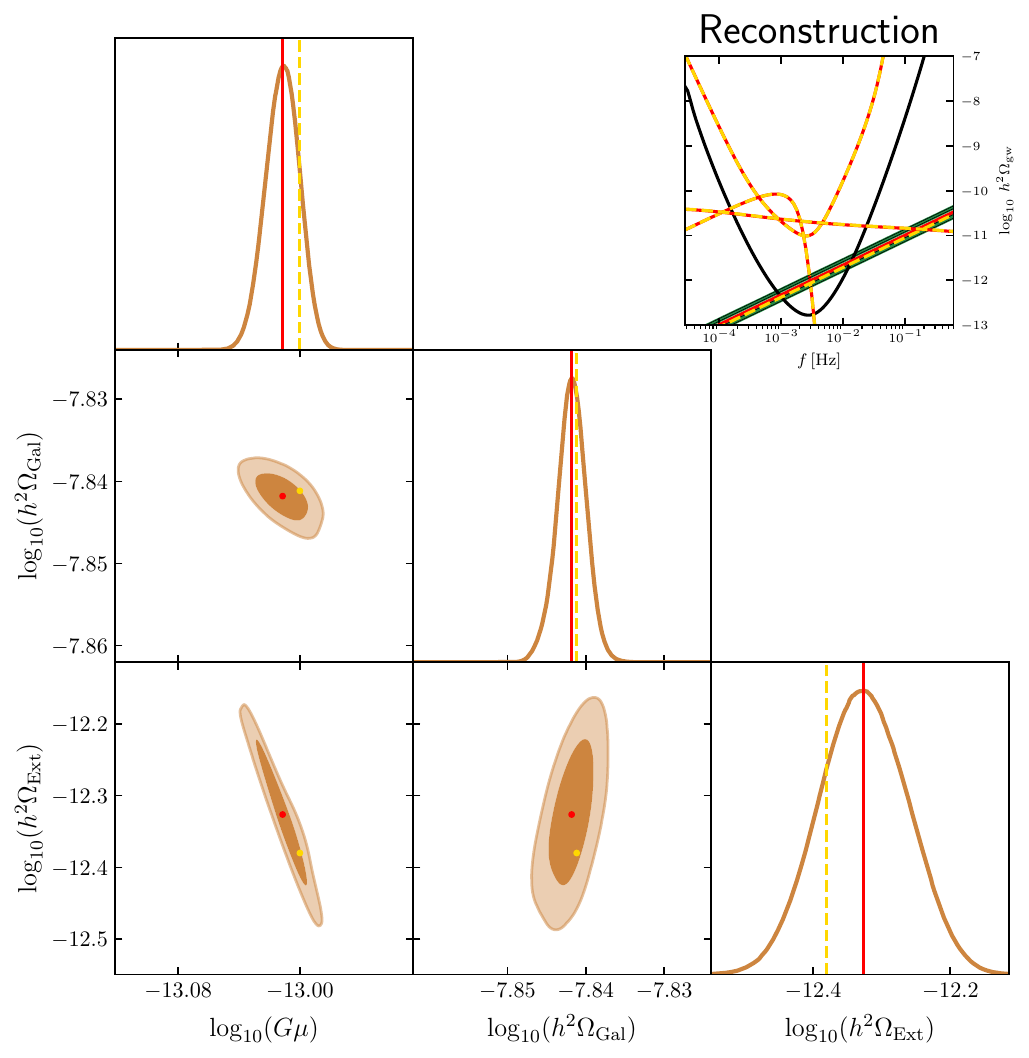}
    \caption{Corner plot of the parameter posterior distributions for Model II using the BOS power spectrum, a logarithmic tension of $\log_{10}(G\mu)=-13.0$, and including foregrounds. The injected and recovered signals are shown inset, upper-right.}
    \label{fig:csII-sample-reconstruction}
\end{figure}

The signal for Model II follows the template presented in \cref{subsec:modelII}.
A sample reconstruction, for a BOS power spectrum
and $\log_{10}(G\mu)=-13.0$, is given in figure~\ref{fig:csII-sample-reconstruction}.

We picked a range in logarithmic tension from $-11.0$ to $-17.0$, with the rationale being that the upper bound is near the maximum tension allowed by PTA non-detections~\cite{NANOGrav:2023hvm}, while the lower bound is near to the minimum tension predicted to be detectable by LISA in the absence of any foregrounds ~\cite{Auclair:2019wcv}. When discussing the goodness of reconstructions, we will often refer to the \emph{absolute error}, which we define as the difference between the logarithmic tension recovered and the logarithmic tension injected. This has a nice interpretability in that $10^{(absolute\,error)}$ is the ratio of the tension recovered to the tension injected. Thus an absolute error of about $-0.301$ (about $0.301$) indicates we have recovered a tension which is half (double) of the tension injected. These thresholds, of $-50\%$ ($+100\%$) error, will be a typical metric in the following discussion.

For all injected and recovered signals, we use the BOS $P_j$ model, and for each injection/recovery, we collected: the most likely reconstructed parameter value; the 1-$\sigma$ width of the likelihood distribution of the reconstructed parameter value; and the SNR of the reconstructed signal. For the case without foregrounds, we varied the injected $\log(G\mu)$ from $-11.0$ to $-17.0$ in steps of $0.1$; for the case with foregrounds, our lower bound of injected logarithmic tension was $-16.5$. This modified lower bound was taken due to severe degradation in the ability of \texttt{SGWBinner} to reconstruct the signal at lower tensions, the origin of which is discussed later. We collected 30 data points for each value of the injected tension.

\begin{figure}
    \centering
    \includegraphics[scale=0.6]{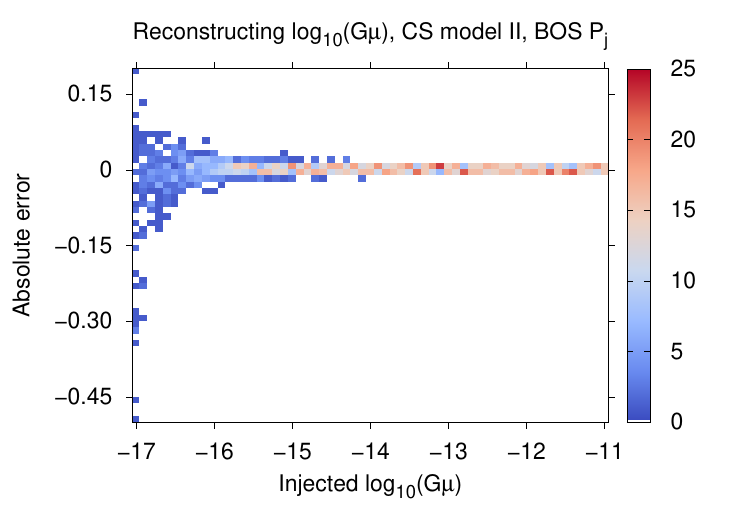}~\includegraphics[scale=0.6]{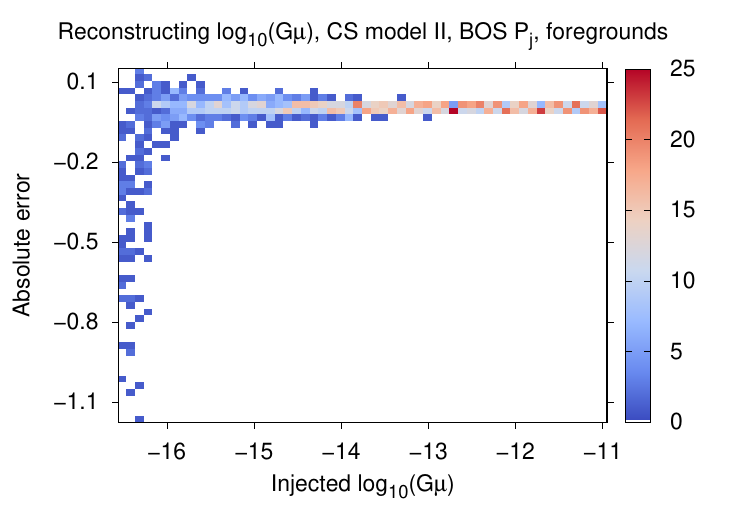}
    \caption{Density histogram of the absolute error made in the reconstruction of $\log_{10}(G\mu)$ of an injected Model II signal, using the BOS $P_j$, (left) in the absence of foregrounds and (right) with foregrounds. The injected logarithmic tensions range from $-11.0$ to $-17.0$ ($-16.5$) for the case without (with) foregrounds in steps of $0.1$, and there were $30$ trials performed at each tension value.}
    \label{fig:csII-BOS-abs-histogram}
\end{figure}

Figure~\ref{fig:csII-BOS-abs-histogram} shows the dependence of the most likely reconstructed tension on the injected tension. At higher tensions, there is very little error made in the reconstruction, but that error grows as the tension decreases. The spread in absolute error is not symmetric about zero: there is a slight bias towards more negative absolute error, corresponding to a preference to reconstruct lighter strings than are injected.

Let us discuss the case without foregrounds first. At the lowest tensions, we find a maximum absolute error (at $\log_{10}(G\mu)=-17.0$) of about -0.45, and thus an error in the actual tension of around $-68\%$. The largest positive absolute error (again at $\log_{10}(G\mu)=-17.0$) is close to 0.188, or a $+54\%$ error in the actual tension. Thus the reconstructed actual tensions, at worst, are within a factor $\sim 6$ of each other. If we wish to restrict ourselves to tensions at which the reconstruction makes no worse than a $-50\%/+100\%$ error in the actual tension, we should consider only $\log_{10}(G\mu)\gtrsim -16.9$, which represents only a slight adjustment to the $\log_{10}(G\mu)\gtrsim -17.0$ predicted in ref.~\cite{Auclair:2019wcv}.

However, it should be noted that even the worst performance, with the ratio of largest/smallest reconstructed tension being $\sim 5$, is acceptable in light of other theoretical uncertainties. Specifically, the relationship between the scale of symmetry breaking $\eta$ and the tension $\mu$ is $\mu = \lambda\eta^2$, with $\lambda$ some $\mathcal{O}(1)$ coupling set by the theory which produced the symmetry breaking. In effect, reconstruction at lowest tensions simply adds another $\mathcal{O}(1)$ uncertainty.

Now let us turn to the case with foregrounds. The right panel of figure~\ref{fig:csII-BOS-abs-histogram} shows that the impact of including foregrounds, while observable at all tensions, is most significant at lower tensions, and drastically so below $\log_{10}(G\mu) \simeq -15.8$. We see a rapid broadening in the reconstructed tensions with a strong bias towards reconstructing lighter strings (more negative $\log_{10}(G\mu)$s); note that the range of absolute errors here is twice greater than that of the case without foregrounds. The maximum absolute error of about $-1.16$ (at ($\log_{10}(G\mu)=-16.3$) corresponds to an error in the actual tension of around $-93.1\%$. This is significantly larger than the size of the error made at logarithmic tension $-17.0$ without foregrounds. As there, if we were instead to look for the lowest tension we could reconstruct while making less than $-50\%/+100\%$ error, we would estimate it to be in the vicinity of $-16.0$.\footnote{None of our trials made an error in the actual tension greater than $100\%$; the most positive error found is $\approx 37.1\%$ in the actual tension (at a logarithmic tension of $-16.3$).}

\begin{figure}
    \centering
    \includegraphics[scale=0.8]{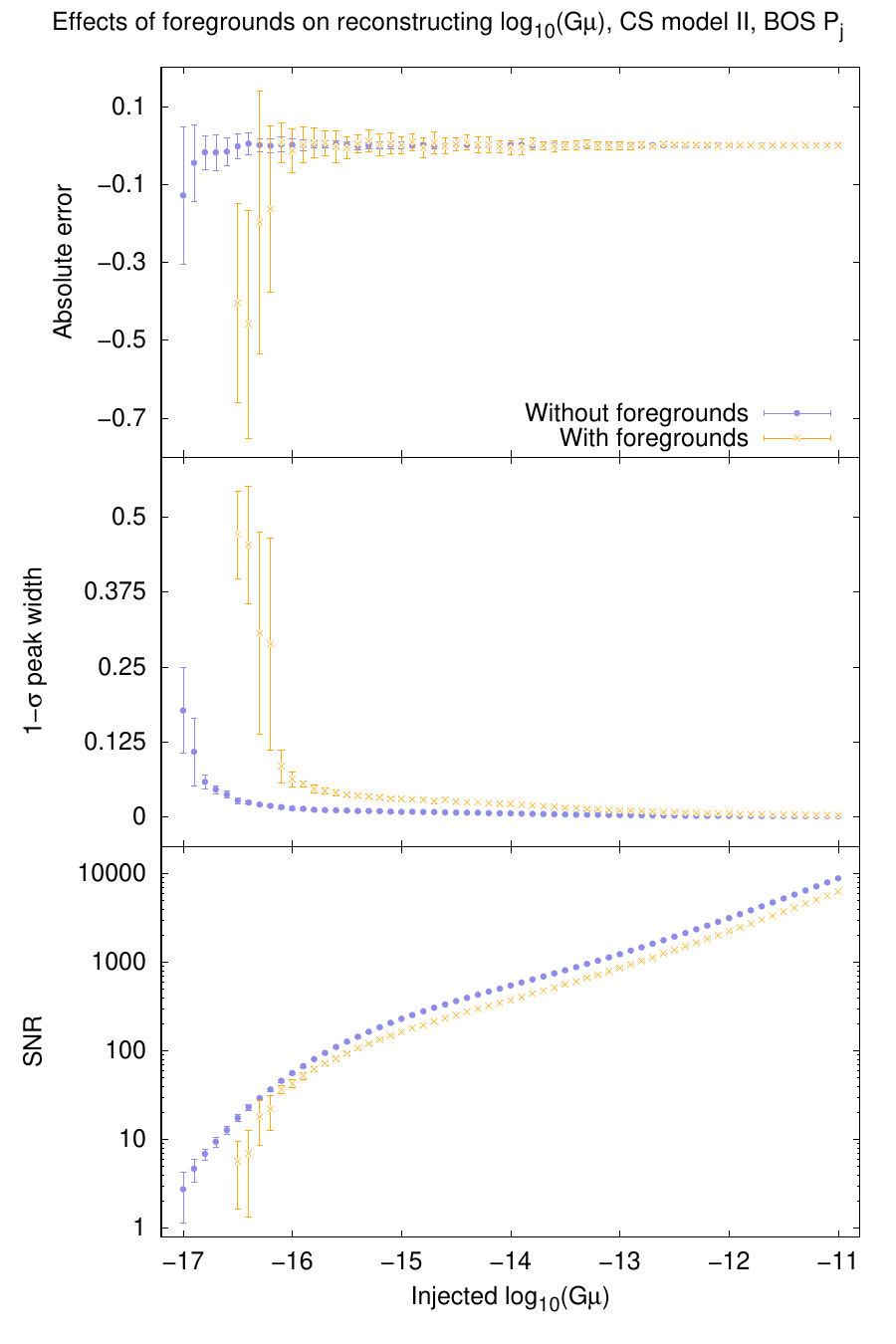}
    \caption{Various performance metrics comparing reconstructions with and without foregrounds. All points represent the mean of the 30 values collected for that tension with or without foregrounds, and all error bars are at one standard deviation of those 30 values, for both data sets. The ability to reliably reconstruct a signal degrades below $\log_{10}(G\mu)$ of $\approx -16.0$ and $\approx -16.7$ for the cases with and without foregrounds, respectively.}
    \label{fig:csII-BOS-performance}
\end{figure}

Further details on the fitting performance can be found in figure~\ref{fig:csII-BOS-performance}. There, we show the dependence on the injected logarithmic tension of, in order top-to-bottom: the absolute error in the reconstructed logarithmic tension; the likelihood distribution width; and the SNR of the reconstructed signal.

The points are at the mean values of the 30-trial data set for each value of injected tension we studied, and the error bars are the standard deviations of that data set. The absolute error plot may therefore be reasonably compared to figure~\ref{fig:csII-BOS-abs-histogram}. The likelihood distribution width's mean values show the general increasing uncertainty in the reconstructed parameter with decreasing tension, consistent with figure~\ref{fig:csII-BOS-abs-histogram}. The SNR is effectively power-law decreasing, with a slight bump beginning around $\log_{10}(G\mu)=-14.0$.
 
Particularly in this visualization, we see a noticeable decay in the quality of the reconstruction around $\log_{10}(G\mu)=-16.7$ for the case without foregrounds, and around $\log_{10}(G\mu)=-16.0$ for the case with foregrounds. However, a separation between the results of reconstructions with and without foregrounds can be seen as low as $\log_{10}(G\mu)\simeq -13.0$, most clearly in the points of the $1\sigma$ width plot and in the error bars of the absolute error plot. The SNR is very consistently lower for reconstructions with foregrounds, $\approx 70\%$ of the SNR without foregrounds, until degradation at the lowest tensions as discussed above.

\begin{figure}
    \centering
    \includegraphics[scale=0.8]{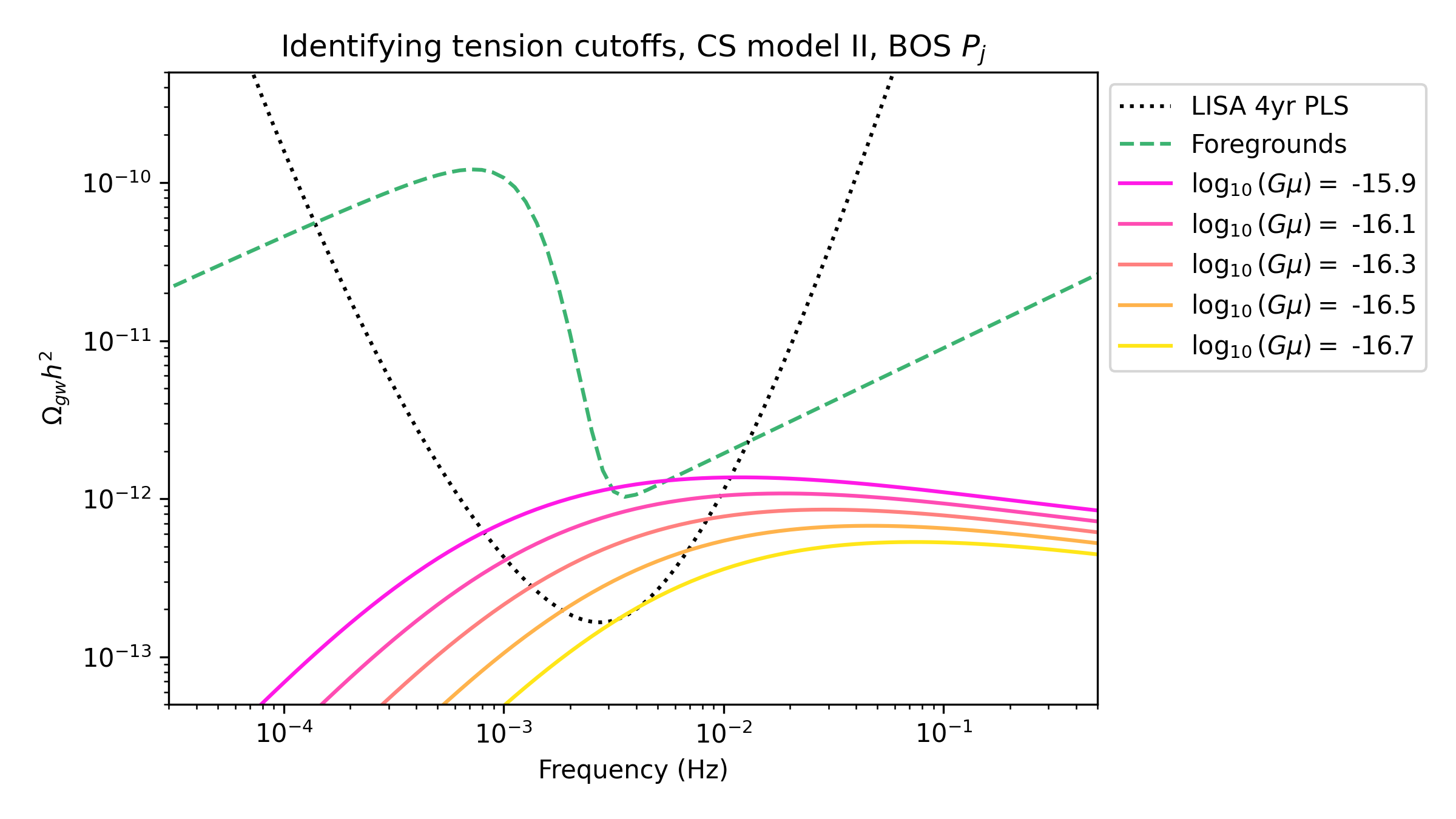}
    \caption{At a tension of just above $\log_{10}(G\mu)=-15.9$, the cosmic string Model II (BOS $P_n$) signal drops below the combined galactic and extragalactic foregrounds. At just below $\log_{10}(G\mu)=-16.7$, the signal drops below the LISA power-law sensitivity curve. This explains the noticeable decay in fitting performance around those tensions, as seen in figures~\ref{fig:csII-BOS-abs-histogram} and \ref{fig:csII-BOS-performance}.}
    \label{fig:csII-BOS-cutoffs}
\end{figure}

The decrease in fitting ability at lower tension can be explained by examining the position of the cosmic string SGWB relative to both the combined galactic and extragalactic foregrounds and the LISA PLS curve in the relevant range of tensions, as shown in figure~\ref{fig:csII-BOS-cutoffs}.

The importance of $\log_{10}(G\mu)=-16.0$ is made clear here, as this is (approximately) the tension at which the cosmic string SGWB drops entirely below the combined foreground curve. By the time we reach $\log_{10}(G\mu)=-16.5$, the cosmic string SGWB is, at best, about three
times smaller in amplitude than the combined foreground. As a consequence, the presence of foregrounds reduces the effective constraints LISA can place on the cosmic string tension.

\subsection{Probing the pre-BBN expansion rate with reconstruction}\label{ssec:mod-pre-bbn}

We can also consider the ability of LISA to detect modifications to the string SGWB due to non-standard pre-BBN histories. First, let us consider a generic case of new DoF BSM. For simplicity, we add $\Delta g$ new DoF all at once at a temperature of $T_\Delta$ which modifies the expansion rate briefly around the corresponding time. We can then add these new DoF and temperature of injection as parameters for \texttt{SGWBinner} to use in reconstructing the signal. As this significantly increases the parameter space the minimizer needs to search, we will only consider a narrow range of tensions as proof-of-concept to determine the feasibility of this approach. For this pilot study, we used ranges of $-10.1\leq\log_{10}(G\mu)\leq -9.9$, $0\leq\Delta g\leq 120$, and $0.005\,\text{GeV}\leq T_\Delta\leq 50\,\text{GeV}$.

\begin{figure}
    \centering
    \includegraphics[scale=0.85]{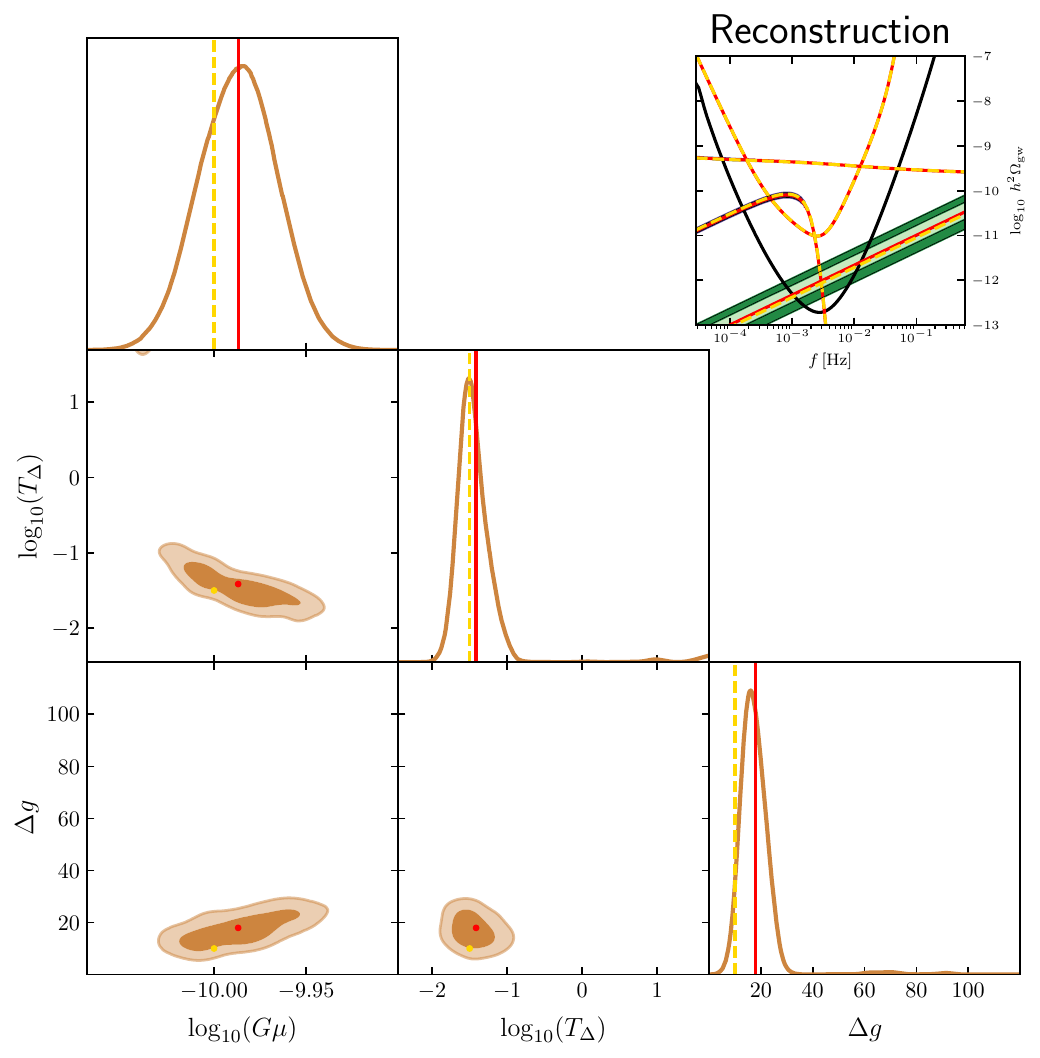}
        \caption{The reconstruction of a Model IIb cosmic string SGWB, in the presence of foregrounds (not shown in the corner plots), at $\log_{10}(G\mu)=-10.0$ with modifications due to $10$ new DoF added at $10^{-1.5}\,\text{GeV}$. Repeated fits of the same injected signal consistently produce closed contours that contain the correct values, although the biases of the returned means can change.}
    \label{fig:csII-dof-example}
\end{figure}

\begin{figure}
    \centering
    \includegraphics[width=0.99\textwidth]{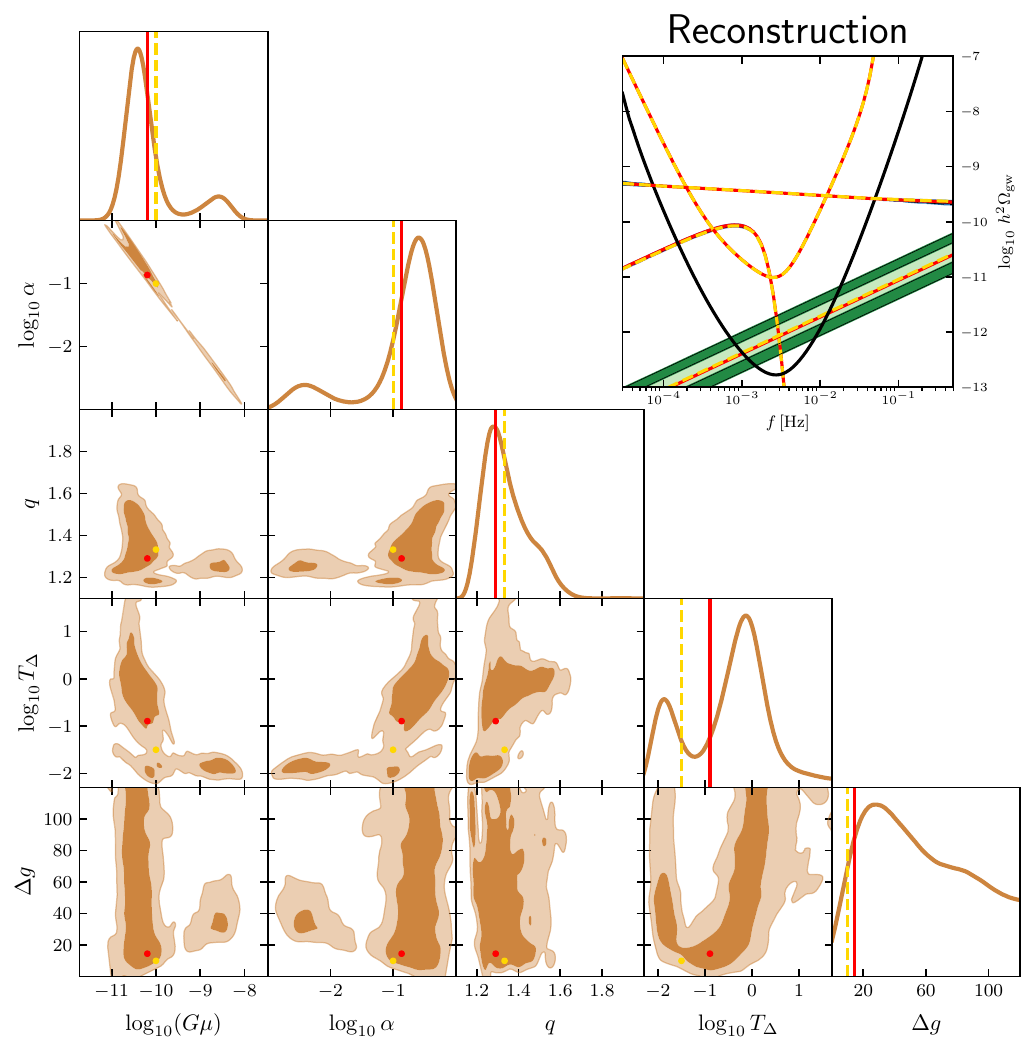}
        \caption{The reconstruction of a Model Ib cosmic string SGWB, in the presence of foregrounds, at $\log_{10}(G\mu)=-10.0$ with modifications due to $10$ new DoF added at $10^{-1.5}\,\text{GeV}$. }
    \label{fig:csI-dof-example}
\end{figure}

\begin{figure}
    \centering
    \includegraphics[width=\textwidth]{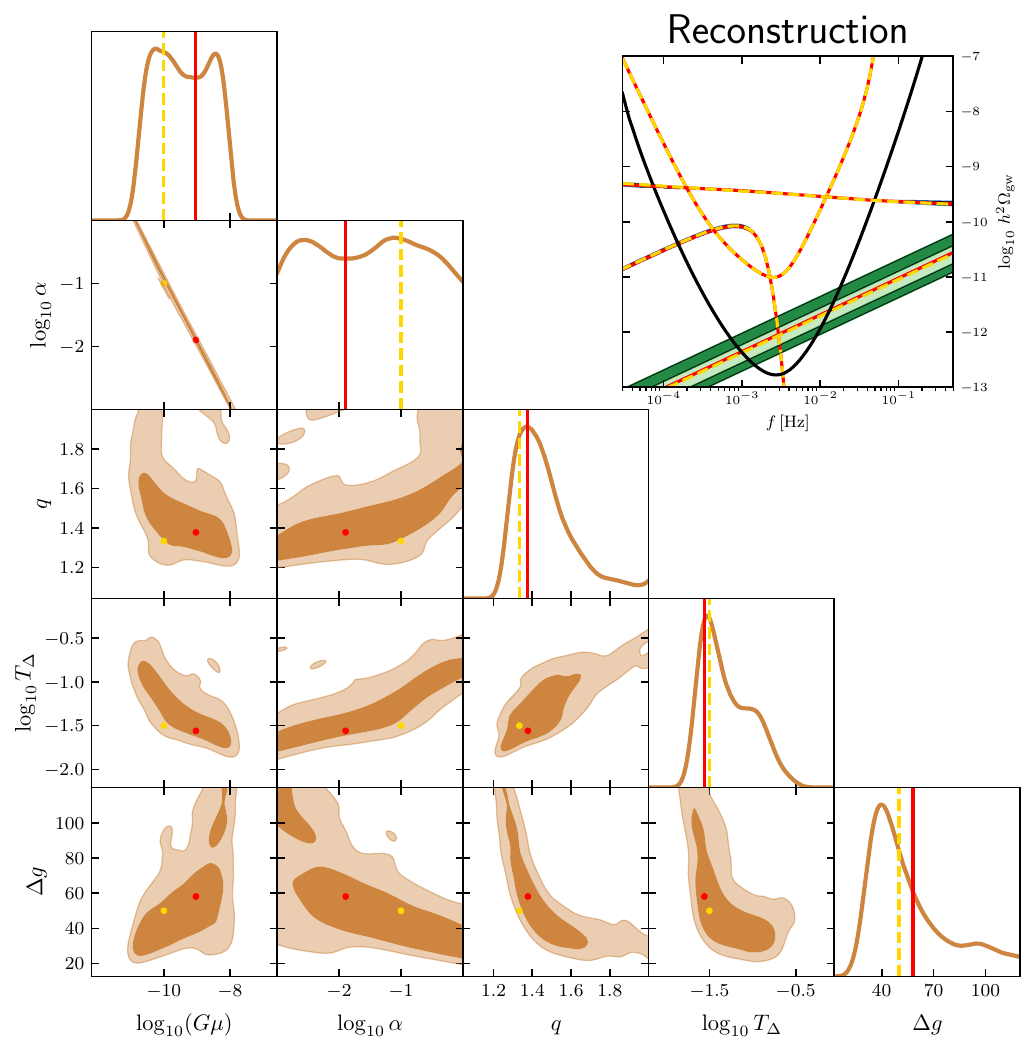}
        \caption{The reconstruction of a Model Ib cosmic string SGWB, in the presence of foregrounds, at $\log_{10}(G\mu)=-10.0$ with modifications due to $50$ new DoF added at $10^{-1.5}\,\text{GeV}$. }
    \label{fig:csI-dof-example2}
\end{figure}

Unsurprisingly, the quality of the reconstruction depends strongly on the values of $\Delta g$ and $T_\Delta$. The result for the Model IIb signal with $\log_{10}(G\mu)=-10.0$, $\Delta g=10$, and $T_\Delta = 10^{-1.5}\,\text{GeV}$ in the  presence of foregrounds is shown in figure~\ref{fig:csII-dof-example}. In this case, the $\Delta g=0$ is significantly disfavored, and thus this cosmic string signal would allow us to also confirm the presence of new DoF from beyond the SM. We also test the reconstruction of this same new DoF setting in the case of Model Ib signal in figure~\ref{fig:csI-dof-example} (with $\mathcal{F}=0.1$ as well). We observe that the reconstruction is more challenging, as the injected value of $\Delta g=10$ is compatible with $\Delta g=0$ at 2\,$\sigma$. The degradation in the reconstruction is partly due to the two additional free parameters involved in Model I. This model also includes the evolution of the loop number density due to the modified expansion. Given the relatively slow relaxation of the network, the modification is then spread over a longer time, resulting in a smoother spectrum which also makes the reconstruction more challenging. 
While finding the imprint of new degrees of freedom is more difficult, in the spectra of Model I it is not impossible. In figure~\ref{fig:csI-dof-example2} we show that a successful reconstruction is possible with the number of new degrees of freedom increased to $\Delta g=50$.

These are only single examples, although we tested the performance of the reconstruction in all corners of the relevant parameter spaces in order to validate \texttt{SGWBinner}'s ability to study a range of new DoF scenarios. Signals with lower $T_\Delta$ are generally more reliably reconstructed, as a larger fraction of the SGWB in the LISA band is modified; as the temperature increases, the modified region is pushed to higher frequencies, making all reconstructions less precise and less accurate (and, for high $T_\Delta\sim 10\,\text{GeV}$, potentially leading to contours which do not close). Larger numbers of new DoF are likewise more reliably reconstructed, as the magnitude of the change to the SGWB is greater. For very low temperatures, $\Delta g\sim\mathcal{O}(1)$ remains distinguishable from no new DoF.

These results are something of a best-case scenario, as they rely on a string network with $\log_{10}(G\mu)\approx -10.0$, at the upper limit of what's been constrained by current measurements. We expect that for very low tension strings, reconstructing non-standard pre-BBN scenarios would be significantly more difficult, especially given the fact that lowering the tension causes a shift of these features towards higher frequencies and away from the LISA window. However, making any definitive statement requires significant additional data and investigation with a broader range of $G\mu$ and $T_\Delta$.

We can also consider more dramatic modifications where for a period of time the expansion is driven by some new energy constituent parameterized by the equation-of-state parameter $w$, instead of radiation. Examples here would be an early period of matter domination with $w=0$ or kination with $w=1$ active between inflationary reheating and BBN~\cite{Allahverdi:2020bys}. 

\begin{figure}
    \centering
    \includegraphics[scale=0.9]{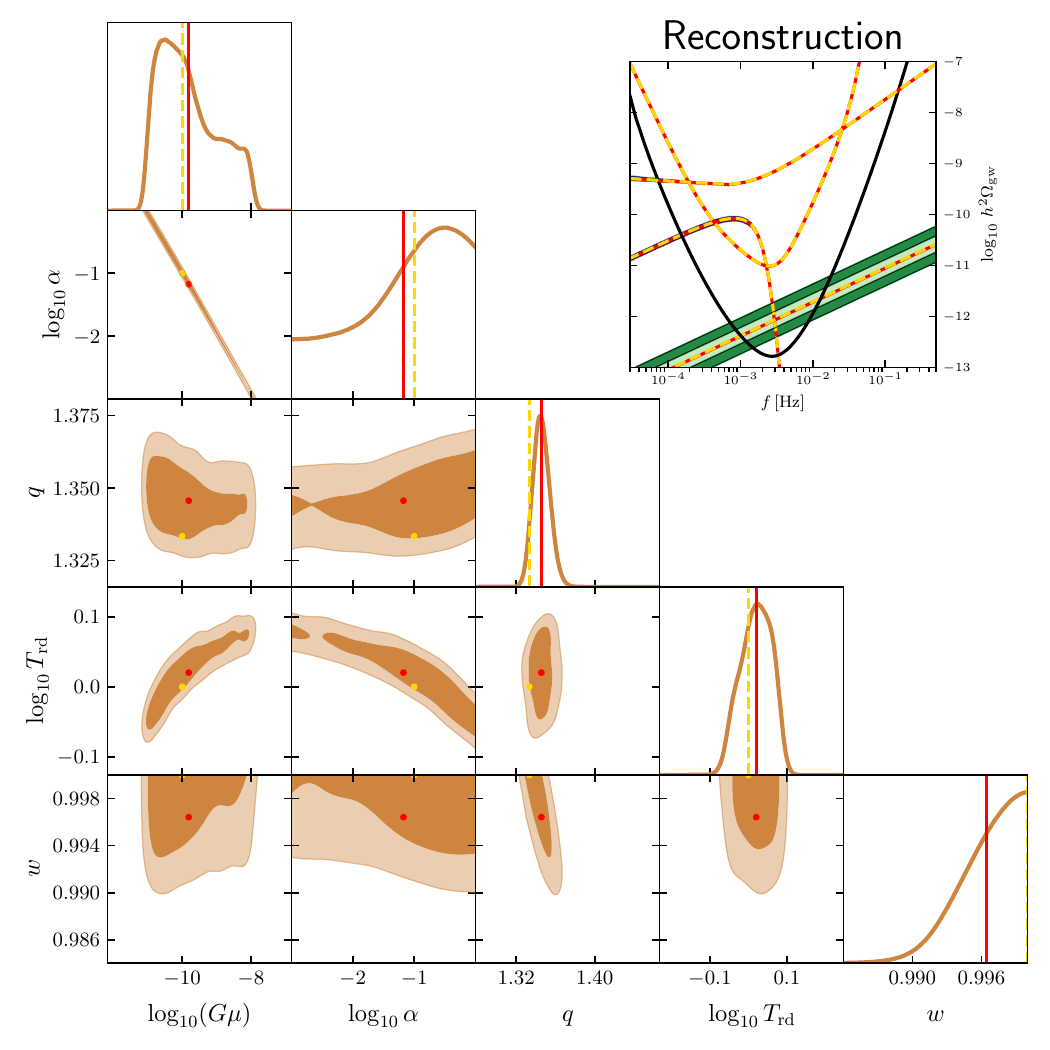}
        \caption{The reconstruction of a Model Ia cosmic string SGWB, in the presence of foregrounds, at $\log_{10}G\mu=-10.0$ with non-standard equation of state $w=1$.}
    \label{fig:eos}
\end{figure}

In figure~\ref{fig:eos}, we show an example of the parameter estimation for Model Ia,  with its two additional free parameters: the equation-of-state parameter $w$, which is set to 1 assuming a kination phase, and the temperature of the universe when the non-standard evolution ends and switches to the radiation-dominated universe, denoted as $T_{rd}$ (we also took $\mathcal{F}=0.1$). The reconstruction is quite successful in this example, but, as before, the quality of the results will necessarily be better for lower $T_{rd}$ and will also depend strongly on the value of $w$. In particular, as shown in section~\ref{ssec:mod-pre-bbn}, for $w<1/3$ the spectrum develops a negative slope for $f>f_{rd}$ and its amplitude quickly drops below that of the foregrounds and below the LISA sensitivity window. Also, as previously discussed, for all $w\le 1/9$ the predicted spectrum is the same and very similar to that of strings created during inflation, which makes the reconstruction of $w$ more challenging.

\subsection{Different injection and recovery templates}\label{sec:fit-I-II}

One can also ask if the LISA SGWB search needs to include both Model I and Model II in the template library. Model I with fixed parameters $\alpha=0.1$, $\mathcal{F}=0.1$, and $q=4/3$ is typically taken as a good approximation of Model II. However, while the agreement at the high-frequency plateau is exact, there are important, frequency-dependent differences between these models. Since the power spectrum is modelled in different ways in these models, the shape of the peak of the SGWB can be slightly different. Moreover, Model I, unlike Model II, includes the impact of the evolution of the effective number of DoF in the dynamics of the network and in loop production, which results in a smoothing of these features. As a result, the matching between these two models for this set of parameters, although quite good, is not perfect.
Nevertheless, such a mismatch is not necessarily relevant for LISA. If the mismatch between the two templates was largely within the reconstruction error bars in the whole cosmic string parameter space, one could disregard one of the templates in the frequency shape reconstruction required to isolate the SGWB signal in the global fit.

Moreover, our tests up to this point have been done by injecting a signal using some model with fixed parameters, and then fitting the signal using that same model. This is a good test of the reconstruction capabilities, but may not be entirely accurate to the real situation of fitting an unknown cosmic string background with a template. Theoretical uncertainties on the cosmic string modelling might indeed introduce systematics, meaning that the true string signal might be slightly different from our templates --- the question of the effect of gravitational backreaction on the SGWB has already been mentioned, as one example.

To investigate both of these questions, we take advantage of one of \texttt{SGWBinner}'s features: the template used to inject the signal and the template used to recover (or fit) the signal does not have to be the same. For our example, we will inject a Model II signal at a fixed tension and fit it twice, once using Model II and once using Model I. We will not fix any of the parameters of Model I, as we wish to see if it naturally selects the typical values of $\alpha$, $\mathcal{F}$, and $q$ mentioned above. We will then investigate both the recovered parameters of the fit using Model I, and compare the evidences for the fit using Model II versus using Model I.

We inject Model II in the presence of foregrounds with two different tensions: $\log_{10}(G\mu)=-11.0$ and $-14.0$. The first is done to test the reconstruction in a regime where most of the LISA band sees the high-frequency plateau, where the agreement should be exact; the second is done to see how the reconstruction is affected for the tensions where the foreground's overlap with the string GWB becomes significant (see Fig.~\ref{fig:csII-BOS-performance} and nearby discussion). Thirty injections are performed for each tension, and the result of the reconstructions are summarized in table~\ref{tbl:model-ii-by-i}. Larger tensions run into issues of degeneracy, whereas at lower tensions the foregrounds cause a degradation of fitting quality regardless of what template recovers what signal.

\begin{table}
    \begin{tabular}{c|c|cccc}
        \multicolumn{2}{c|}{~} & \multicolumn{4}{c}{Recovered parameters} \\
        \multicolumn{2}{c|}{~} & $\log_{10}(G\mu)$ & $\log_{10}(\alpha)$ & $q$ & $\log_{10}(\mathcal{F})$ \\\hline        
        \multirow{2}{*}{Injected $\log_{10}(G\mu)=-11.0$} & Mean & $-12.1$ & $-1.05$ & $1.33$ & $-0.484$ \\
        & Std. dev. & $0.0842$ & $0.608$ & $0.00130$ & $0.297$\\
        \multirow{2}{*}{Injected $\log_{10}(G\mu)=-14.0$} & Mean & $-14.0$ & $-1.12$ & $1.53$ & $-0.896$ \\
        & Std. dev. & $0.538$ & $0.563$ & $0.149$ & $0.318$
    \end{tabular}
    \caption{The results of 30 recoveries of a Model II signal by a four-parameter Model I template for large ($10^{-11}$) and moderate ($10^{-14}$) tensions. Degeneracies in the Model I parameters lead to poor fitting at large tensions, which are improved at moderate tensions. The presence of foregrounds confounds lowering the tension further.}\label{tbl:model-ii-by-i}
\end{table}

The recovery for injected $\log_{10}(G\mu)=-11.0$ is shown in the first two rows of Table~\ref{tbl:model-ii-by-i}. The mean recovered tension is many standard deviations below the injected tension; $\log_{10}(\alpha)$ and $q$ are within one standard deviation of the typical values used when approximating a Model II SGWB using Model I, but the log-fuzziness is just over two standard deviations too high. This indicates that it would not be correct to perform searches using only Model I templates, as a Model II-like signal would be misidentified as coming from a network with different parameters and thus different physics.

Here, much of the mismatch between the expected outcomes and the recovered values can be explained by the degeneracy between the Model I parameters at large tension values. In the case studied, the log-fuzziness parameter is higher than the expected $-1.0$, raising the height of the SGWB, but the tension is lesser, lowering the height of the SGWB. Because an SGWB at this tension is high above the LISA noise curve and foregrounds, \texttt{SGWBinner} reports very low uncertainties on these fits, making the difference between the recovered and injected parameters (in a number-of-standard-deviations sense) much worse than the lower-tension case we first studied.

For $\log_{10}(G\mu)=-14.0$, the recovered tension, log-fuzziness, and log-loop size all fall within one standard deviation of their expected or typical values; the power spectral index about one and a half standard deviations high. For intermediate tensions, the recovery of a near-but-not-exact signal by a Model I template can be given moderate credence. However, further testing and optimization of the code is necessary to obtain robust and reliable results---note, for example, the significant increase in the standard deviation of the fit $\log_{10}(G\mu)$ and $q$ values between log-tensions of $-11.0$ and $-14.0$.

\begin{figure}
    \centering
    \includegraphics[width=1.00\linewidth]{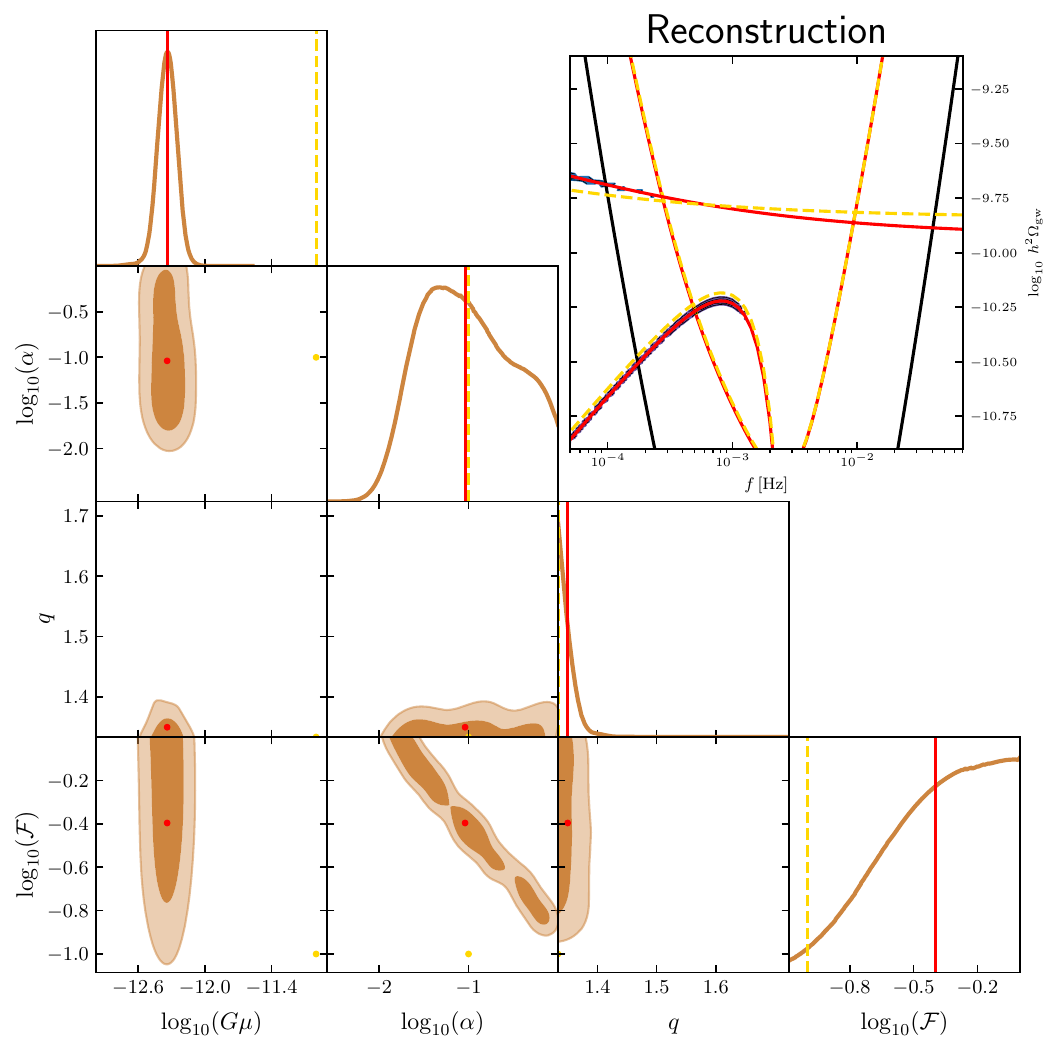}
    \caption{The reconstruction of a Model II cosmic string SGWB with $\log_{10}(G\mu)=-11.0$ by Model I with all four parameters free. In the high-frequency plateau region, degeneracies between $\alpha$ and $\mathcal{F}$ make recovering the correct tension difficult. We have zoomed in on the reconstruction plot to indicate that over a wide range of the LISA band, the injected SGWB (dashed gold) does not fall within the 1$\sigma$ bands of the recovered SGWB (blue regions surrounding the red solid line). This mismatch has caused the quality of fit of the foregrounds to degrade as well.}
    \label{fig:model-ii-by-i}
\end{figure}

Examining the reconstructed Model I SGWB in comparison to the injected Model II SGWB, as shown in fig.~\ref{fig:model-ii-by-i} for a $\log_{10}(G\mu)=-11.0$ scenario, shows the impact of the degeneracy between $\alpha$ and $\mathcal{F}$, both of which have large variance according to table~\ref{tbl:model-ii-by-i}. In the inset ``Reconstruction'' plot, we have zoomed in to emphasize an important outcome of the large mismatch between the injected signal and the reconstructed signal: over a significant range frequencies, the injected SGWB does not fall within the 1$\sigma$ bands of the reconstructed SGWB.

This has important consequences for the rest of the reconstructions. Observe that, at frequencies below 1\,mHz or so, the reconstructed signal is higher than the injected signal. In the total reconstruction, this mismatch is compensated for in the reconstruction of the foreground ``bump'' in the bottom-left of the plot: it is consistently lower than the injected value. Close examination shows that here as well the injected foreground signal falls outside of the 1$\sigma$ bands of the reconstructed signal over some range of frequencies, if just barely. The mismatch between the injected and recovered signal has been absorbed into the global fit, causing issues with other reconstructions. This would cause significant knock-on effects in a fully-integrated \texttt{SGWBinner}, including, e.g., SGWBs from first-order phase transitions and inflation~\cite{LISACosmologyWorkingGroupPT,LISACosmologyWorkingGroupInflation}.

Finally, we may quantify which model is a better fit for any given reconstruction using the evidences reported by \texttt{polychord}, used by \texttt{SGWBinner} to find best-fit parameters in each run. For each of the 30 runs mentioned above, we found the log-Bayes factor with the convention that $\log_{10}(BF)>0$ 
indicates support for fitting the signal with Model II. For the $\log_{10}(G\mu)=-11.0$ injections, we found the mean of $\log_{10}(BF)$ to be $7.61$, with a range of $5.32$ to $10.3$. In all cases, this is decisive evidence for reconstructing the signal with Model II. For the $\log_{10}(G\mu)=-14.0$ injections, we found the mean of $\log_{10}(BF)$ to be $3.74$, with a range of $0.113$ to $5.99$; while Model II was still preferred in every scenario, in one realization the evidence is not worth mentioning, and only 27 of the 30 realizations contain decisive evidence for Model II.

While we have not attempted a reconstruction of an injected Model I signal using both Models I and II, we anticipate similar results, if not stronger: the additional parameters in Model I permit a greater range of SGWB morphologies, and a Model II template should never match, e.g., a Model I signal with $\alpha=10^{-3}$ (except, perhaps, in the very-high-tension plateau case, as discussed above). This underscores the necessity of having both Model I and Model II in the \texttt{SGWBinner} template bank.


\subsection{Models with non-gravitational radiation}\label{sec:fng_model}

The preceding sections looked at string models where the only channel for string decay is into GWs. Outside of these minimally-coupled strings, several models exist which have other, sometimes dominant, modes of energy loss.

\begin{figure}
    \centering
    \includegraphics[width=1.00\textwidth]{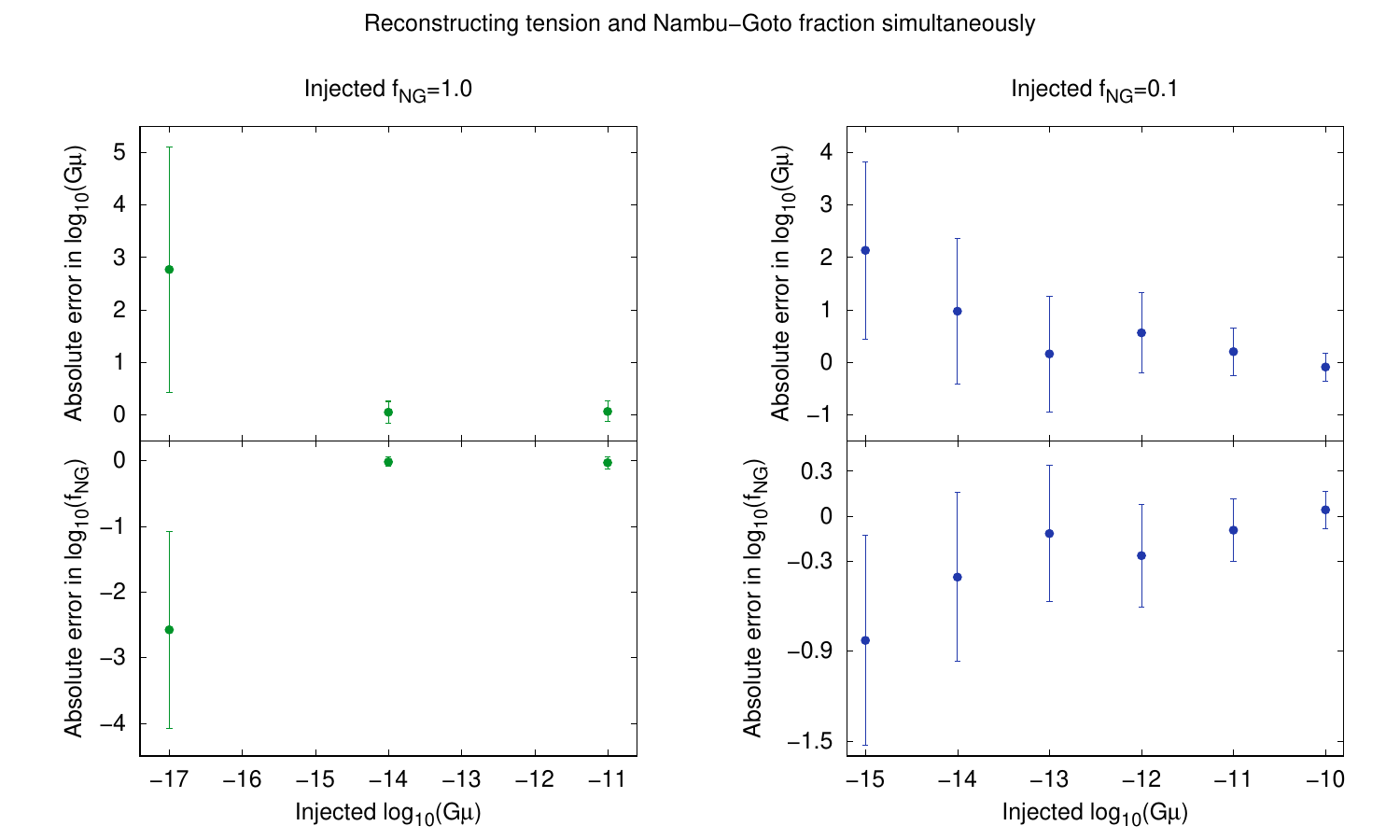}
    \caption{The reconstruction of a model for an Abelian--Higgs string network for both the case where all radiation is into GWs (left) and for the case where only 10\% of radiation is into GWs (right). The template used is a direct modification of Model II with a prefactor $f_\text{NG}$ representing the fraction of gravitational-wave radiation. The inclusion of an additional parameter to fit, combined with the degeneracy of $f_\text{NG}$ with $G\mu$, leads to lower accuracy and precision in the reconstructions.}
    \label{fig:fng-results}
\end{figure}

Consider the work of refs.~\cite{Matsunami:2019fss, Auclair:2019jip,  Hindmarsh:2021mnl, Hindmarsh:2022awe}, which suggests modeling an Abelian--Higgs (AH) network of strings by multiplying a Nambu--Goto (NG) SGWB by a parameter $f_\text{NG}$, representing the non-gravitational radiation:
\begin{equation}
  \Omega^{(\text{AH})}_\text{\rm gw} = f_\text{NG}\Omega^{(\text{NG})}_\text{\rm gw}\,.
\end{equation}
This can be straightforwardly implemented in \texttt{SGWBinner} by multiplying any existing template by a new parameter. As ref.~\cite{Hindmarsh:2022awe} use the data of Model II, so shall we; our two parameters are now $\log_{10}(G\mu)$ and $\log_{10}(f_\text{NG})$, where we choose the logarithm of the NG fraction because it can vary over many decades. We conduct two sets of runs on this model. The first has $f_\text{NG}=1.0$ (corresponding to all of the emission into GWs) at logarithmic tensions of $-11.0$, $-14.0$, $-17.0$ to cover the same range as Model II in \ref{ssec:model-ii-results} to give us a baseline for comparison. The second has $f_\text{NG}=0.1$ at logarithmic tensions from $-10.0$ down to $-15.0$ in steps of $-1.0$, chosen because ref.~\cite{Hindmarsh:2021mnl} sets this as the 95\%-confidence upper bound. As with the runs from section~\ref{ssec:model-ii-results}, we perform 30 trials for each choice of parameters and find the average and standard deviation of the best-fit results. These trials are all done in the absence of foregrounds to focus on the effect of introducing the Nambu--Goto fraction parameter on reconstructions. The results are shown in figure~\ref{fig:fng-results}.

The accuracy and precision of the reconstructions when $f_\text{NG}=1.0$ is lower than reconstructions done at the same tension for the base Model II, but this is to be expected. An additional parameter should allow for a larger range of tensions which \texttt{SGWBinner} finds to be good fits to the injected signal. The fits are still fairly accurate for the $-11.0$ and $-14.0$ logarithmic tensions; they bias slightly towards lighter strings, with larger standard deviations compared to the base case. The $-17.0$ case is a much worse fit, but this is explained as follows. This tension is technically below the LISA PLS, and so \texttt{SGWBinner} will accept most marginal signals as fitting the injected signal. For every tension, there is an NG fraction which places the resulting SGWB just below the LISA PLS. Thus, \texttt{SGWBinner} reports a very wide range of tension/fraction combinations. 

\begin{figure}
    \centering
    \includegraphics[width=0.5\textwidth]{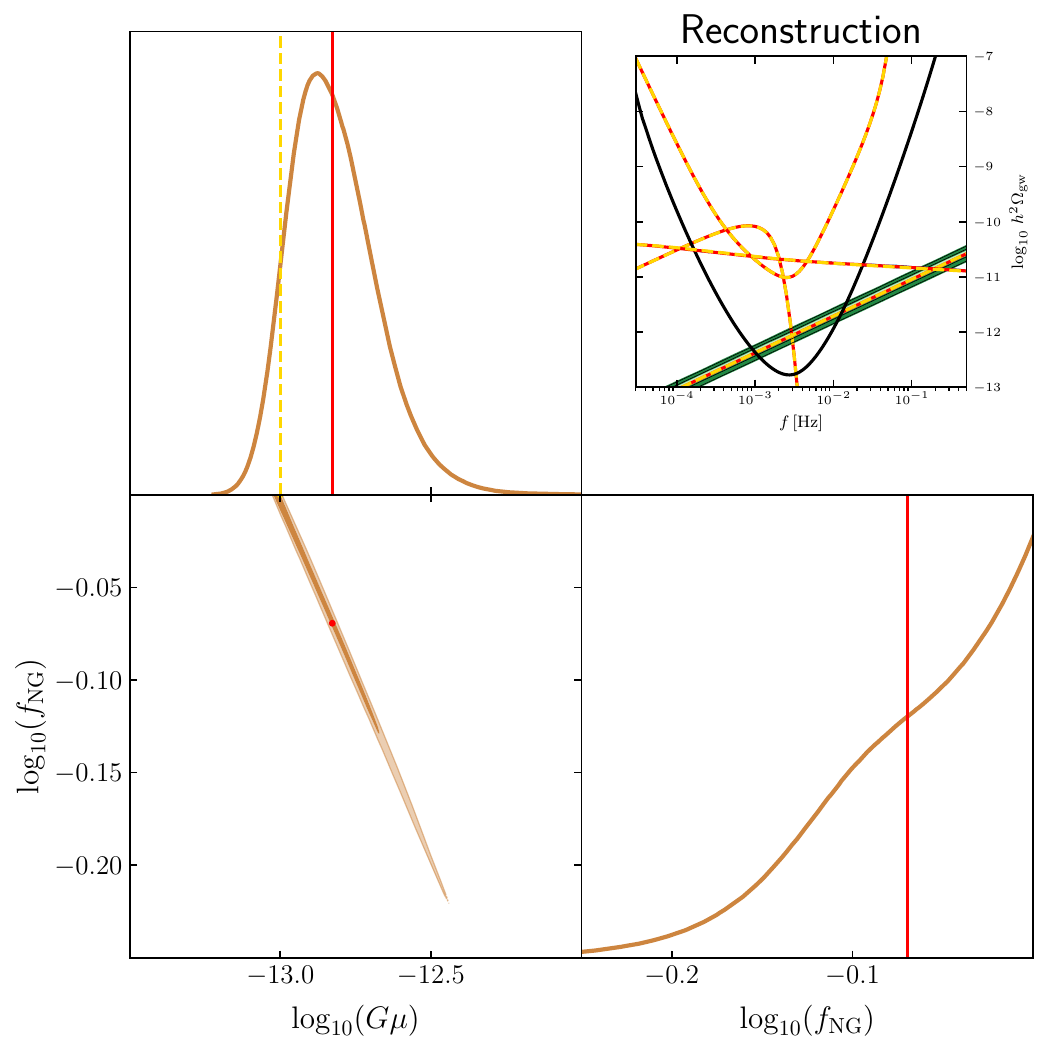}~\hspace{0.5em}~\includegraphics[width=0.5\textwidth]{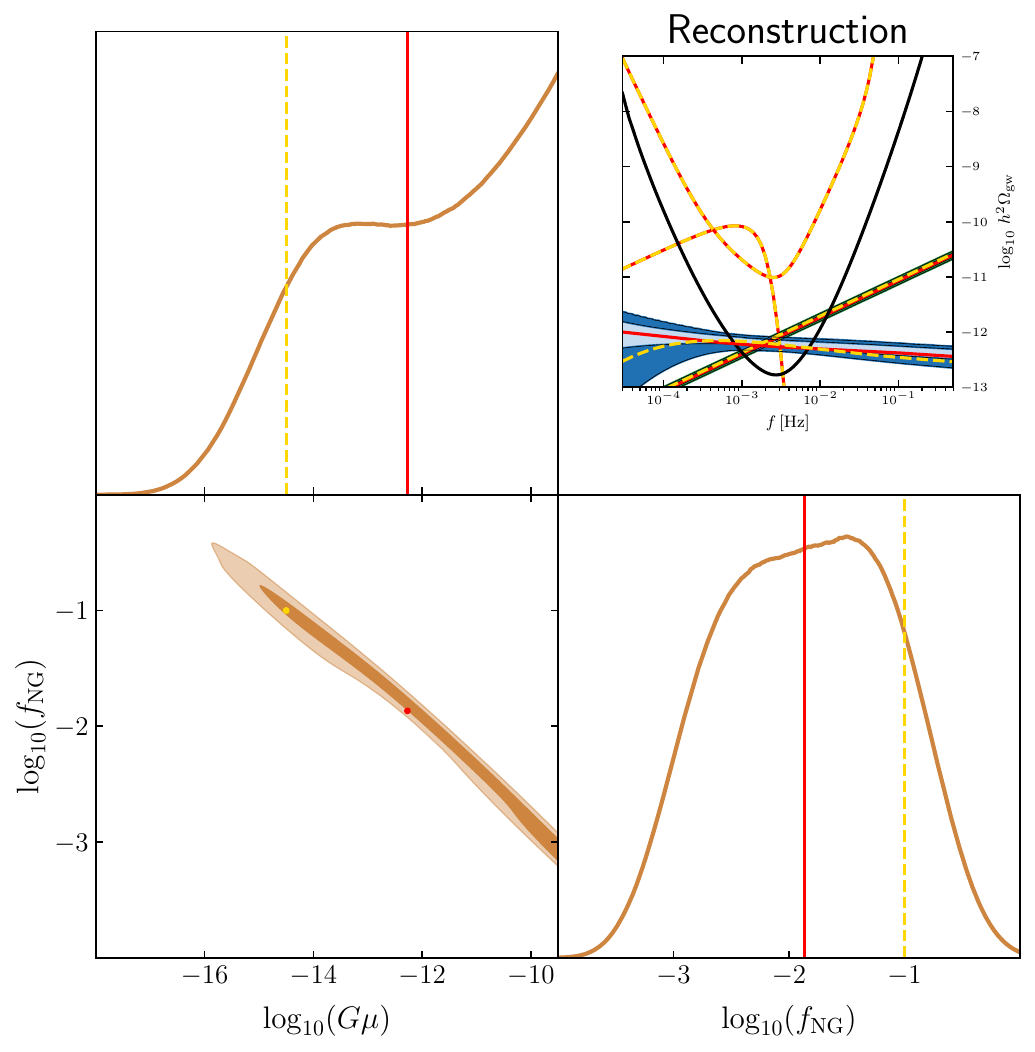}
    \caption{(Left) The result of a single reconstruction of the Abelian--Higgs model with input parameters of $\log_{10}(G\mu)=-13.0$, $f_\text{NG}=1$, illustrating the strong degeneracy between the two parameters. (Right) The result of a single reconstruction of the Abelian--Higgs model with input parameters of $\log_{10}(G\mu)=-14.5$, $f_\text{NG}=0.1$, illustrating the impact of the initial tension on the reconstruction.
    }
    \label{fig:fng-spread}
\end{figure}

The accuracy and precision of the reconstructions when $f_\text{NG}=0.1$ show a fairly consistent uncertainty of about a factor ten in the tension (and thus a factor $\sqrt{10}$ in the Nambu--Goto fraction). The error is greater than the base case. To explain further the larger errors and uncertainties, we recall: as the tension varies, the location (frequency value) and amplitude ($\Omega_{\rm gw}$ value) of a point on the string SGWB shift roughly like $1/G\mu$ and $\sqrt{G\mu}$, respectively. For example, we might expect the SGWB for $\log_{10}(G\mu)=-10.0$, $f_\text{NG}=0.1$ to be comparable in amplitude to the SGWB for $\log_{10}(G\mu)=-12.0$ in the base case. This leads to a very strong degeneracy in the two parameters illustrated in figure~\ref{fig:fng-spread}. 

Interestingly, the uncertainty in the case of $\log_{10}(G\mu)=-12.0$, $f_\text{NG}=0.1$ is less than the other points with $f_\text{NG}=0.1$, but its absolute error is greater when compared to the points adjacent to it. The reconstructions of this point prefer heavier strings with lower Nambu--Goto fraction than the injected values. While it is not immediately clear why this is the case, the answer may lie in which parts of the string SGWB are in the LISA window. Much of the structure in the string SGWB is in the radiation-to-matter-transition `bump' and in the changing-degree-of-freedom `steps'. If the portion of the SGWB in the LISA window lacks much of this structure (as one might get by simply lowering a high-tension SGWB), then a slightly heavier string which emits less of its energy into GWs might be able to provide a better fit by \texttt{SGWBinner}'s metrics.

\section{Science interpretation}\label{sec:science-interpretation}

Finding a stochastic background of GWs consistent with one of the templates described in this paper would have a huge impact on the study of physics beyond the standard model. 

Perhaps the most important piece of information one can obtain from such an observation is the scale of the symmetry breaking that leads to the string formation. This scale is directly related to one of the central parameters of our reconstructed spectrum, the tension of the string. Our work shows that we should be able to recover the value of this tension with a precision that in many cases is within one order of magnitude of the real one. Assuming a mild dependence  of the value of the tension on the coupling constants one can use the best fit value obtained from the reconstructed SGWB to estimate the scale of new physics.

The dependence of the scale of symmetry breaking on the coupling constants diminishes the significance of achieving an extremely precise reconstruction of the string tension. However, discovering conclusive evidence of a phase transition at a certain energy scale provides invaluable information for constructing phenomenological models for physics BSM, even if the exact value of this scale can not be pinpointed.

Note that, given the results we presented in section \ref{sec:results}, we expect LISA to be sensitive to 
tensions as low as $G\mu =10^{-16.0} $, which translates into energy scales roughly of $10^{11}~\text{GeV}$. Therefore, even including foregrounds, we will be able to probe new physics in energy scales not accessible to terrestrial accelerators. This intermediate scale falls below the typical energy range of Grand Unification Theory (GUT) and it could so far only be probed indirectly via, for example, proton decay searches in certain models.

On the other hand, numerical studies of Nambu-Goto cosmic string networks that we used in this paper (Model II) seem to indicate
that the scaling distribution of non-intersecting loops peaks at around $\alpha \sim 0.1$ \cite{Blanco-Pillado:2013qja}. A substantial departure from this regime could be, in principle, identified using a Model I reconstruction. Finding a large departure from the preferred value of $\alpha$ would likely point to a more complex cosmic string model than the one discussed here. Numerical simulations of strings with reduced intercommutation probability~\cite{Avgoustidis:2005nv}, for example, suggest that the production of small loops may be favored in cosmic superstring networks. Such a departure from the expected values of $\alpha$ could also hint at the possibility of a much richer string micro-physics than that of the strings studied in this paper. Models with such new internal structure have been suggested in the literature ~\cite{Witten:1984eb,Nielsen:1987fy,Carter:1990nb,Vilenkin:1990mz}, but the details of loop production and their implications in terms of their GW signatures are much less understood at this point. Alternatively, interpretations of such a deviation could potentially result from the possibility known in the literature as Model III \cite{Ringeval:2005kr, Lorenz:2010sm, Auclair:2019zoz, Auclair:2020oww}.

Observation of the spectrum would also help to settle the debate on possible emission other than GWs from the cosmic string network~\cite{Vachaspati:1984yi, Vincent:1997cx, Moore:1998gp, Olum:1998ag, Olum:1999sg, Moore:2001px, Hindmarsh:2008dw, Hindmarsh:2017qff, Matsunami:2019fss, Hindmarsh:2021mnl, Blanco-Pillado:2022rad, Hindmarsh:2022awe, Blanco-Pillado:2023sap}. In Sec.~\ref{sec:fng_model} we find that a mild reduction of the GW emission would still allow for the observation of the spectrum although the degeneracy of the emission efficiency with $G\mu$ makes extraction of the value more difficult in much of the parameter space.

Within our Models I and II, there are still theoretical uncertainties. The biggest one
at the moment continues to be the form of the power spectrum from individual strings, the so-called $P_j$ spectrum. Future investigations using an accurate model of gravitational backreaction would undoubtedly improve the situation before the arrival of the LISA data. On the other hand, these models also assume a somewhat minimalistic version of the string microphysics, therefore, obtaining a good handle on the slope of this
spectrum (our $q$ parameter) could be helpful in identifying possible deviations from these simple models. It is therefore interesting to point out the ability of our reconstructions to limit the possible values of the exponent of a power-law fit to this spectrum. 

Note, however, that our capacity to obtain a clear estimate of these other parameters depends on the value of $G \mu$. The reason for this is that for large enough values of the string tension the SGWB is almost flat and without many features.
In such a situation, the amplitude of the spectrum is therefore susceptible to degeneracies 
between the parameters that make it difficult to extract their preferred value from the data.

We have also explored the possibility of making precision measurements of the cosmological expansion history using the observation of the SGWB from strings. In particular, our analysis concludes that we should be able to identify an additional contribution to the number of DoF BSM of the order of $\Delta g ~\sim \mathcal{O}(1)-\mathcal{O}(100) $ for $G\mu \sim 10^{-10}$ provided that this deviation from the standard cosmology happens at a low enough background temperature, roughly around $T_{\Delta} ~\sim 0.005- 0.05 \,\text{GeV}$. This clearly opens up the possibility of using these observations to probe the matter content of
any extension of the SM involving new states provided their masses are low enough.
Similarly, we could also constrain possible deviations of the equation of state of the universe from the standard radiation domination. This would again require these to last until low enough temperatures up to a few GeV. However, our capability of using this SGWB as a precision probe of cosmological backgrounds is reduced for lower values of the string tension as well as higher values of the transition temperature. This is to be expected since the distinctive features of the signal would be pushed towards higher frequencies where its reconstruction will be harder. In fact, for $G\mu \sim \mathcal{O}(10^{-14})$, these signatures would already be outside of the LISA frequency range.

The observation of any of these extra features on the SGWB spectrum would give us additional
information not only about the origin of strings but also of the subsequent
thermal history of the universe. Here the presence of degeneracy would also need to be addressed since different models may give similar spectra. Some of these
 models may be quite different in nature as a model of strings produced during inflation
 and a model with a post-inflationary formation of strings with a period of 
non-conventional cosmology deep in the usual radiation era.

Even in the case of no detection, there is still very important information to be obtained from these observations.
The present results reported in the nanohertz frequency band impose constraints on models with strings for energy scales of the order of $G\mu \sim 10^{-10}$ \cite{NANOGrav:2023hvm}. Even if these constraints get tighter over the next decade or so, we will expect that by the time of the LISA launch a large portion of the parameter space for strings will remain unexplored. The LISA observatory will therefore be able to constrain a huge section of this parameter space in the intermediate region of scales, below the GUT scale and above the Electroweak (EW) scale. It is normally stated that cosmic strings are an expected by-product of phase transitions in the GUT scale~\cite{Jeannerot:2003qv}. It would also be interesting to see whether one can find a similar result at these lower scales. Moreover, we can also imagine some other scenario not linked to GUT with an intermediate scale phase transition that leads to strings like, for example, models associated with a dark sector~\cite{Long:2014mxa,Long:2014lxa}. Similarly, a recent set of models has been proposed
wherein the symmetry breaking takes place at this intermediate scale \cite{Dror:2019syi,King:2020hyd,Lazarides:2021uxv,Fu:2023mdu}. LISA will also be able to put important constraints on these models. Finally, it is interesting to note that the exploration of new physics we describe in this paper is complementary to the capabilities of LISA regarding the possible observation of the SGWB from phase transitions, which are focused on a much narrower band of energies around the EW scale \cite{LISACosmologyWorkingGroup:2022jok}.

While the main focus of this work is the characterization of the SGWB from CSs, it is worth commenting on the impact that a large signal might have on the measurements of the SGWBs of astrophysical origin. While for $\log_{\rm 10} (G \mu) \sim -16$, the SGWB due to CS is, in the LISA band, roughly of the same order as the foreground due to extragalactic mergers in eq.~\eqref{eq:Ext_template}, for larger values of $\log_{\rm 10} (G \mu)$, the latter gets completely masked by the CS signal. A similar behavior occurs for the galactic signal in eq.~\eqref{eq:SGWB_gal}, for values of $\log_{\rm 10} (G \mu) \gtrsim -11.5$. If these conditions are met, such a large SGWB of primordial origin would jeopardize the measurement of the astrophysical components, which are among the main targets of the LISA mission.

As mentioned in the introduction, there exist more intricate models associated with cosmic strings. Notable examples include models grounded in field theory objects with more internal structure, such as superconducting strings ~\cite{Witten:1984eb, Nielsen:1987fy} or metastable strings~\cite{Preskill:1992ck,Leblond:2009fq, Buchmuller:2020lbh, Buchmuller:2023aus,Chitose:2023dam}, as well as models derived from String Theory which give rise to cosmic superstring networks~\cite{Dvali:2003zj, Copeland:2003bj}. While some of these models' gravitational signatures have been investigated in the literature~\cite{Pourtsidou:2010gu, Buchmuller:2021mbb, Auclair:2022ylu, Rybak:2022sbo, Yamada:2022aax, Yamada:2022imq, Ellis:2023tsl}, further research is required to establish robust predictions in these scenarios. We leave the exploration of these models for future publications.

Finally, it is important to stress that even though we have considered the presence of astrophysical backgrounds, all our reconstructions assume that cosmic strings are the sole SGWB source of cosmological origin. In a realistic situation, we will also have to consider other possible sources from the early universe and evaluate our ability to extract the correct parameters of the string model in their presence. This will require us to combine all these different templates and perform a completely agnostic reconstruction in order to have a better idea of what kind of source fits the data better or alternatively put stringent constraints on scenarios that lead to the formation of cosmic string networks.

\section{Conclusions}
\label{sec:conclusions}

The initial stage of the LISA data analysis pipeline will possibly consist of an agnostic search for an SGWB with minimal assumptions. However, in order to fully utilize the findings and proceed to their interpretation an analysis using templates corresponding to individual sources will be necessary. With this project, we initiated a template bank which will be a first step in the efforts of the community to prepare for analysis of the data coming from the LISA mission in search for cosmic strings. We have included an analytical template based on the semi-analytical model in ref.~\cite{Sousa:2013aaa} (Model I) and the simulation-inferred BOS model~\cite{Blanco-Pillado:2013qja} (Model II). We used the \texttt{SGWBinner} code capable of simulating the data including the instrumental noise and astrophysical foregrounds together with the SGWB from cosmic strings in the form of our templates. This allows us to perform a state-of-the-art reconstruction of the injected signals from cosmic strings and an estimation of the possible impact of the mission for the models that predict them.

The two models agree on the range of cosmic string tensions that will be within reach of LISA. We find that tensions as low as $G\mu=10^{-16}$ can be probed in the presence of astrophysical foregrounds. With progress on methods dealing with their separation from the signal, values down to $G\mu=10^{-17}$ could be within reach. Beyond that threshold, further improvements are limited by the low amplitude of the signal falling below the sensitivity of LISA. While the spectra predicted in both models lead to very similar predictions for the same assumptions on the network, there are some key differences impacting the parameter reconstruction. Firstly, the initial sizes of cosmic string loops
produced by the network is a free parameter in Model I, while in Model II it is fixed based on the results of numerical simulations. This induces a degeneracy between the loop-size parameter $\alpha$ and $G\mu$ in Model I for strong signals which makes accurate determination of both challenging. We also verified the interplay between the two models using different ones for injection of the signal and reconstruction. Again we find a good agreement between the results for weaker signals where the small discrepancy in the reconstruction can be traced to the fitting of the free parameters of Model I. This is again not true for stronger signals with a larger tension, where the degeneracy of Model I parameters leads, in some instances, to results incompatible with the injected Model II signal. Nevertheless, Model I enables more agnostic searches for the cosmic string SGWB and may allow us to identify potential departures of the signal from the simulation-based expectations.

We have also directly shown that the LISA capability to perform cosmography provided a strong cosmic string signal is present in its frequency band. In the optimistic case of $G\mu=10^{-10}$, we find that ten additional DoF BSM can be probed at least up to temperatures around $T_{\Delta}\approx 0.05$ GeV. Although the reconstruction of this scenario is challenging, we find it is possible to constrain the number of new DoF from below and prove then the presence of new physics. More drastic modifications involving new exotic energy constituents would result in more pronounced modifications of the SGWB and may lead to more optimistic detection prospects and to a larger available parameter space in terms of higher temperature and lower tension.

It is important to stress once more that the conclusions of this paper are based on simple models of
field theory local cosmic strings which have been the focus of the most detailed studies so far in the literature.
We hope that the prospects of detection of a SGWB from strings by the LISA observatory will inspire new detailed
studies of some of the longstanding open issues in these models like the effects of gravitational backreaction in
these networks or the possible particle production by high curvature regions on the strings. Another possible issue with the templates in modified cosmology scenarios is that only simulations with single-component backgrounds have been performed and the evolution of the string network in dynamically evolving cosmologies remains somewhat uncertain.

Finally, as we have already mentioned there are plenty of ways one can enrich the phenomenology of these simple models, either by adding new DoF that could alter the dynamics and stability of the string or adding a new coupling of the string to other light fields (apart from gravity) that can be easily
radiated. All these modifications could potentially alter the expected SGWB presented here in a dramatic way. In
the coming years we anticipate an increase in dedicated analytic and numerical studies of these alternative models
that will allow us to obtain robust predictions for them to the extent that their SGWB templates could be integrated
with those presented in this paper.

LISA has the potential to revolutionize particle physics and cosmology with the discovery of an SGWB from the early universe. Much more work, however, will be needed for that potential to be fully realized. With the recent adoption of the mission by ESA the production of the device will now commence. Preparation for the data analysis and interpretation is now the main goal of the scientific community. With this work, we started the push towards that goal with respect to cosmic strings. However, the current results still rely on theoretical assumptions and could be improved with progress on the issues explained above. It is our hope that this paper will energise the scientific community in pursuit of our common goal of a groundbreaking discovery.

\paragraph*{Contributions}
We highlight individual contributions to this
manuscript.  
Conceptualization: JJBP, GN,  JMW, MP, LS.
Coordination: ML, GN.
Templates for Model I: LS, IR, and Model II: JMW, JJBP.
Software development: JMW, SK, MP.
Statistical analysis of Model I: MP, SK, and Model II: JMW.
Validation and writing: JJBP, YC, ML, LS, GN, JMW, SK, MP, IR.

\acknowledgments
We acknowledge the LISA Cosmology Group members for helpful discussions. We are especially grateful to the authors of Refs.~\cite{Caprini:2024hue, LISACosmologyWorkingGroupInflation} for the collaborative developments of the \texttt{SGWBinner} code shared among this and those works. We extend our gratitude to Pierre Auclair for helpful discussions and contributions during the initial stages of this paper.

JJBP is supported in part by the PID2021-123703NB-C21 grant funded by MCIN/ AEI /10.13039/501100011033/ and by ERDF; ``A way of making Europe'', the Basque Government grant (IT-1628-22) and the Basque Foundation for Science (IKERBASQUE). 
YC is supported by the US Department of Energy under award number DE-SC0008541.
SK is supported by the Spanish Atracci\'on de Talento contract no. 2019-T1/TIC-13177 granted by Comunidad de Madrid, the Spanish Research Agency (Agencia Estatal de Investigaci\'on) through the Grant IFT Centro de Excelencia Severo Ochoa No CEX2020-001007-S and the I+D grant PID2020-118159GA-C42 funded by MCIN/AEI/10.13039/501100011033, the Consolidaci\'on Investigadora 2022 grant CNS2022-135211 of CSIC, and Japan Society for the Promotion of Science (JSPS) KAKENHI Grant no. JP20H05853, and JP23H00110, JP24K00624. 
ML is supported by the Polish National Agency for Academic Exchange within the Polish Returns Programme under agreement PPN/PPO/2020/1/00013/U/00001 and the Polish National Science Center grant 2018/31/D/ST2/02048.
GN is partly supported by the grant Project. No. 302640 funded by the Research Council of Norway. 
MP acknowledges the hospitality of Imperial College London, which provided office space during some parts of this project. 
LS is supported by FCT -- Funda\c{c}\~{a}o para a Ci\^{e}ncia e a Tecnologia through contract No. DL 57/2016/CP1364/CT0001. Funding for this work has also been provided by FCT through the research grants UIDB/04434/2020 and UIDP/04434/2020 and through the R \& D project 2022.03495.PTDC -- \textit{Uncovering the nature of cosmic strings}.  
IR acknowledges support from the Grant PGC2022-126078NB-C21 funded by MCIN/AEI/ 10.13039/501100011033 and ``ERDF A way of making Europe'', as well as Grant DGA-FSE grant 2020-E21-17R from the Aragon Government and the European Union - NextGenerationEU Recovery and Resilience Program on `Astrof\'{\i}sica y F\'{\i}sica de Altas Energ\'{\i}as' CEFCA-CAPA-ITAINNOVA.

\appendix

\section{Complete template for Model I}
\label{sec:app_modelI}

In this appendix, we provide a detailed derivation of the analytical template for Model I, including the contributions of all relevant modes of emission. Although analytical approximations to the contribution of a single harmonic mode of emission were derived in ref.~\cite{Sousa:2020sxs}, to fully characterize the SGWB generated by cosmic strings, we need to sum the contributions of all relevant harmonic modes, which implies computing the summation
\begin{equation}
    \Omega ^a_{{\rm gw} }(f,q,G\mu,\alpha)=\Gamma \sum_{j=1}^N {\Omega}^{a \, j }_{{\rm gw} }(f,G\mu,\alpha)\,,
    \label{eq:summation}
\end{equation}
where $a=r$, $rm$ or $m$ label respectively the contribution of the loops that are created and decay in the radiation era, that of radiation-era loops that survive into the matter era and the contribution of the loops that are created in the matter era. This summation needs to be performed, as explained in Sec.~\ref{subsec:modelI}, up to $N(f)=f/f_{min,r}$ for $\mathrm{\Omega}_{\rm gw}^r$ and $\mathrm{\Omega}_{\rm gw}^{rm}$ and up to $N(f)=f/f_{min,m}$ in the case of $\mathrm{\Omega}_{\rm gw}^m$ . Since performing it explicitly is time-costly, here we resort to the Euler–Maclaurin summation formula, as it allows us to obtain fast and accurate results without performing the summation explicitly. According to this approximation~\cite{knopp1990theory} we should have that:
\begin{equation}
\label{E-M_app}
\sum_{j=1}^{N} s_j = \int_1^{N} s(j) dj + \frac{s(N)+s(0)}{2} + \sum_{k=1}^{\infty} \frac{B_{2k}}{(2k)!} \left( s^{(2k-1)}(N)-s^{(2k-1)}(0) \right)\,,
\end{equation}
where $B_k$ is a Bernoulli number and $s_j$ and $s_j^{(k)}$ represent respectively the terms of the sum (\ref{eq:summation}) and their $k$-th derivative. Note that, depending on the required precision, the infinite sum on the right-hand side of eq.~(\ref{E-M_app}) can be truncated at a given $k$. Here we include terms up to the third derivative $s^{(3)}(N)$. 

We will then resort to the analytical approximation to the contribution of the $j$-th mode of emission to the SGWB derived in ref.~\cite{Sousa:2020sxs} to construct a novel template that includes all harmonic modes of emission using eq.~\eqref{E-M_app}. The essential physical ingredients of this approximation are:
\begin{itemize}
    \item String loop parameters: from Model I (see Table~\ref{tbl:parameter-ranges}) we have tension $G\mu$, loop size $\alpha$, power spectral index $q$, and fuzziness $\mathcal{F}$; in addition, we require the loop rate of GW radiation $\Gamma$.
    \item Parameters of the velocity one-scale (VOS) model~\cite{Martins:1996jp,Martins:2000cs}: the RMS velocity of the cosmic string network, $\bar v$; the efficiency of the energy-loss mechanism of the cosmic string network, $\tilde{c}$; and the charateristic lengthscale of the network, $L=\xi t$.
    \item Cosmological parameters: the Hubble constant $H_0$; the relative energy density in matter and radiation, $\Omega_m$ and $\Omega_r$; the Planck time $t_{pl}$; and the power $\nu$ in the scale factor relation $a\propto t^\nu$.
\end{itemize}
From these physical parameters, it is helpful to define many auxiliary terms to simplify the eventual expressions for $\Omega_\text{gw}$. With $f$ being the frequency, these auxiliary terms are:
\begin{equation}
\begin{gathered}
    \epsilon_r=\frac{ \alpha}{\Gamma G\mu}\,,\quad\epsilon_m=\epsilon_r\frac{\xi_m}{\xi_r}\,,\quad \gamma_m=1+\epsilon_m^{-1}\,,\quad \beta_{m}=\frac{1}{\epsilon_m}\left(1+\frac{1}{\gamma_m}\right)\,,\\
     \mathcal{C}_i=\frac{\tilde c}{\sqrt{2}}\mathcal{F}\frac{v_i}{\xi_i^3}\,,\quad C_m=162\pi\mathcal{C}_m \frac{(1+\epsilon_m)}{ G\mu \epsilon_m}\left(\frac{H_0\Omega_m}{\Gamma f}\right)^2 \,,\quad D_i=\frac{2H_0\Omega_i^{1/2}}{ \nu_i\Gamma G\mu}\,,\\
    C_r=\frac{128}{9}\pi\mathcal{C}_r\Omega_r \left(1+\epsilon_r\right)^{3/2} \frac{G\mu}{\epsilon_r},\quad C_{rm}= 32\sqrt{3}\pi\mathcal{C}_r\left(1+\epsilon_r\right)^{3/2}\frac{H_0 \left(\Omega_m \Omega_r\right)^{3/4}}{\epsilon_r \Gamma  f} \,,\\
    {\tilde A_r}= \frac{D_r}{f}\frac{\Omega_m}{\Omega_r}\,,\quad {\tilde A_{rm}}= \frac{D_m}{f} \sqrt{\frac{\Omega_m}{\Omega_r}} \,,\quad {\tilde A_m}=\frac{D_m}{f}\,,\quad {\tilde A}_0=\frac{2\sqrt{2}}{f\Gamma}\sqrt{\frac{H_0\Omega_r^{1/2}}{t_{pl}\left(1+\epsilon_r\right)}}\,.
\end{gathered}
\end{equation}
Labels $i=r,m$ are used to refer to the values of the corresponding variables in the radiation and matter eras, respectively. In the radiation era, for standard cosmic strings, we have $\nu_r=1/2$, $\xi_r=0.271$, $v_r=0.662$ and $\mathcal{C}_r=5.4\mathcal{F}$, while in the matter era, $\nu_m=2/3$, $\xi_m=0.625$, $v_m=0.583$ and $\mathcal{C}_m=0.39\mathcal{F}$. Moreover, $\tilde{c}=0.23$~\cite{Martins:2000cs} in both radiation and matter eras. Note also that $f_{min,r}=D_r\Omega_m/(\epsilon_r\Omega_r)$ and $f_{min,m}=D_m / \epsilon_m$.

With these definitions, the results of ref.~\cite{Sousa:2020sxs} may be re-written as~\footnote{Note that here, as explained in section~\ref{subsec:modelI}, we re-express the results of ref.~\cite{Sousa:2020sxs} in terms of $\alpha$ instead of using the natural parameter of the model $\alpha_L$. Moreover, the constants in eq.~(\ref{eq:omegaj}) have been absorbed into the definition of $C_r$, $C_m$ and $C_{rm}$, as in ref.~\cite{Sousa:2020sxs}.}
\begin{equation}
    \label{Rad}
    \mathrm{\Omega}^{r \, j }_{{\rm gw} } = C_r \frac{j^{-q}}{\zeta(q)} \left[  \frac{1}{ \left( \tilde{A}_{r} j + 1 \right)^{3/2} }  -  \frac{1 }{ \left(\tilde{A}_{0} j + 1\right)^{3/2}}  \right],
\end{equation}
\begin{equation}
\begin{gathered}
    \label{Rad-Mat}
    \mathrm{\Omega}_{{\rm gw}}^{rm \, j } = C_{rm} \frac{j^{1-q}}{\zeta(q)} \Bigg( \frac{\left( \Omega_m / \Omega_r \right)^{1/4} }{ \left( \tilde{A}_{rm} j + 1 \right)^{1/2} } \left[ 2 + \frac{1}{ \tilde{A}_{rm} j + 1 } \right] \\
    -\frac{ 1 }{ \left( \tilde{A}_{m} j + 1 \right)^{1/2} } \left[ 2 + \frac{1}{ \tilde{A}_{m} j + 1 } \right]  \Bigg),
\end{gathered}
\end{equation}
\begin{equation}
\begin{gathered}
    \label{Mat}
    \mathrm{\Omega}_{{\rm gw} }^{m \, j } = C_m \frac{j^{2-q}}{\zeta(q)} \left[ \frac{1}{ \tilde{A}_m j + 1} + \frac{1}{ \tilde{A}_m j} - \beta_m  +  2 \log \left( \frac{ \gamma_m \tilde{A}_m j }{ \tilde{A}_m j + 1 } \right)  \right],
\end{gathered}
\end{equation}
where $\mathrm{\Omega}_{{\rm gw}}^{r \, j }$ represents the contribution of the loops that are created and decay in the radiation era, $\mathrm{\Omega}_{{\rm gw}}^{rm \, j }$ that of radiation-era loops that survive into the matter era and $\mathrm{\Omega}_{{\rm gw}}^{m \, j }$ corresponds to the contribution of the loops that are created in the matter era. 

Note that most of the terms in eqs.~(\ref{Rad})-(\ref{Mat}) may be written in the form $j^a (1+Aj)^{-b}$ for different values of $a$, $b$ and $A$ (e.g. the first term in eq.~\eqref{Rad} corresponds to $a=-q$, $b=3/2$ and $A=\tilde{A}_{r}$). The computation of the first term in eq.~\eqref{E-M_app} then involves the computation of integrals of the form:
\begin{equation}
    \label{Int}
    \mathcal{I}(a,b,A,x) \equiv \int \frac{x^{a}}{(1+A x)^b} dx = \frac{x^{1+a}}{1+a} \; {}_2F_1(1+a,b;2+a;-A x),
\end{equation}
where ${}_2F_1(\dots)$ is the Gauss hypergeometric function. Moreover, the derivatives needed to compute the last term in eq.~\eqref{E-M_app} yield:
\begin{equation}
    \label{D1}
    \mathcal{D}_1(a,b,A,x) \equiv \partial_x \left( \frac{x^{a}}{(1+A x)^b} \right) = x^{a-1} \frac{a + A x (a - b)}{(1+A x)^{1+b}},
\end{equation}
and
\begin{equation}
\begin{gathered}
    \label{D3}
    \mathcal{D}_3(a,b,A,x) \equiv \partial^3_x \left( \frac{x^{a}}{(1+A x)^b} \right) = \frac{x^{a}}{(1+A x)^b} \times \\ 
    \left( \frac{ a (1-a) (2-a) }{ x^3 } - \frac{ A^3 b (1+b) (2+b) }{ (1+A x)^3 } + \frac{3 a A b ( 1 - a + A x (2 - a + b) ) }{ x^2 (1 + A x)^2 }  \right).
\end{gathered}
\end{equation}
Combining these results, one finds that, including terms up to the third derivative, terms of this form lead to a contribution to the full spectrum of the form:
\begin{equation}
\begin{gathered}
    \label{Sum1App}
 \mathcal{M}(a,b,A,N) \equiv \sum_{j=1}^{N} \frac{j^a}{(A j + 1)^b} \approx \Delta \mathcal{I}(a,b,A,N,1)+\frac{1}{2 (1+A)^{b}} \\
    + \frac{N^a}{2 (1+A N)^{b} } + \frac{1}{12} \Delta \mathcal{D}_1(a,b,A,N,1) - \frac{1}{720} \Delta \mathcal{D}_3(a,b,A,N,1), 
\end{gathered}
\end{equation}
where we use the notation $\Delta F (a,b,A,N,1) = F (a,b,A,N) - F (a,b,A,1)$, for any function $F$.

There is only one term with a distinct form in $\mathrm{\Omega}_{\rm gw}^m$ (the last term in eq.~\eqref{Mat}, which involves a logarithm), but in this case the integral may also be performed analytically if we use (\ref{Int}) and integrate by parts:
\begin{equation}
\begin{gathered}
    \label{Int2}
      \mathcal{I}_l (a,b,A,x) \equiv \int x^{a} \log \left( \frac{b x}{A x + 1} \right) dx\\
      =\frac{1}{1+a}\left( x^{1+a} \log \left( \frac{b x}{A x + 1} \right) - \mathcal{I}(a,1,A,x)\right)\,.
\end{gathered}
\end{equation}

As to its derivatives, we have
\begin{equation}
\begin{gathered}
    \label{Dl1}
    \mathcal{D}_{l1}(a,b,A,x) \equiv \partial_x \left( x^{a} \log \left( \frac{b x}{1 + A x } \right) \right) \\
    =x^{a-1} \left( \frac{1}{1 + A x} + a \log \left( \frac{b x}{1 + A x} \right) \right),
\end{gathered}
\end{equation}
and
\begin{equation}
\begin{gathered}
    \label{Dl3}
    \mathcal{D}_{l3}(a,b,A,x) \equiv \partial^3_x \left( x^{a} \log \left( \frac{\alpha_{\epsilon} x}{1 + A x } \right) \right) = x^{a-3} \Big( \frac{ 2 }{(1 + A x)^3} \\
    + 3 \frac{ (a-2)}{(1 + A x)^2} + 3 \frac{ (2-a) (1-a) }{1 + A x} + (2-a) (1-a) a \log \left( \frac{b x}{1 + A x} \right) \Big).
\end{gathered}
\end{equation}

This term then gives rise to a contribution of the form:
\begin{equation}
\begin{gathered}
    \label{Sum1App2}
 \mathcal{N}(a,b,A,N) \equiv \sum_{j=1}^{N} j^a \log \left( \frac{b j}{1+A j} \right) \approx \Delta \mathcal{I}_l(a,b,A,N,1)+\frac{1}{2}\log \left( \frac{b}{1+A} \right)  \\
    + \frac{N^a}{2} \log \left( \frac{b N}{1+A N} \right) + \frac{1}{12} \Delta \mathcal{D}_{l1}(a,b,A,N,1) - \frac{1}{720} \Delta \mathcal{D}_{l3}(a,b,A,N,1).
\end{gathered}
\end{equation}

We then find that the radiation, radiation-matter and matter contributions to the SGWB --- including the contribution of all relevant harmonic modes --- may be expressed respectively as
\begin{equation}
\begin{gathered}
    \label{OmRad}
    \Omega_{\text{\rm gw}}^{r} (f, q, G\mu, \alpha) = \frac{C_r}{\zeta(q)} \left[ \mathcal{M} \left( -q,\frac{3}{2},\tilde{A}_r,N \right) - \mathcal{M} \left( -q,\frac{3}{2},\tilde{A}_{0},N \right) \right]\,, \\
     \Omega_{\text{\rm gw}}^{rm} (f, q, G\mu, \alpha) = \frac{C_{rm}}{\zeta(q)} \Bigg[  \frac{ 2 \mathcal{M} \left( 1-q,\frac{1}{2},\tilde{A}_{rm},N \right) +  \mathcal{M} \left( 1-q,\frac{3}{2},\tilde{A}_{rm},N \right) }{ \Omega_r^{1/4} / \Omega_m^{1/4} } \\
     - 2 \mathcal{M} \left( 1-q,\frac{1}{2},\tilde{A}_{m},N \right) - \mathcal{M} \left( 1-q,\frac{3}{2},\tilde{A}_{m},N \right) \Bigg]\,,\\
      \Omega_{\text{\rm gw}}^{m} (f, q, G\mu, \alpha) = \frac{C_m}{\zeta(q)} \Big[ \mathcal{M}\left(2-q,1,\tilde{A}_m,N \right) + \frac{\mathcal{H}^{(q-1)}_{N}}{\tilde{A}_m} - \beta_m \mathcal{H}^{(q-2)}_{N} \\
      + 2 \mathcal{N} \left( 2-q,\gamma_m \tilde{A}_m, \tilde{A}_m, N \right) \Big]\,,
\end{gathered}
\end{equation}
where $\mathcal{H}^{(q)}_N=\sum_{k=1}^N k^{-q}$ is the generalized (or hyper) harmonic number.

The approximation in eqs.~(\ref{Rad})-(\ref{Mat}) is only valid for $\alpha>\Gamma G\mu$. However, for $\alpha\le\Gamma G\mu$, the contribution of the first harmonic mode of emission to the SGWB simplifies to~\cite{Sousa:2014gka,Sousa:2020sxs}:

\begin{equation}
\mathrm{\Omega}^1_{\rm gw}(f, G\mu, \alpha)=\frac{64\pi}{3} G\mu \Omega_r \mathcal{C}_r+54\pi\frac{H_0\Omega_m^{3/2}}{\epsilon_m\Gamma}\frac{\mathcal{C}_m}{f}\left[1-\frac{D_m}{\epsilon_m}\frac{1}{f}\right]\,,
\end{equation}
for $f>f_{min,m}$.
In this case, the summation in eq.~(\ref{eq:summation}) can be computed analytically and yields
\begin{eqnarray}
\label{eq:smallalpha}
\mathrm{\Omega}_{\rm gw}(f, q, G\mu, \alpha)=\frac{64\pi}{3} G\mu \Omega_r \mathcal{C}_r+54\pi\frac{H_0\Omega_m^{3/2}}{\epsilon_m\Gamma}\frac{\mathcal{C}_m}{\zeta(q) f}\left\{\mathcal{H}^{(q-1)}_N-\frac{D_m}{\epsilon_m}\frac{\mathcal{H}^{(q-2)}_N}{f}\right\}\,,\nonumber
\end{eqnarray} 
with $N=f/f_{min,m}$ and for $f>f_{min,m}$.

Note however that eq. (\ref{OmRad}) does not yet include the impact of the change of the effective number of massless DoF as the universe expands. We can take this effect into consideration by breaking the integration in ref.~\cite{Sousa:2020sxs} into smaller intervals in which the measure of the number of massless DoF, $\Delta_r(a)$ (defined in eq.~(\ref{eq:deltarhoR})), remains roughly constant. In this case, \eqref{Rad} may be rewritten as:
\begin{equation}
\begin{gathered}
    \label{SGWB_rad2}
    \mathrm{\Omega}_{\rm gw}^{r\,1}(f,\alpha,G\mu) = \frac{128}{9} \pi \left( \frac{G \mu}{\Gamma} \right)^{1/2} \Omega_r C_r(\alpha) \Bigg[ \Delta_{r\,1} \left( \frac{1}{ B_1(a) + 1} \right)^{3/2} \Bigg|_{a = a_0^*}^{a = a^*_{1}} \\ 
    + \Delta_{r\,2} \left( \frac{1}{B_2(a) + 1} \right)^{3/2} \Bigg|_{a = a^*_{1}}^{a = a^*_{2}} + \dots + \Delta_{r\,N_{\text{dof}+1}} \left( \frac{1}{B_{N_{\text{dof}+1}} (a) + 1} \right)^{3/2} \Bigg|_{a = a^*_{N_{\text{dof}}}}^{a = a_{\text{eq}}}  \Bigg],
\end{gathered}
\end{equation}
where we defined $B_n(a) \equiv D_{r}\Delta^{1/2}_{r\,n} (a_0/a)$ and $C_r(\alpha)\equiv \mathcal{C}_r\left(\alpha + \Gamma G \mu \right)^{3/2}/\alpha $. Here we have approximated the changes in the effective number of relativistic DoF as a piecewise constant function with $N_{\text{dof}}$ steps occurring at values of scale factor $a^*_{1}, \cdots, a^*_{N_{\text{dof}}}$. $\Delta_{r\,n}$ is the (constant) value of $\Delta_r(a)$ between $a^*_{n-1}$ and $a^*_n$, while $a^*_{N_{\text{dof}}+1}=a_{\text{eq}}$ and the start of integration corresponds to $a^*_0= a_0 (2 H_0 \Omega_r^{1/2} (\epsilon_r+1) t_{\text{pl}})^{1/2}/( G \mu )$ (corresponding to the start of significant GW emission from cosmic string loops). In this case, the total contribution of loops that decay during the radiation era may be rewritten as:
\begin{equation}
\begin{gathered}
    \label{SGWB_rad4}
    \mathrm{\Omega}_{\rm gw}^r = C_r \sum_{n=1}^{N_{\text{dof}}+1}   \mathcal{M}_{\Delta n} \left( -q, \frac{3}{2}, \Delta^{1/2}_{r \, n} \mathcal{A}_{n}, N \right),
\end{gathered}
\end{equation}
where
\begin{equation}
\label{eq:ModelIdof-full}
    \begin{gathered}
\mathcal{M}_{\Delta n} \left( -q, \frac{3}{2}, \Delta^{1/2}_{r \, n} \mathcal{A}_{n}, N \right) \equiv  \Delta_{r \, n} \mathcal{M} \left( -q, \frac{3}{2}, \Delta^{1/2}_{r \, n} \mathcal{A}_{n}, N \right) \\
- \Delta_{r \, n} \mathcal{M} \left( -q, \frac{3}{2}, \Delta^{1/2}_{r \, n} \mathcal{A}_{n-1}, N \right),    
    \end{gathered}
\end{equation}
and $\mathcal{A}_{n} = D_r  a_0 / a^*_n /f$ . Since the impact of the change of the number of massless DoF on the SGWB is highly non-linear (because it also affects the evolution of the cosmic string network), we have used the full numerical computation of this spectrum as prescribed in~\cite{Sousa:2013aaa} to develop a fit to the numerical for the values of $\Delta_r$ and $a^*_n$ for the SM. This fit involves three steps in $\Delta_r(a)$ and the following seven parameters: $\Delta_{r \, 1} = 0.4065$, $\Delta_{r \, 2} = 0.4853$, $\Delta_{r \, 3} = 0.7085$, $\Delta_{r \, 4} = 1$, $a^*_1 = 5.8897\times 10^{16}G\mu\,a^*_0$, $a^*_2 = 7.2182\times 10^{19} G\mu \,a^*_0$, and  $a^*_3 = 7.5617\times 10^{21}G\mu\, a^*_0$.  Note that, since, for small cosmic string loops with $\alpha<\Gamma G\mu$, the effects of the change of the number of massless DoF only appears in frequencies that are outside of the LISA window, we do not include this effect in that case.

The template for Model I is then given by
\begin{equation}
    \mathrm{\Omega}_\text{\rm gw}(f, q, G\mu, \alpha)=\mathrm{\Omega}_{\rm gw}^r(f, q, G\mu, \alpha)+\Omega_{\text{\rm gw}}^{rm}(f, q, G\mu, \alpha)+\Omega_{\text{\rm gw}}^{m}(f, q, G\mu, \alpha)\,,
\end{equation}
for $\alpha>\Gamma G\mu$, where $\mathrm{\Omega}_{\rm gw}^r$ is given in eq.~(\ref{SGWB_rad4}) and $\Omega_{\text{\rm gw}}^{rm}$ and $\Omega_{\text{\rm gw}}^{m}$ in eq.~(\ref{OmRad}). For smaller values of $\alpha$, it is simply given by eq.~(\ref{eq:smallalpha}). Note that this template may also be used to approximate Model II by setting $\alpha=0.1$, $q=4/3$ and $\mathcal{F}=0.1$. In this case, the contribution of $\Omega_{\text{\rm gw}}^{m}$ is negligible for $G\mu$ below $\sim 10^{-11}$.

\section{Templates for Non-Standard pre-BBN Cosmologies}
\label{sec:app-prebbn}
In this appendix, we briefly outline how we have constructed the templates for non-standard pre-BBN cosmologies, introduced in section~\ref{ssec:mod-pre-bbn}. Let us start by considering the cases in which there are modifications to the equation of state of the early universe before the onset of radiation domination described in section \ref{subsubsec:eos} (or the strings are created during an inflationary era in section \ref{subsubsec:inflation}). In this case, we resort to Model I to develop the templates. As shown in~\cite{Cui:2017ufi}, the spectrum in the fundamental mode is of the form in eq.~(\ref{eq:modomega1}). As in Appendix~\ref{sec:app_modelI}, we then need to sum over the contributions of all relevant harmonic modes of emission to obtain a complete template for the signal. It is straightforward to show that, for $f<f_{\rm rd}$, the spectra is unaffected by the change in the cosmological background and thus the template for model I may be used. For $f>f_{\rm rd}$, however, the full SGWB, including all the harmonic modes of emission of loops, is given by
\begin{eqnarray} \label{eq:modomegafull}
   &  & \Omega_{\rm gw}^*(f>f_{\rm rd},q,G\mu,\alpha,f_{\rm rd},d)  =  \Gamma\sum_{j=1}^{\infty}\frac{j^{-q}}{\zeta(q)}{\Omega_{\rm gw}^*}^1 (f/j,G\mu,\alpha,f_{\rm rd}, d) = \nonumber \\
    & = &  \Gamma\sum_{j=1}^{N_{\rm rd}} \frac{j^{d-q}}{\zeta(q)} \Omega_{\rm gw}^1 (f_{\rm rd},\alpha,G\mu) \left(\frac{f_{\rm rd}}{f}\right)^d +\Gamma\sum_{j=N_{\rm rd}}^{\infty}\frac{j^{-q}}{\zeta(q)}\mathrm{\Omega}_{\rm gw}^1(f/j,G\mu,\alpha) = \nonumber \\
    & = & \frac{\Omega_{\rm rd}}{\zeta(q)} \left(\frac{f_{\rm rd}}{f}\right)^d \mathcal{H}^{(q-d)}_{N_{\rm rd}}+\Omega_{\rm gw}(f,q,G\mu,\alpha) -\Gamma\sum_{j=1}^{N_{\rm rd}}\frac{j^{-q}}{\zeta(q)}{\Omega_{\rm gw}}^1(f/j,G\mu,\alpha)\,,\label{eq:modeos-full}
\end{eqnarray}
where $N_{\rm rd}=f/f_{\rm rd}$ and the last term may be computed as described in Appendix~\ref{sec:app_modelI} after substituting $N$ by $N_{\rm rd}$. In the last equality, we introduced the definition of generalized harmonic number in the first term and used the fact that $\sum_{j=N_{\rm rd}}^{+\infty}=\sum_{j=1}^{+\infty}-\sum_{j=1}^{N_{\rm rd}}$ and \eqref{eq:summation} in the last term.

Note that the generalized harmonic numbers may be approximated by~\footnote{In our templates, we include terms up to order $N^{-r-1}$ in the computation of generalized harmonic numbers.}

\begin{equation}\label{hyperharmonics}
\mathcal{H}^{(r)}_N\approx \zeta(r)+\frac{N^{1-r}}{1-r}+\mathcal{O}\left(N^{-r}\right)\,.
\end{equation}
 For $f\gg f_{\rm rd}$, the last two terms in eq.~(\ref{eq:modomegafull}) roughly cancel each other and thus $\Omega_{\rm gw}^*$ is determined by the behaviour of $\mathcal{H}^{(q-d)}_{N_{\rm rd}}$. As a matter of fact, for $d< q-1$, in this limit, this yields simply $\zeta(q-d)$. The full spectrum then has the same slope as the fundamental mode: $\Omega_{\rm gw}^*\propto f^{-d}$. However, for $d\ge q-1$, the second term in eq.~(\ref{hyperharmonics}) becomes dominant and the slope of the full spectrum is different from that of the fundamental mode and asymptotically approaches $\Omega_{\rm gw}^*\propto f^{q-1}$, as shown in eq.~(\ref{eq:modeos-app}).

 As to the scenario in which there are additional relativistic DoFs, we assume that these cause a modification of $g_*$ of the form in eq.~(\ref{eq:dofmod}), corresponding to a rapid change in $\Delta_r$ at $T_\Delta$, and that this does not have a significant impact on the large-scale dynamics of cosmic string networks. For Model I, this corresponds to including an additional step in the integration in eq.~(\ref{SGWB_rad4}) at $T_\Delta$ and recalculating the contribution of the radiation era loops to the spectrum in eq.~(\ref{eq:ModelIdof-full}). The procedure is identical for Model II --- an additional step is also included in $\Delta_r$ at $T_\Delta$ --- but, in this case, the re-computation of the $\Omega_{\rm gw}$ is carried out numerically.

\bibliographystyle{JHEP}
\bibliography{references.bib}

\end{document}